\documentclass[sigconf, screen]{acmart}

\AtBeginDocument{%
  }

\setcopyright{acmlicensed}
\copyrightyear{2025} 
\acmYear{2025} 
\setcopyright{acmlicensed}\acmConference[MM '25]{Proceedings of the 33rd ACM International Conference on Multimedia}{October 27--31, 2025}{Dublin, Ireland}
\acmBooktitle{Proceedings of the 33rd ACM International Conference on Multimedia (MM '25), October 27--31, 2025, Dublin, Ireland}
\acmDOI{10.1145/3746027.3755468}
\acmISBN{979-8-4007-2035-2/2025/10}

\acmSubmissionID{3958}

\usepackage{booktabs}
\usepackage{bookmark}
\usepackage{calc}
\usepackage{multicol}
\usepackage{multirow}%
\usepackage{color}
\usepackage{siunitx}
\usepackage{subcaption} % 

\usepackage{algorithm}
\usepackage{algorithmic}

\usepackage{tocloft}

\newcommand{\cmmnt}[1]{}

\usepackage{amsmath,amsfonts}
\usepackage{amsopn,bm}

\usepackage{nicefrac}       % compact symbols for 1/2, etc.

\usepackage{epsfig}
\usepackage{graphicx}
\usepackage{float}

\usepackage{bm,xspace}
\usepackage{comment}
\usepackage{verbatim}
\usepackage{multirow}
\usepackage{makecell}
\usepackage{array}
\usepackage{url}
\usepackage{booktabs}
\usepackage{etoolbox,siunitx}
\usepackage{calc}
\usepackage{pifont,hologo}
 
\usepackage{nicefrac}

\usepackage{arydshln}
\usepackage[utf8]{inputenc} % allow utf-8 input
\usepackage[T1]{fontenc}    % use 8-bit T1 fonts
\usepackage{url}            % simple URL typesetting
\usepackage{amsfonts}       % blackboard math symbols
\usepackage{nicefrac}       % compact symbols for 1/2, etc.
\usepackage{microtype}      % microtypography
\usepackage[american]{babel}
\usepackage{enumitem}
\usepackage{siunitx}

\usepackage{enumitem}

\usepackage[super]{nth}
\usepackage{nicefrac}
\robustify\bfseries

\usepackage{dsfont}

\usepackage{listings}
\usepackage{xcolor}
\definecolor{codegreen}{rgb}{0,0.6,0}
\definecolor{codegray}{rgb}{0.5,0.5,0.5}
\definecolor{codepurple}{rgb}{0.58,0,0.82}
\definecolor{backcolour}{rgb}{1.0,1.0,1.0}

\lstdefinestyle{mystyle}{
    commentstyle=\color{codegreen},
    keywordstyle=\color{magenta},
    numberstyle=\tiny\color{codegray},
    stringstyle=\color{codepurple},
    basicstyle=\ttfamily\footnotesize,
    breakatwhitespace=false,         
    breaklines=true,                 
    captionpos=b,                    
    keepspaces=true,                 
    numbersep=0pt,                  
    showspaces=false,                
    showstringspaces=false,
    showtabs=false,                  
    tabsize=2,
    linewidth=.99\textwidth,
    xleftmargin=0.01cm
}
\usepackage{mathtools}

\newcommand{\vct}[1]{\boldsymbol{#1}} % vector
\newcommand{\mat}[1]{\boldsymbol{#1}} % matrix
\newcommand{\field}[1]{\mathbb{#1}}
\newcommand{\R}{\field{R}} % real domain
\newcommand{\ProbOpr}[1]{\mathbb{#1}}
\newcommand{\expect}[2]{%
\ifthenelse{\equal{#2}{}}{\ProbOpr{E}_{#1}}
{\ifthenelse{\equal{#1}{}}{\ProbOpr{E}\left[#2\right]}{\ProbOpr{E}_{#1}\left[#2\right]}}} % Expectation: syntax: E{1}{2} = E_1[2], E{}{2}=E[2], E{1}{} = E_1
\newcommand{\var}[2]{%
\ifthenelse{\equal{#2}{}}{\ProbOpr{VAR}_{#1}}
{\ifthenelse{\equal{#1}{}}{\ProbOpr{VAR}\left[#2\right]}{\ProbOpr{VAR}_{#1}\left[#2\right]}}} % Expectation: syntax: V{1}{2} = V_1[2], V{}{2}=V[2], V{1}{} = V_1
\newcommand{\ones}{\vct{1}}

\newcommand{\va}{\vct{a}}
\newcommand{\vb}{\vct{b}}

\newcommand{\vf}{\vct{f}}

\newcommand{\vq}{\vct{q}}

\newcommand{\vt}{\vct{t}}

\newcommand{\mH}{\mat{H}}

\newcommand{\mS}{\mat{S}}

\newcommand{\sB}{\mathcal{B}}

\newcommand{\sL}{\mathcal{L}}

\newcommand{\sQ}{\mathcal{Q}}
\newcommand{\sS}{\mathcal{S}}
\newcommand{\sT}{\mathcal{T}}

\newcommand{\mxi}{\mat{\xi}}
\newcommand{\vpi}{\vct{\pi}}
\newcommand{\mPi}{\mat{\Pi}}

\newcommand{\valpha}{\vct{\alpha}}
\newcommand{\vbeta}{\vct{\beta}}

\makeatletter
\DeclareRobustCommand\onedot{\futurelet\@let@token\@onedot}
\def\@onedot{\ifx\@let@token.\else.\null\fi\xspace}

\makeatother

\newcommand{\symtext}[2]{\textsc{#1}#2}
\newcommand{\eat}[1]{{}}

\newcommand\mypara[1]{\vspace{1mm}\noindent\textbf{#1}}

\newtheorem{proposition}{Proposition}

\usepackage{array}
\usepackage{xcolor, colortbl}

\definecolor{Gray}{gray}{0.5}

\newlength\savewidth

\newcolumntype{x}[1]{>{\centering\arraybackslash}p{#1pt}}
\newcolumntype{y}[1]{>{\raggedright\arraybackslash}p{#1pt}}
\newcolumntype{z}[1]{>{\raggedleft\arraybackslash}p{#1pt}}

\definecolor{grey}{rgb}{0.8, 0.8, 0.8}

\begin{document}
%%
%% The "title" command has an optional parameter,
%% allowing the author to define a "short title" to be used in page headers.
\title{Hubness Reduction with Dual Bank Sinkhorn Normalization for Cross-Modal Retrieval}

%%
%% The "author" command and its associated commands are used to define
%% the authors and their affiliations.
%% Of note is the shared affiliation of the first two authors, and the
%% "authornote" and "authornotemark" commands
%% used to denote shared contribution to the research.
% \author{Ben Trovato}
% \authornote{Both authors contributed equally to this research.}
% \email{trovato@corporation.com}
% \orcid{1234-5678-9012}
% \author{G.K.M. Tobin}
% \authornotemark[1]
% \email{webmaster@marysville-ohio.com}
% \affiliation{%
%   \institution{Institute for Clarity in Documentation}
%   \city{Dublin}
%   \state{Ohio}
%   \country{USA}
% }

\author{Zhengxin Pan}
\affiliation{%
  \institution{Zhejiang Key Lab of Accessible Perception \& Intelligent Systems, Zhejiang University}
  \city{Hangzhou}
  \country{China}
  }
\email{panzx@zju.edu.cn}

\author{Haishuai Wang}
\authornote{Corresponding author: Haishuai Wang.}
\affiliation{%
  \institution{Zhejiang Key Lab of Accessible Perception \& Intelligent Systems, Zhejiang University}
  \city{Hangzhou}
  \country{China}
  }
\email{haishuai.wang@zju.edu.cn}

\author{Fangyu Wu}
\affiliation{%
  \institution{School of Advanced Technology, Xi'an Jiaotong-Liverpool University}
  \city{Suzhou}
  \country{China}
  }
\email{Fangyu.wu02@xjtlu.edu.cn}

\author{Peng Zhang}
\affiliation{%
  \institution{Cyberspace Institute of Advanced Technology, Guangzhou University}
  \city{Guangzhou}
  \country{China}
  }
\email{p.zhang@gzhu.edu.cn}

\author{Jiajun Bu}
\affiliation{%
  \institution{Zhejiang Key Lab of Accessible Perception \& Intelligent Systems}
  \city{Zhejiang University}
  \country{China}
  }
\email{bjj@zju.edu.cn}

% \author{Valerie B\'eranger}
% \affiliation{%
%   \institution{Inria Paris-Rocquencourt}
%   \city{Rocquencourt}
%   \country{France}
% }

% \author{Aparna Patel}
% \affiliation{%
%  \institution{Rajiv Gandhi University}
%  \city{Doimukh}
%  \state{Arunachal Pradesh}
%  \country{India}}

% \author{Huifen Chan}
% \affiliation{%
%   \institution{Tsinghua University}
%   \city{Haidian Qu}
%   \state{Beijing Shi}
%   \country{China}}

% \author{Charles Palmer}
% \affiliation{%
%   \institution{Palmer Research Laboratories}
%   \city{San Antonio}
%   \state{Texas}
%   \country{USA}}
% \email{cpalmer@prl.com}

% \author{John Smith}
% \affiliation{%
%   \institution{The Th{\o}rv{\"a}ld Group}
%   \city{Hekla}
%   \country{Iceland}}
% \email{jsmith@affiliation.org}

% \author{Julius P. Kumquat}
% \affiliation{%
%   \institution{The Kumquat Consortium}
%   \city{New York}
%   \country{USA}}
% \email{jpkumquat@consortium.net}

%%
%% By default, the full list of authors will be used in the page
%% headers. Often, this list is too long, and will overlap
%% other information printed in the page headers. This command allows
%% the author to define a more concise list
%% of authors' names for this purpose.
% \renewcommand{\shortauthors}{Anonymous Authors}

%%
%% The abstract is a short summary of the work to be presented in the
%% article.
\begin{abstract}
The past decade has witnessed rapid advancements in cross-modal retrieval, with significant progress made in accurately measuring the similarity between cross-modal pairs. However, the persistent hubness problem, a phenomenon where a small number of targets frequently appear as nearest neighbors to numerous queries, continues to hinder the precision of similarity measurements. Despite several proposed methods to reduce hubness, their underlying mechanisms remain poorly understood. To bridge this gap, we analyze the widely-adopted Inverted Softmax approach and demonstrate its effectiveness in balancing target probabilities during retrieval. Building on these insights, we propose a probability-balancing framework for more effective hubness reduction. We contend that balancing target probabilities alone is inadequate and, therefore, extend the framework to balance both query and target probabilities by introducing Sinkhorn Normalization (SN). Notably, we extend SN to scenarios where the true query distribution is unknown, showing that current methods, which rely solely on a query bank to estimate target hubness, produce suboptimal results due to a significant distributional gap between the query bank and targets. To mitigate this issue, we introduce Dual Bank Sinkhorn Normalization (DBSN), incorporating a corresponding target bank alongside the query bank to narrow this distributional gap. Our comprehensive evaluation across various cross-modal retrieval tasks, including image-text retrieval, video-text retrieval, and audio-text retrieval, demonstrates consistent performance improvements, validating the effectiveness of both SN and DBSN. All codes are publicly available at \url{https://github.com/ppanzx/DBSN}.
\end{abstract} 

%%
%% The code below is generated by the tool at http://dl.acm.org/ccs.cfm.
%% Please copy and paste the code instead of the example below.
%%
  

%%
%% Keywords. The author(s) should pick words that accurately describe
%% the work being presented. Separate the keywords with commas.
%% A "teaser" image appears between the author and affiliation
%% information and the body of the document, and typically spans the
%% page.
% \begin{teaserfigure}
%   \includegraphics[width=\textwidth]{sampleteaser}
%   \caption{Seattle Mariners at Spring Training, 2010.}
%   \Description{Enjoying the baseball game from the third-base
%   seats. Ichiro Suzuki preparing to bat.}
%   \label{fig:teaser}
% \end{teaserfigure}

% \received{20 February 2007}
% \received[revised]{12 March 2009}
% \received[accepted]{5 June 2009}

%%
%% This command processes the author and affiliation and title
%% information and builds the first part of the formatted document.
\maketitle

\begin{figure}[tbp]
    \centering
    \begin{subfigure}[b]{0.235\textwidth}
        \centering
        \includegraphics[width=\textwidth]{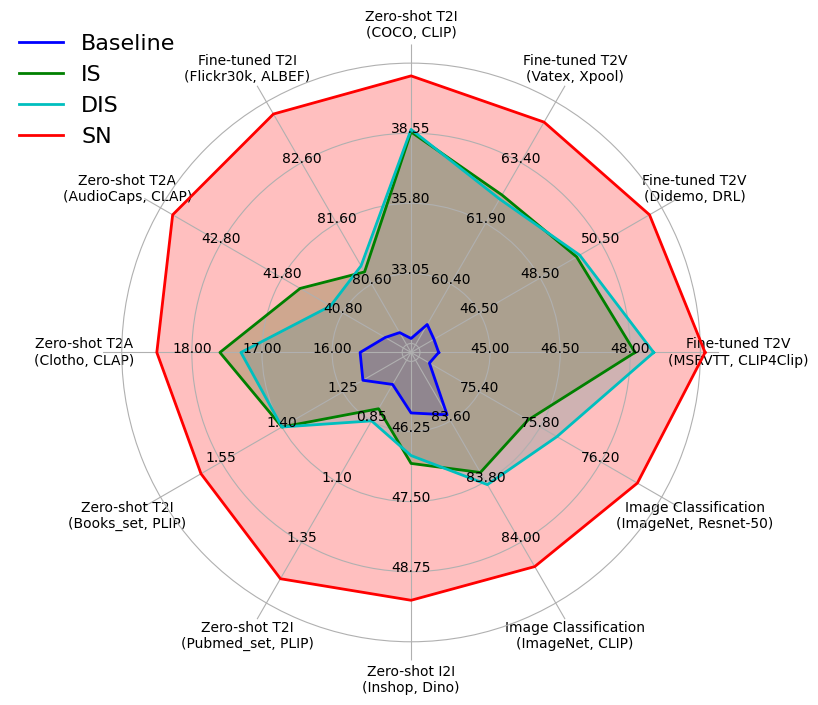}
    \end{subfigure}
    \begin{subfigure}[b]{0.235\textwidth}
        \centering
        \includegraphics[width=\textwidth]{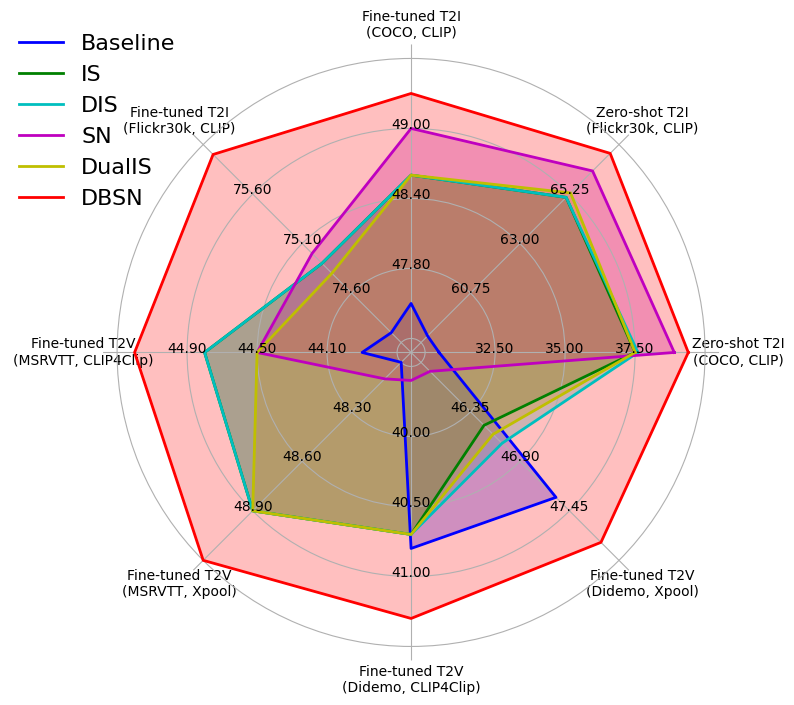}
    \end{subfigure}
    \caption{Retrieval comparisons under different query conditions. \textbf{Left:} SN vs current methods in query-aware scenarios (directly taking testing queries as query bank); \textbf{Right:} DBSN vs current methods in query-agnostic scenarios (leveraging training queries as query bank). Quantitative results can be found in \S~\ref{sec:4-exper} and our Appendix.}
    \label{fig:radar}
\end{figure}

\section{Introduction}
\label{sec:intro}

Cross-modal retrieval involves identifying the nearest target from a target gallery in one modality based on a query from another modality. The primary challenge lies in accurately measuring the similarity between cross-modal pairs, which necessitates bridging both the heterogeneous and semantic gaps. Significant progress has been made over the past decade, particularly with advancements in visual-language pre-training~\cite{radford2021learning, li2021align, jia2021scaling, li2022blip}. Despite these developments, the hubness problem, a critical issue that undermines the precision of similarity measurements and is pervasive in current cross-modal retrieval methods, remains unresolved.

\begin{figure*}[tbp]
    \centering
    \includegraphics[width=1\linewidth]{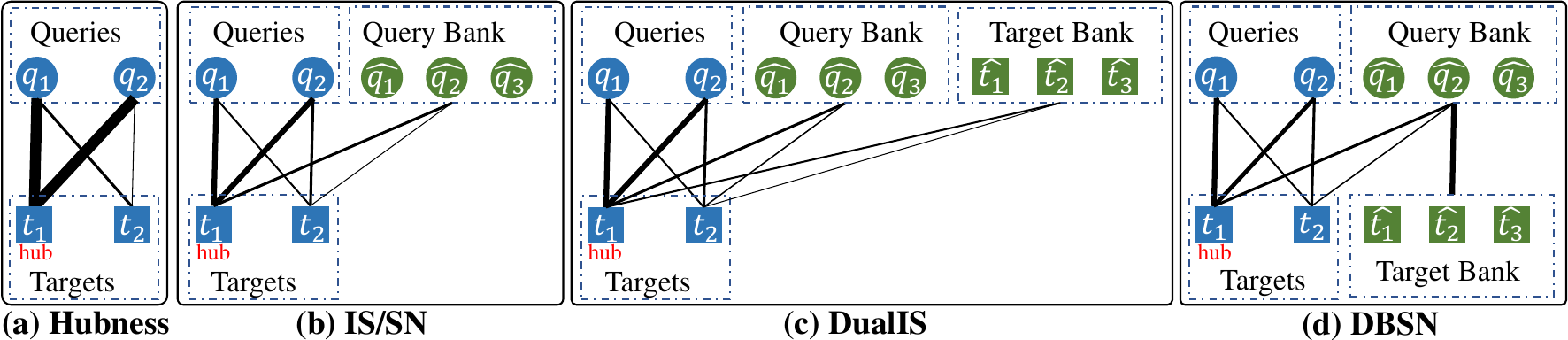}\vspace{-0.2cm}
    \caption{Structural comparisons between our SN/DBSN and current methods: \textbf{(a)} The hubness problem where the \textit{hub}($t_1$) is the nearest neighbor of multiple queries($q_1$, $q_2$), compromising retrieval precision. \textbf{(b)} IS/SN alleviates the hubness problem through query bank normalization. \textbf{(c)} DualIS expands queries to narrow the query-target gap but enlarges the query-query gap. \textbf{(d)} DBSN narrows the query-target gap while preserving the query-query gap by expanding targets instead of queries.}
    \label{fig:variants}
\end{figure*}

The hubness problem refers to a phenomenon where a small subset of \textit{hub} targets frequently emerge as nearest neighbors to numerous queries, whereas some \textit{non-hub} targets are rarely selected during retrieval~\cite{radovanovic2010hubs}, as depicted in Figure~\ref{fig:variants}(a). This problem arises from the spatial centrality~\cite{hara2016flattening} and the asymmetric nearest neighbor relations~\cite{schnitzer2012local}. To mitigate hubness, existing methods fall into two paradigms: centering and scaling. Centering approaches address spatial centrality, whereas scaling approaches correct asymmetric relations. In cross-modal retrieval, scaling methods have demonstrated superior performance over centering methods~\cite{bogolin2022cross}; however, their underlying mechanisms remain poorly understood. In this work, we analyze Inverted Softmax (IS)~\cite{smith2017offline}, a widely adopted scaling approach, to investigate its operational dynamics. Our analysis reveals that IS effectively balances the retrieval probabilities of targets. Inspired by this insight, we propose a probability-balancing framework for hubness reduction, which replaces the original similarity matrix with a doubly stochastic matrix that ensures uniform retrieval probabilities across all targets.

In this framework, we establish that IS can be formulated as a specialized instance of balancing target probabilities through injecting hubness-compensation terms on the target dimension. However, we reveal that solely balancing target probabilities while leaving query probabilities unconstrained is fundamentally limited. Neglecting the query distribution may introduce systematic bias in hubness estimation. To address this limitation, we propose to simultaneously balance both query and target distributions, seeking to derive a doubly stochastic matrix. Within our probability-balancing framework, this objective corresponds to solving an entropy-constrained optimal transport problem. We leverage the Sinkhorn-Knopp algorithm~\cite{cuturi2013sinkhorn} to obtain the solution, thus terming this approach Sinkhorn Normalization (SN). Consistent with IS, we prove that the core mechanism of SN functions through dual hubness compensation on queries and targets, enforcing balanced joint probabilities.

We further investigate the efficacy of SN in query-agnostic scenarios, where only a single query is available under unknown distribution conditions, emulating real-world search engine deployments. An intuitive solution for these cases involves constructing a query bank to approximate the ground-truth query distribution for target hubness estimation~\cite{bogolin2022cross}. However, composing such a query bank is not only laborious but also not effective enough due to the nontrivial query-target distribution gap between the query bank and targets. DualIS~\cite{wang2023balance} attempts to bridge this gap via a target bank, yet yields limited improvements as it amplifies the query-query divergence between ground-truth queries and the augmented bank. Our proposed Dual Bank Sinkhorn Normalization~(DBSN) strategically integrates the target bank on the target side instead of the query side, effectively reducing the query-target gap without enlarging the query-query gap. 

As shown in Figure~\ref{fig:radar}, consistent improvements across various retrieval tasks demonstrate the effectiveness of our SN and DBSN. Our contributions are summarized as follows:

(1) We propose a probability-balancing framework for hubness reduction in cross-modal retrieval. Our framework reveals that IS operates by balancing target probabilities, thereby mitigating the hubness problem.

(2) Within this framework, we identify the limitation of balancing target probabilities exclusively and resolve this through joint balancing query and target probabilities. To this end, we introduce Sinkhorn Normalization (SN).

(3) We extend SN to query-agnostic scenarios and demonstrate that single-bank SN is suboptimal due to a significant query-target gap. We further propose Dual Bank Sinkhorn Normalization (DBSN), enhancing single-bank SN by narrowing the query-target gap with a target bank.

\section{Related Works}
\label{sec:2-related}

\subsection{Cross-Modal Retrieval.} 

In this work, we investigate cross-modal retrieval, which involves identifying the most relevant target from a target gallery in one modality based on queries from distinct modalities. The key challenge lies in precise cross-modal similarity computation, manifesting through three primary aspects: the heterogeneous gap, the semantic gap, and the hubness problem. Notably, the first two challenges have been largely mitigated by Visual Semantic Embedding (VSE)~\cite{frome2013devise}, which pioneers dual encoders to project raw multimodal data into a shared space to bridge the heterogeneous gap and employs contrastive learning to align semantic representations.

Despite VSE's introduction over a decade ago, it still remains a cornerstone framework for multi-modal pre-training~\cite{radford2021learning,li2021align}, with fine-tuning on downstream datasets becoming the de facto standard for tasks including text-to-image~\cite{chun2023improved}, text-to-video~\cite{bai2025bridging, shen2024tempme}, and text-to-audio retrieval~\cite{wu2023large}. While recent advances focus on the adaptation of pre-trained models~\cite{fu2024linguistic, shen2024tempme, yang2024dgl}, the hubness problem remains largely unaddressed, indicating substantial improvements.

\subsection{The Hubness Problem.} 
Formally, the hubness problem refers to the phenomenon where a small proportion of targets appear as nearest neighbors to numerous queries, becoming \textit{hubs}, while some targets, termed \textit{non-hubs}, are rarely selected during retrieval~\cite{radovanovic2010hubs}. It emerges as an intrinsic property of the data distribution in high-dimensional space under the widely used assumptions such as 1) independent and identically distributed (i.i.d.) data and 2) the data follows a symmetric distance metric. Due to the non-linearity of neural networks, embeddings of i.i.d. data encoded by these networks tend to cluster within a narrow core of the hyperspace, resulting in the phenomenon of spatial centrality~\cite{hara2016flattening}. Under the influence of the symmetric metrics, spatial centrality causes samples near the center of the dataset to appear closer to all other samples, resulting in asymmetric nearest neighbor relationships~\cite{schnitzer2012local}, exacerbating the hubness problem. 

% Current hubness reduction methods can be broadly categorized into two paradigms: centering and scaling. The centering paradigm aims to alleviate spatial centrality by promoting uniform data distributions in high-dimensional spaces. However, these methods often require task-specific designs, such as specialized network architectures~\cite{fei2021z}, objective functions~\cite{trosten2023hubs}, or embedding configurations~\cite{ramasinghe2024accept}, limiting their applicability. In contrast, the scaling paradigm addresses the hubness problem by introducing an asymmetric metric that assigns adaptive weights to targets, thereby counteracting asymmetry in nearest neighbor relationships. This metric is typically implemented by compensating for the estimated target hubness using query information, without requiring architectural modifications or task-specific training. Notably, scaling methods demonstrate superior effectiveness over centering methods in cross-modal tasks~\cite{bogolin2022cross, guo2025metanerv, chen2024deepasd, chen2025multi}.

Current hubness reduction methods can be broadly categorized into two paradigms: centering and scaling. The centering paradigm aims to alleviate spatial centrality by promoting uniform data distributions in high-dimensional spaces. However, these methods often require task-specific designs, such as specialized network architectures~\cite{fei2021z}, objective functions~\cite{trosten2023hubs}, or embedding configurations~\cite{ramasinghe2024accept}, limiting their applicability. In contrast, the scaling paradigm addresses the hubness problem by introducing an asymmetric metric that assigns adaptive weights to targets, thereby counteracting asymmetry in nearest neighbor relationships. This metric is typically implemented by compensating for the estimated target hubness using query information, without requiring architectural modifications or task-specific training. Notably, scaling methods demonstrate superior effectiveness over centering methods in cross-modal retrieval tasks~\cite{bogolin2022cross}.

\subsection{Scaling Methods for Hubness Reduction.} 
Previous work~\cite{feldbauer2019comprehensive} comprehensively compares classical scaling methods, such as local scaling~\cite{zelnik2004self} and global scaling~\cite{schnitzer2012local}. However, these methods suffer from quadratic complexity, making them impractical for large datasets. For large-scale retrieval tasks, \cite{bogolin2022cross} evaluates scaling approaches, including Globally-Corrected (GC)~\cite{dinu2014improving}, Cross-Domain Similarity Local Scaling (CSLS)~\cite{lample2018word} and Inverted Softmax~\cite{smith2017offline} for similarity normalization. Although these methods demonstrate empirical effectiveness, their underlying mechanisms remain inadequately understood. Moreover, they focus solely on mitigating target hubness while overlooking query hubness, thereby confining their applicability to query-aware scenarios.

Closely related to our work are DIS~\cite{bogolin2022cross} and DualIS~\cite{wang2023balance}. While DIS attempts to enhance IS through query pruning, it fails to narrow the query-target gap. DualIS partially mitigates the query-target gap but inadvertently amplifies query-query divergence. In contrast, SN achieves simultaneous query-target probability balancing, demonstrating superior performance in query-aware scenarios over IS, while DBSN effectively narrows the query-target gap without enlarging the query-query gap, thus improving SN in query-agnostic scenarios. The architectural distinctions between SN/DBSN and other methods are illustrated in Figure~\ref{fig:variants}. 

Notably, while both serving as post-processing techniques, scaling-based methods (e.g., IS/SN) differ fundamentally from reranking approaches~\cite{nogueira2019passage} by avoiding iterative query-reconstruction requirements. The plug-and-play capability of scaling-based methods boosts efficiency over reranking approaches.

\section{Method}
\label{sec:3-method}

\subsection{Preliminary.}
Taking text-to-image retrieval as an example, state-of-the-art retrieval models like CLIP~\cite{radford2021learning} encode $m$ textual queries into a normalized query embedding set $\sQ=\{\vq_i\in\R^d\mid\|\vq_i\|_2=1,i\in[1,\cdots,m]\}$ and project $n$ candidate images into a target embedding set $\sT=\{\vt_j\in\R^d\mid\|\vt_j\|_2=1,j\in[1,\cdots,n]\}$. In this $d$-dimensional hyperspherical space, pairwise text-image similarities are calculated through matrix multiplication $\mathop{sim}(\sQ,\sT)=\sQ^\top\sT$, resulting in a similarity matrix $\sS\in\R^{m\times n}$ where $\sS_{i,j}=\vq_i^\top\vt_j$. The nearest-neighbor image $\vt_k$ for query $\vq_i$ is retrieved by ranking the $i$-th row of $\sS$, with $k = \mathop{arg~max}\limits_{j=[1,\cdots,n]} \sS_{i,j}$.

To mitigate the hubness problem in $\sS$, IS employs query-wise normalization with a temperature $\tau\geq 0$:
\begin{align}
    \centering
    \begin{split}
        \widehat{\mathcal{S}_{i,j}} = \frac{\exp\left(\frac{\mathcal{S}_{i,j}}{\tau}\right)}{\sum_{i} \exp\left(\frac{\mathcal{S}_{i,j}}{\tau}\right)}
    \end{split}
    \label{eq:is}
\end{align}

To explain how IS operates in query-agnostic scenarios, we decompose Equation~\ref{eq:is} into equivalent components via:
\begin{align}
    \centering
    \begin{split}
        \widehat{\sS_{i,j}} = \exp\left(\frac{\sS_{i,j}+\hbar(\vt_j)}{\tau}\right)
    \end{split}
    \label{eq:is_decomp}
\end{align}
where the target-specific hubness scalar $\hbar(\vt_j)$ is defined as:
\begin{align}
    \centering
    \begin{split}
        \hbar(\vt_j)&=-\tau\operatorname{LogSumExp}\limits_{\vq_i\in\sQ}\left(\frac{\vq_i^\top\vt_j}{\tau}\right)
    \end{split}
    \label{eq:hub_item} % disentangle inverted softmax
\end{align}

Notably, the exponential function $\exp(\frac{\cdot}{\tau})$ preserves relative ranking order due to its monotonicity. Thus, ranking based on $\sS_{i,j} + \hbar(\vt_j)$ produces identical retrieval results to ranking using $\exp(\frac{\sS_{i,j} + \hbar(\vt_j)}{\tau})$. This reveals IS's mechanism as injecting a hubness compensation scalar estimated by the weighted-sum similarity between target $\vt_j$ and query set $\sQ$.

In query-agnostic scenarios where $\sQ$ is inaccessible during inference, practical implementations substitute $\sQ$ in Equation~\ref{eq:hub_item} with a query bank $\sB_q\in\R^{|\sB_q| \times d}$ (e.g., using the query set in training data as the query bank), as adopted in~\cite{bogolin2022cross}, to estimate the target hubness $\hat{\hbar}(\vt_j)$ for hubness reduction. 

\subsection{The Probability-Balancing Framework.} 
In essence, scaling methods differ primarily in their modeling of $\hbar(\vt_j)$. However, existing approaches lack a systematic design framework grounded in theoretical principles, relying instead on empirical validation of retrieval performance gains to evaluate such modeling. We posit that the hubness problem fundamentally arises from the non-uniform probabilities of targets being retrieved. Effectively mitigating this issue requires balancing the probabilities of all targets. To translate this principle into a computational solution, we formalize probability balancing through a unified optimization framework. Specifically, we propose to project the original affinity $\sS$ onto a convex set $\mPi$ enforcing uniform target probability constraints, thereby obtaining an optimized probability matrix $\vpi^\star$ that satisfies:
\begin{align}
    \centering
    \begin{split}
        \vpi^\star &= \mathop{arg~min}\limits_{\vpi\in\mPi(\vb)} \| \sS-\vpi \|_{\mathrm{F}} \\
        \text{s.t.~} \mPi(\vb)&=\{\vpi\in\R_+^{m\times n}\mid\vpi^\top\ones_m=\vb\}
    \end{split}
    \label{eq:proj}
\end{align}
where $\|\cdot\|_{\mathrm{F}}$ denotes the Frobenius norm and $\vb=\frac{1}{n}\ones_n$ enforces uniform target probability constraints. Problem~\ref{eq:proj} can be equivalently reformulated as a maximization problem:
\begin{align}
    \centering
    \begin{split}
        \vpi^\star = \mathop{arg~max}\limits_{\vpi\in\mPi(\vb)} \langle \sS,\vpi \rangle -\frac{1}{2}\|\vpi\|_{\mathrm{F}}
    \end{split}
    \label{eq:sot}
\end{align}

Here we omit the constant term $\|\sS\|_{\mathrm{F}}$ independent of $\vpi$. As discussed in~\cite{blondel2018smooth}, the quadratically-constrained formulation in Equation~\ref{eq:sot} can yield a solution $\vpi^\star$ that is extremely sparse. This sparsity can be detrimental to retrieval performance, as it results in the exclusion of most targets. To address this sparsity issue, the entropic regularization term can be introduced, leading to the entropy-constrained Optimal Transport problem: 
\begin{align}
    \centering
    \begin{split}
        \vpi^\star = \mathop{arg~max}\limits_{\vpi\in\mPi(\vb)}\langle\sS,\vpi\rangle+\tau\mH(\vpi)
    \end{split}
    \label{eq:eot}
\end{align}
where $\mH(\vpi)=\sum_{i,j}-\vpi_{i,j}(\log(\vpi_{i,j})-1)$ defines the entropy of $\vpi$, and $\tau \geq 0$ controls the amount of entropy regularization. We take the same notation $\tau$ in both Equation~\ref{eq:is} and ~\ref{eq:eot} due to the following proposition:
\begin{proposition}[IS functions to balance target probabilities.]
    \label{prop:is_ot}
    The normalized matrix $\widehat{\sS}$ defined in Equation~\ref{eq:is} serves as a specific solution to the problem formulated in Equation~\ref{eq:eot}, when scaled by a constant factor $\frac{1}{n}$. This demonstrates that IS mitigates the hubness problem by balancing target probabilities.
\end{proposition}
The proof is given in our Appendix. Furthermore, Proposition~\ref{prop:is_ot} shows that IS only balances target probabilities, while leaving query probabilities unconstrained, which limits its effectiveness.

\begin{figure}[tbp]
    \centering
    \includegraphics[width=0.45\textwidth]{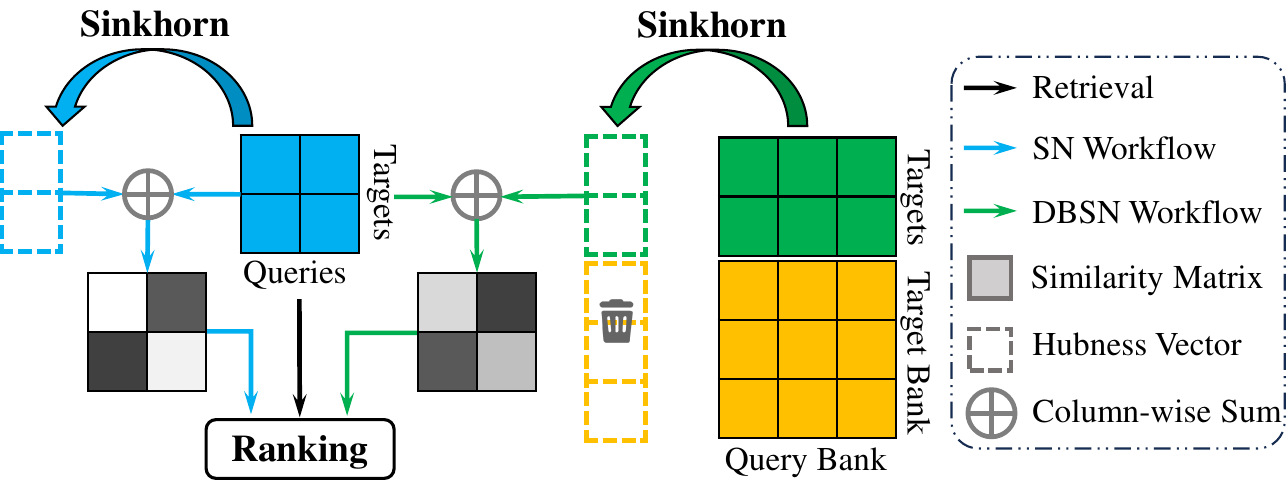}\vspace{-0.2cm}
    \caption{Overview of the proposed methods: \textbf{(a) \textcolor[rgb]{0.0, 0.6875, 0.9375}{SN}} directly estimates the hubness vector using testing queries; \textbf{(b) \textcolor[rgb]{0.0, 0.70, 0.0}{DBSN}} leverages a query bank and a target bank to reduce distribution divergence for hubness estimation.}    
    \label{fig:framework}
\end{figure}

\subsection{Sinkhorn Normalization.} 
\label{sec:sn}
When jointly addressing query and target hubness through uniform probability constraints, we extend Equation~\eqref{eq:eot} to:
\begin{align}
    \centering
    \begin{split}
        \vpi^\star &= \mathop{arg~max}\limits_{\vpi\in\mPi(\va,\vb)}\langle\sS,\vpi\rangle+\tau\mH(\vpi) \\
        \text{s.t.~} \mPi(\va,\vb)&=\{\vpi\in\R_+^{m\times n}\mid\vpi\ones_n=\va,\vpi^\top\ones_m=\vb\} \\
    \end{split}
    \label{eq:uab}
\end{align}
where $\va=\frac{1}{m}\ones_m$ define a uniform query distribution over $\vpi$. Under the marginal constraint $\mPi(\va, \vb)$, the problem in Equation~\ref{eq:uab} admits a unique solution $\vpi^\star$. This solution can be numerically computed via the Sinkhorn-Knopp algorithm~\cite{cuturi2013sinkhorn} applied to the Gibbs kernel $\mxi = \exp(\frac{\mS}{\tau})$, where the algorithm iteratively normalizes the rows and columns of $\mxi$. The optimal transport plan $\vpi^\star$ is formally expressed as:
\begin{align}
    \centering
    \begin{split}
        \vpi^\star=\mathop{diag}(\valpha^{(t)})\mxi\mathop{diag}(\vbeta^{(t)})
    \end{split}
    \label{eq:sinkhorn}
\end{align}
where $t$ denotes the iteration index. In each iteration, the intermediate variables are updated as $\valpha^{(t)}=\frac{\va}{\mxi\vbeta^{(t-1)}}$ and $\vbeta^{(t)}=\frac{\vb}{\mxi^\top\valpha^{(t)}}$, with $\vbeta^{(0)}=\ones_n$ initialized as an all-ones vector. We refer to this normalization as Sinkhorn Normalization (SN). Notably, as shown in Figure~\ref{fig:radar}(a), the ranking results of $\vpi^\star$ significantly outperform those of $\sS$ across various retrieval tasks. These results validate the effectiveness of the Probability-Balancing Framework.

Similar to IS, SN is not directly applicable in query-agnostic scenarios. To address this issue, we first establish that SN operates by adding both query-hubness item $\hbar(\vq_i)$ and target-hubness item $\hbar(\vt_j)$ for $\sS_{i,j}$. Mirroring the decomposition in Equation~\ref{eq:is_decomp}, $\vpi^\star$ in Equation~\ref{eq:sinkhorn} can also be rewritten as:
\begin{align}
    \centering
    \label{eq:dot}% disentangle optimal transport
    \vpi^\star_{i,j} &= \exp(\frac{\sS_{i,j}+\hbar(\vq_i)+\hbar(\vt_j)}{\tau}) \\
    \text{where~}\hbar(\vq_i)&=\tau\mathop{log}(\valpha_i^{(t)}) 
    \text{~and~}  
    \hbar(\vt_j)=\tau\mathop{log}(\vbeta_j^{(t)})
    \notag 
\end{align}

The estimated $\hbar(\vq_i)$ and $\hbar(\vt_j)$ are controlled by $\valpha^{(t)}$ and $\vbeta^{(t)}$ defined in Equation~\ref{eq:sinkhorn}, which are intrinsically related to the distribution divergence between $\sQ$ and $\sT$. Crucially, while SN additionally computes the query-specific $\hbar(\vq_i)$, this item has no impact on rankings; that is, the ranking derived from $\sS_{i,j} + \hbar(\vt_j)$ is equivalent to that produced by $\vpi^\star_{i,j}$. Figure~\ref{fig:framework}(a) illustrates this mechanism in SN.

Analogous to query-agnostic IS, we leverage a query bank $\sB_q$ and replace $\sS$ in Equation~\ref{eq:uab} with an auxiliary similarity matrix $\sS_{bt}=\sB_q^\top\sT$ to estimate $\hbar(\vt_j)$. However, our experiments demonstrate that when significant distribution divergence exists between the query bank $\sB_q$ and the target set $\sT$ (e.g., training texts vs. testing videos in Didemo), both SN and IS exhibit performance degradation, as shown in Figure~\ref{fig:radar}(b). Crucially, SN exhibits more severe degradation than IS, revealing its limitations under large query-target gap.

\subsection{Dual Bank Sinkhorn Normalization.} 
The core objective of scaling methods in query-agnostic scenarios is constructing a query bank $\sB_q$ that sufficiently approximates $\sT$. To achieve this goal, heuristic approaches may be proposed to align $\sB_q$ with $\sT$. For instance, cross-modal generative models like image captioners~\cite{liu2023visual, liu2024improved, li2023blip, gpt4} could generate text queries aligned with images. However, such methods are laborious, computationally expensive, and may yield descriptions of inconsistent quality. Alternatively, training queries can be directly used as $\sB_q$ under the assumption of identical train-test distributions. In practice, the latter strategy is more widely adopted due to its simplicity and efficiency.

As discussed at the end of \S~\ref{sec:sn}, when the query-target gap is large, the estimated target-hubness item $\hbar(\vt_j)$ exhibits substantial bias, thereby inducing significant performance degradation. Several works like DIS~\cite{bogolin2022cross} and DualIS~\cite{wang2023balance} have attempted to narrow the query-target gap through query pruning or expansion. However, they fail to significantly reduce the divergence between query bank $\sB_q$ and ground-truth queries $\sQ$, thus limiting their improvements.

Motivated by DualIS~\cite{wang2023balance}, we propose Dual Bank Sinkhorn Normalization (DBSN), which bridges the gap via target expansion. Specifically, we concatenate $\sT$ with an auxiliary target bank $\sB_t \in \mathbb{R}^{|\sB_t| \times d}$, forming an extended target set $[\sT; \sB_t]$ better aligned with $\sB_q$. Here, $\sB_t$ typically comprises training targets. DBSN estimates joint hubness $[\tilde{\hbar}(\sT); \tilde{\hbar}(\sB_t)]$ via SN and ranks using $\sS_{i,j} + \tilde{\hbar}(\vt_j)$, as visualized in Figure~\ref{fig:framework}(b).

\begin{proposition}[DBSN narrows the query-target gap.]
    \label{prop:dbsn}
    DBSN expands the target set $\sT$ to $[\sT; \sB_t]$, reducing the divergence between query bank $\sB_q$ and extended targets $[\sT; \sB_t]$, thereby improving SN.
\end{proposition}
A detailed proof of this proposition is given in our Appendix. Notably, our dual bank setting is effective only for SN, not for IS. This is because the hubness scalar $\hbar(\vt_j)$ estimated by IS is controlled by the discrepancy between the target $\vt_j$ and the query set $\sQ$ (as shown in Equation~\ref{eq:hub_item}), independent of the overall distribution divergence between $\sT$ and $\sQ$. Consequently, adding a target bank $\sB_t$ does not alter the IS-estimated $\hbar(\vt_j)$.

\section{Experiments}
\label{sec:4-exper}

\subsection{Datasets, Metrics, and Comparison Methods.} 
We evaluate SN on three cross-modal retrieval tasks and DBSN on two tasks, demonstrating the effectiveness of both methods. For each task, we conduct experiments on two most popular benchmarks, i.e., MSR-VTT~\cite{xu2016msr} and Didemo~\cite{anne2017localizing} for text-to-video retrieval; Flickr30k~\cite{young2014image} and MS-COCO~\cite{lin2014microsoft} for text-to-image retrieval; and AudioCaps~\cite{kim2019audiocaps} and Clotho~\cite{drossos2020clotho} for text-to-audio retrieval. We adopt the commonly used recall at rank $K$(R@K, where $K \in \{1,5,10\}$) to evaluate all tasks. Additionally, we report two supplementary metrics to assess overall performance: mean rank (MnR) and median rank (MdR). For fair comparison, We evaluate SN against IS~\cite{smith2017offline} and DIS~\cite{bogolin2022cross} in query-aware scenarios. For DBSN, we additionally include comparisons with DualIS~\cite{wang2023balance} in query-agnostic scenarios.

\subsection{Comparisons of Sinkhorn Normalization.} 
\label{sec:4.1-sn}

Tables~\ref{tab:msrvtt_didemo}-\ref{tab:sn_atr} provide a detailed comparison between SN and counterparts across three retrieval tasks. Both SN and IS variants demonstrate substantial improvements over baselines, achieving an average 5\% improvement in R@1 scores across all datasets. Notably, SN consistently outperforms IS and DIS by a significant margin across all baselines and datasets, highlighting the advantages of simultaneously addressing both query and target hubness. While DIS shows no substantial improvements over IS, as evidenced in~\cite{wang2023balance}, demonstrating the limitation of query pruning alone. In particular, SN achieves state-of-the-art (SOTA) performance when equipped with advanced baselines such as X-Pool and ALBEF, e.g., attaining R@1 scores of 52.7 on MSR-VTT and 56.7 on MS-COCO. We also present results for additional tasks (e.g., image-to-image retrieval on InShop~\cite{liu2016deepfashion} and image classification on ImageNet~\cite{deng2009imagenet}, and for more datasets including ActivityNet~\cite{caba2015activitynet} and VATEX~\cite{wang2019vatex}, MSVD~\cite{wu2017msvd}, LSMDC~\cite{rohrbach2015LSMDC}) in the appendix. As summarized in Figure~\ref{fig:radar}(a), these results further demonstrate the broad applicability of SN.

\subsection{Comparisons of Dual-Bank Sinkhorn Normalization.} 
\label{sec:4.3-dbsn}

In query-agnostic scenarios where testing queries are unavailable, we replace the testing query set $\sQ$ with a query bank $\sB_q$ (constructed from training/validation data) to evaluate single-bank methods like IS and SN. For dual-bank methods such as DualIS~\cite{wang2023balance} and DBSN, we additionally employ a target bank~($\sB_t$). As shown in Table~\ref{tab:dbsn}, single-bank normalization yields only marginal improvements—and sometimes even degrades performance—compared to the baseline. While expanding the query bank (from validation to training set size) provides minor gains, DIS and DualIS show negligible improvements over IS, as they fail to address query-side hubness. In contrast, DBSN effectively utilizes the dual-bank structure, achieving significant improvements over SN and approaching optimal performance. Notably, SN underperforms IS on MSR-VTT when using the training query bank, which we attribute to the distributional gap between training and test queries. This hypothesis is confirmed by experiments with a lower-discrepancy query bank~\cite[jsfusion]{yu2018joint}, demonstrating that query-target alignment critically affects hubness reduction efficacy. Figure~\ref{fig:radar}(b) provides a visual summary highlighting DBSN's superiority in query-agnostic scenarios (see Appendix for full details).

\begin{table*}[h]
    \centering
    \small
    \caption{Video-Text retrieval performance comparison on MSR-VTT and Didemo. All methods employ the same \emph{CLIP VIT-B/32} backbone. Bold denotes the best performance. $\ddag$ marks our reproduced results.}
    % We refer to the \SM for extensions of this table with more baselines and COCO 5K results.
    % \resizebox{\textwidth}{!}
    \label{tab:msrvtt_didemo}
    {
        \tabcolsep 5 pt
	    \begin{tabular}{ lc cccc cccc  cccc cccc}
	        \addlinespace
	        \toprule
                \multicolumn{1}{c}{\multirow{3}{*}{\symtext{Method}}} & \multicolumn{1}{c}{\multirow{3}{*}{\symtext{Norm}}}  & \multicolumn{8}{c}{MSR-VTT 1k test} & \multicolumn{8}{c}{Didemo} \\ \cmidrule(lr){3-10} \cmidrule(lr){11-18}
                & & \multicolumn{4}{c}{Text$\rightarrow$Video} & \multicolumn{4}{c}{Video$\rightarrow$Text} & \multicolumn{4}{c}{Text$\rightarrow$Video} & \multicolumn{4}{c}{Video$\rightarrow$Text} \\ \cmidrule(lr){3-6} \cmidrule(lr){7-10} \cmidrule(lr){11-14} \cmidrule(lr){15-18}
                 &  &  R@1 & R@5 & R@10 & MnR$\downarrow$  &  R@1 & R@5 & R@10 & MnR$\downarrow$ &  R@1 & R@5 & R@10 & MnR$\downarrow$ &  R@1 & R@5 & R@10 & MnR$\downarrow$ \\ 
                \midrule
                \multicolumn{2}{l}{\textcolor{gray}{CLIP4Clip~\cite{luo2022clip4clip}}} & \textcolor{gray}{43.1} & \textcolor{gray}{70.4} & \textcolor{gray}{80.8} & \textcolor{gray}{16.2} & \textcolor{gray}{43.1} & \textcolor{gray}{70.5} & \textcolor{gray}{81.2} & \textcolor{gray}{12.4} & \textcolor{gray}{43.4} & \textcolor{gray}{70.2} & \textcolor{gray}{80.6} & \textcolor{gray}{17.5} & \textcolor{gray}{42.5} & \textcolor{gray}{70.6} & \textcolor{gray}{80.2} & \textcolor{gray}{11.6} \\
                \multicolumn{2}{l}{\textcolor{gray}{CLIP2Video~\cite{fang2021clip2video}}} & \textcolor{gray}{45.6} & \textcolor{gray}{72.6} & \textcolor{gray}{81.7} & \textcolor{gray}{14.6} & \textcolor{gray}{43.5} & \textcolor{gray}{72.3} & \textcolor{gray}{82.1} & \textcolor{gray}{10.2} & \textcolor{gray}{-} & \textcolor{gray}{-} & \textcolor{gray}{-} & \textcolor{gray}{-} & \textcolor{gray}{-} & \textcolor{gray}{-} & \textcolor{gray}{-} & \textcolor{gray}{-} \\
                \multicolumn{2}{l}{\textcolor{gray}{X-CLIP~\cite{ma2022x}}} & \textcolor{gray}{46.1} & \textcolor{gray}{73.0} & \textcolor{gray}{83.1} & \textcolor{gray}{13.2} & \textcolor{gray}{46.8} & \textcolor{gray}{73.3} & \textcolor{gray}{84.0} & \textcolor{gray}{9.1} & \textcolor{gray}{45.2} & \textcolor{gray}{74.0} & \textcolor{gray}{-} & \textcolor{gray}{14.6} & \textcolor{gray}{43.1} & \textcolor{gray}{72.2} & \textcolor{gray}{-} & \textcolor{gray}{10.9} \\
                \multicolumn{2}{l}{\textcolor{gray}{DRL~\cite{wang2022disentangled}}} & \textcolor{gray}{47.4} & \textcolor{gray}{74.6} & \textcolor{gray}{83.8} & \textcolor{gray}{-} & \textcolor{gray}{45.3} & \textcolor{gray}{73.9} & \textcolor{gray}{83.3} & \textcolor{gray}{9.1} & \textcolor{gray}{47.9} & \textcolor{gray}{73.8} & \textcolor{gray}{82.7} & \textcolor{gray}{-} & \textcolor{gray}{45.4} & \textcolor{gray}{72.6} & \textcolor{gray}{82.1} & \textcolor{gray}{-} \\
                \multicolumn{2}{l}{\textcolor{gray}{X-Pool~\cite{gorti2022x}}}& \textcolor{gray}{46.9} & \textcolor{gray}{72.8} & \textcolor{gray}{82.2} & \textcolor{gray}{14.3} & \textcolor{gray}{44.4} & \textcolor{gray}{73.3} & \textcolor{gray}{84.0} & \textcolor{gray}{9.0} & \textcolor{gray}{-} & \textcolor{gray}{-} & \textcolor{gray}{-} & \textcolor{gray}{-} & \textcolor{gray}{-} & \textcolor{gray}{-} & \textcolor{gray}{-} & \textcolor{gray}{-} \\
                \multicolumn{2}{l}{\textcolor{gray}{TS2-Net~\cite{liu2022ts2}}} & \textcolor{gray}{47.0} & \textcolor{gray}{74.5} & \textcolor{gray}{83.8} & \textcolor{gray}{-} & \textcolor{gray}{45.3} & \textcolor{gray}{74.1} & \textcolor{gray}{83.7} & \textcolor{gray}{9.2} & \textcolor{gray}{41.8} & \textcolor{gray}{71.6} & \textcolor{gray}{82.0} & \textcolor{gray}{14.8} & \textcolor{gray}{-} & \textcolor{gray}{-} & \textcolor{gray}{-} & \textcolor{gray}{-} \\
                % \textcolor{gray}{CAMoE}       &         & \textcolor{gray}{47.3} & \textcolor{gray}{74.2} & \textcolor{gray}{84.5} & \textcolor{gray}{2.0}   & \textcolor{gray}{11.9}        \\
                \multicolumn{2}{l}{\textcolor{gray}{UATVR~\cite{fang2023uatvr}}} & \textcolor{gray}{47.5} & \textcolor{gray}{73.9} & \textcolor{gray}{83.5} & \textcolor{gray}{12.3} & \textcolor{gray}{46.9} & \textcolor{gray}{73.8} & \textcolor{gray}{83.8} & \textcolor{gray}{8.6} & \textcolor{gray}{43.1} & \textcolor{gray}{71.8} & \textcolor{gray}{82.3} & \textcolor{gray}{15.1} & \textcolor{gray}{-} & \textcolor{gray}{-} & \textcolor{gray}{-} & \textcolor{gray}{-} \\
                \multicolumn{2}{l}{\textcolor{gray}{ProST~\cite{li2023progressive}}} & \textcolor{gray}{48.2} & \textcolor{gray}{74.6} & \textcolor{gray}{83.4} &  \textcolor{gray}{12.4} & \textcolor{gray}{46.3} & \textcolor{gray}{74.2} & \textcolor{gray}{83.2} & \textcolor{gray}{8.7} & \textcolor{gray}{44.9} & \textcolor{gray}{72.7} & \textcolor{gray}{82.7} & \textcolor{gray}{13.7} & \textcolor{gray}{-} & \textcolor{gray}{-} & \textcolor{gray}{-} & \textcolor{gray}{-} \\
                \multicolumn{2}{l}{\textcolor{gray}{T-MASS~\cite{wang2024text}}} & \textcolor{gray}{50.2} & \textcolor{gray}{75.3} & \textcolor{gray}{85.1} &  \textcolor{gray}{11.9} & \textcolor{gray}{47.7} & \textcolor{gray}{78.0} & \textcolor{gray}{86.3} & \textcolor{gray}{8.0} & \textcolor{gray}{50.9} & \textcolor{gray}{77.2} & \textcolor{gray}{85.3} & \textcolor{gray}{12.1} & \textcolor{gray}{-} & \textcolor{gray}{-} & \textcolor{gray}{-} & \textcolor{gray}{-} \\
                \multicolumn{2}{l}{\textcolor{gray}{NarVid~\cite{hong2025narrating}}} & \textcolor{gray}{51.0} & \textcolor{gray}{76.4} & \textcolor{gray}{85.2} &  \textcolor{gray}{11.6} & \textcolor{gray}{50.0} & \textcolor{gray}{75.4} & \textcolor{gray}{83.8} & \textcolor{gray}{7.9} & \bf \textcolor{gray}{53.4} & \bf \textcolor{gray}{79.1} & \textcolor{gray}{86.3} & \textcolor{gray}{-} & \textcolor{gray}{-} & \textcolor{gray}{-} & \textcolor{gray}{-} & \textcolor{gray}{-} \\
                \midrule
                \multicolumn{2}{l}{CLIP4Clip~\cite{luo2022clip4clip}$\ddag$} & 43.9 & 70.6 & 80.7 & 16.0 & 44.7 & 71.6 & 81.5 & 11.1 & 40.8 & 69.7 & 80.3 & 18.4 & 41.4 & 70.6 & 79.4 & 11.7 \\
                & + IS & 48.1 & 73.5 & 83.3 & 12.1 & 48.0 & 74.0 & 83.7 & 9.9 & 46.0 & 72.8 & 81.1 & 15.8 & 48.2 & 72.9 & 82.5 & 9.8 \\
                & + DIS & 48.5 & 73.9 & 83.2 & 12.0 & 48.1 & 73.9 & 83.2 & 10.0 & 46.1 & 73.1 & 82.6 & 15.8 & 48.0 & 73.1 & 82.6 & 9.8\\
                \rowcolor{gray} & + SN & 49.6 & 75.5 & 84.2 & 11.6 & 50.8 & 75.4 & 84.8 & 9.2 & 48.0 & 74.6 & 82.9 & 13.6 & 50.4 & 73.7 & 83.8 & 9.6 \\
                \midrule
                \multicolumn{2}{l}{DRL~\cite{wang2022disentangled}$\ddag$} & 45.4 & 74.0 & 83.1 & 13.0 & 45.3 & 73.8 & 82.6 & 9.1 & 45.0 & 73.2 & 83.9 & 14.2 & 43.1 & 72.6 & 82.0 & 9.6  \\
                                      & + IS & 49.7 & 76.7 & 84.8 & 11.5 & 51.0 & 76.0 & 85.3 & 8.7 & 49.7 & 77.1 & 84.2 & 11.8 & 53.9 & 77.9 & 86.1 & 8.1  \\
                                      & + DIS & 49.8 & 75.9 & 85.1 & 11.5 & 50.7 & 75.9 & 85.1 & 8.7 & 49.8 & 78.0 & 86.4 & 11.8 & 54.2 & 78.0 & 86.4 & 8.1 \\
                                      \rowcolor{gray} & + SN & 51.4 & 78.2 & 86.3 & 10.3 & 52.9 & 78.2 & 85.6 & 7.8 & 52.1 & 79.1 & 86.2 & 10.5 & 55.3 & 79.8 & 86.1 & 7.7 \\
                \midrule
                \multicolumn{2}{l}{X-Pool~\cite{gorti2022x}$\ddag$} & 48.0 & 73.1 & 83.2 & 14.0 & 47.1 & 75.6 & 84.6 & 8.8 & 47.3 & 73.5 & 82.8 & 14.8 & 44.2 & 72.8 & 82.1 & 9.0 \\
                                        & + IS & 50.8 & 77.2 & 86.5 & 10.5 & 51.3 & 78.5 & 86.0 & 7.8 & 50.9 & 76.3 & 85.1 & 11.4 & 51.2 & 77.6 & 86.7 & 7.3  \\
                                        & + DIS & 51.1 & 78.2 & 86.0 & 10.5 & 51.4 & 78.2 & 86.0 & 7.8 & 50.7 & 77.3 & 86.5 & 11.4 & 51.0 & 77.3 & 86.5 & 7.3\\
                                        \rowcolor{gray} & + SN & \bf 52.7 & \bf 78.4 & \bf 86.5 & \bf 10.1 & \bf 53.4 & \bf 78.8 & \bf 86.9 & \bf 7.4 & 53.1 & 78.6 & \bf 86.8 & \bf 9.6 & \bf 54.3 & \bf 79.5 & \bf 87.1 & \bf 7.0 \\
	       \bottomrule
	    \end{tabular}
	}
\end{table*}

\begin{table*}[h]
    \centering
    \small
    \caption{Image-Text retrieval performance comparison on Flickr30k and MS-COCO. Bold indicates the best performance. $\ddag$ marks our reproduced results. "zs" denots zero-shot, "ft" denotes fine-tuned. \cmmnt{Our released ALBEF baseline is lower than the results reported in the original ALBEF paper. The reason for this is that ALBEF incorporates an additional reranking network, which significantly boosts retrieval performance, while we ignore it.}}
    % We refer to the \SM for extensions of this table with more baselines and COCO 5K results.
    % \resizebox{\textwidth}{!}
    \label{tab:coco_f30k}
    {
        \tabcolsep 5 pt
	    \begin{tabular}{ lc cccc cccc  cccc cccc}
	        \addlinespace
	        \toprule
                \multicolumn{1}{c}{\multirow{3}{*}{\symtext{Method}}} & \multicolumn{1}{c}{\multirow{3}{*}{\symtext{Norm}}}  & \multicolumn{8}{c}{Flickr30k} & \multicolumn{8}{c}{MS-COCO} \\ \cmidrule(lr){3-10} \cmidrule(lr){11-18}
                & & \multicolumn{4}{c}{Text$\rightarrow$Image} & \multicolumn{4}{c}{Image$\rightarrow$Text} & \multicolumn{4}{c}{Text$\rightarrow$Image} & \multicolumn{4}{c}{Image$\rightarrow$Text} \\ \cmidrule(lr){3-6} \cmidrule(lr){7-10} \cmidrule(lr){11-14} \cmidrule(lr){15-18}
                 &  &  R@1 & R@5 & R@10 & MnR$\downarrow$  &  R@1 & R@5 & R@10 & MnR$\downarrow$ &  R@1 & R@5 & R@10 & MnR$\downarrow$ &  R@1 & R@5 & R@10 & MnR$\downarrow$ \\ 
                \midrule
                \multicolumn{2}{l}{\textcolor{gray}{CUSA~\cite{huang2024cross}}} & \textcolor{gray}{67.5} & \textcolor{gray}{89.6} & \textcolor{gray}{93.9} & \textcolor{gray}{-} & \textcolor{gray}{82.1} & \textcolor{gray}{95.3} & \textcolor{gray}{97.9} & \textcolor{gray}{-} & \textcolor{gray}{44.2} & \textcolor{gray}{72.7} & \textcolor{gray}{82.1} & \textcolor{gray}{-} & \textcolor{gray}{57.3} & \textcolor{gray}{83.1} & \textcolor{gray}{90.3} & \textcolor{gray}{-} \\
                \multicolumn{2}{l}{\textcolor{gray}{LAPS~\cite{fu2024linguistic}}} & \textcolor{gray}{80.6} & \textcolor{gray}{95.5} & \textcolor{gray}{-} & \textcolor{gray}{-} & \textcolor{gray}{92.9} & \textcolor{gray}{99.3} & \textcolor{gray}{-} & \textcolor{gray}{-} & \textcolor{gray}{54.3} & \textcolor{gray}{80.0} & \textcolor{gray}{-} & \textcolor{gray}{-} & \textcolor{gray}{69.8} & \textcolor{gray}{90.4} & \textcolor{gray}{-} & \textcolor{gray}{-} \\
                \midrule
                \multicolumn{2}{l}{zs CLIP~\cite{li2022blip}} & 58.8 & 83.4 & 90.1 & 6.0 & 79.3 & 95.0 & 98.1 & 2.1 & 30.5 & 56.0 & 66.8 & 24.5 & 50.0 & 75.0 & 83.5 & 8.9 \\
                            & + IS & 66.8 & 88.7 & 93.5 & 4.5 & 85.3 & 96.5 & 98.6 & 1.7 & 38.6 & 64.0 & 74.1 & 20.0 & 57.2 & 78.9 & 85.9 & 8.2 \\
                            & + DIS & 66.7 & 88.8 & 93.6 & 4.5 & 85.2 & 96.5 & 98.6 & 1.7 & 38.7 & 64.2 & 74.2 & 20.0 & 57.1 & 78.7 & 85.9 & 8.3\\
                            \rowcolor{gray} & + SN & 69.3 & 90.2 & 94.5 & 3.7 & 87.9 & 98.1 & 99.4 & 1.4 & 40.8 & 66.4 & 76.1 & 17.8 & 60.4 & 81.7 & 88.7 & 6.3 \\
                \midrule
                \multicolumn{2}{l}{ft CLIP$\ddag$} & 74.2 & 93.4 & 96.7 & 2.8 & 88.1 & 98.1 & 99.3 & 1.5 &  47.5 & 74.1 & 83.2 & 11.3 & 65.0 & 85.9 & 92.2 & 4.8 \\
                                    & + IS & 76.6 & 93.9 & 97.1 & 2.5 & 93.9 & 99.1 & 99.6 & 1.3 & 50.1 & 75.9 & 84.3 & 10.8 & 72.1 & 89.3 & 93.9 & 3.6 \\
                                    & + DIS & 76.5 & 93.8 & 97.1 & 2.5 & 94.1 & 99.1 & 99.6 & 1.3 & 50.1 & 75.9 & 84.4 & 10.8 & 72.1 & 89.3 & 94.0 & 3.6 \\
                                    \rowcolor{gray} & + SN & 79.2 & 94.8 & 97.5 & 2.2 & 94.6 & 99.4 & 99.8 & 1.2 &  51.6 & 77.0 & 85.3 & 10.0 & 73.3 & 89.8 & 94.3 & 3.4  \\
                \midrule
                \multicolumn{2}{l}{zs ALBEF~\cite{li2021align}} & 79.8 & 95.3 & 97.7 & 2.4 & 92.6 & 99.3 & 99.9 & 1.2 & 51.8 & 78.5 & 86.6 & 11.2 & 71.2 & 91.1 & 95.7 & 2.7 \\
                & + IS &  80.8 & 95.4 & 97.7 & 2.3 & 96.7 & 99.7 & 100.0 & 1.1 & 54.5 & 80.1 & 87.7 & 10.5 & 76.9 & 93.3 & 96.7 & 2.4 \\
                & + DIS &  80.9 & 95.5 & 97.7 & 2.3 & 96.8 & 99.7 & 100.0 & 1.1 & 54.5 & 80.1 & 87.7 & 10.4 & 76.8 & 93.3 & 96.6 & 2.3 \\
                \rowcolor{gray} & + SN & \bf 83.4 & \bf 96.6 & \bf 98.2 & \bf 1.9 & \bf 97.1 & \bf 99.7 & \bf 99.9 & \bf 1.1 & \bf 56.7 & \bf 81.5 & \bf 88.8 & \bf 9.5 & \bf 77.2 & \bf 93.3 & \bf 96.8 & \bf 2.3 \\
	       \bottomrule
	    \end{tabular}
	}
\end{table*}

\begin{table}[h]
    \centering
    \small
    \caption{Text-to-Audio retrieval performance comparison on AudioCaps and Clotho.}
    % \vspace{-10pt}
    \label{tab:sn_atr}
    {
        \tabcolsep 5 pt
	    \begin{tabular}{ lc ccccc}
	        \addlinespace
	        \toprule
                Method &  Normalization &  R@1$\uparrow$ & R@5$\uparrow$ & R@10$\uparrow$ & MdR$\downarrow$ & MnR$\downarrow$ \\ 
                \midrule
                \multicolumn{7}{@{\;}c}{\bf AudioCaps~~Text$\rightarrow$Audio} \\
                \midrule
                \multicolumn{2}{l}{\textcolor{gray}{ML-ACT~\cite{mei2022metric}}} & \textcolor{gray}{33.9} & \textcolor{gray}{69.7} & \textcolor{gray}{82.6} & \textcolor{gray}{-} & \textcolor{gray}{-} \\ 
                \midrule
                \multicolumn{2}{l}{zs CLAP\cite{wu2023large}} & 40.1 & 76.0 & 87.9 & 2.0 & 6.2 \\
                                            & + IS & 41.5 & 77.1 & 88.2 & 2.0 & 6.0 \\
                                            & + DIS & 41.4 & 77.0 & 88.1 & 2.0 & 6.1 \\
                                            \rowcolor{gray}& + SN & \textbf{43.6} & \textbf{79.4} & \textbf{89.2} & \textbf{2.0} & \textbf{5.5} \\
                \midrule
                \multicolumn{7}{@{\;}c}{\bf Clotho~~Text$\rightarrow$Audio} \\
                \midrule
                \multicolumn{2}{l}{\textcolor{gray}{ML-ACT~\cite{mei2022metric}}} & \textcolor{gray}{14.4} & \textcolor{gray}{36.6} & \textcolor{gray}{49.9} & \textcolor{gray}{-} & \textcolor{gray}{-} \\ 
                \midrule
                \multicolumn{2}{l}{zs CLAP\cite{wu2023large}} & 15.6 & 39.8 & 53.1 & 9.0 & 38.8 \\
                                      & + IS & 17.6 & 44.5 & 57.8 & 7.0 & 31.2 \\
                                      & + DIS & 17.7 & 44.6 & 57.8 & 7.0 & 31.0 \\
                                      \rowcolor{gray}& + SN & \textbf{18.5} & \textbf{46.2} & \textbf{59.2} & \textbf{7.0} & \textbf{30.1} \\
	       \bottomrule
	    \end{tabular}
	}
     % \vspace{-10pt}
\end{table}

\begin{table}[t]
    \centering
    \small
    \caption{Comparisons between DBSN and other querybank normalization methods on Flickr 30K and MSR-VTT.}
    % \vspace{-10pt}
    \label{tab:dbsn}
    {
        \tabcolsep 3 pt
	    \begin{tabular}{ ll ccccc}
	        \addlinespace
	        \toprule
                $\sB_q$~\&~$\sB_t$ & Normalization & R@1$\uparrow$ & R@5$\uparrow$ & R@10$\uparrow$ & MdR~$\downarrow$ & MnR~$\downarrow$ \\ 
                \midrule
                \multicolumn{7}{@{\;}c}{\bf zero-shot CLIP on Flicr30K for Text$\rightarrow$Image} \\
                \midrule
                \multirow{3}{*}{test~\&~-}  & -    & 58.8 & 83.4 & 90.1 & 1.0 & 6.0 \\
                                        & + IS & 66.8 & 88.7 & 93.5 & 1.0 & 4.5 \\
                                        & + SN & \textbf{69.3} & \textbf{90.2} & \textbf{94.5} & \textbf{1.0} & \textbf{3.7} \\
                \midrule
                \multirow{3}{*}{val~\&~-} & + IS & 63.5 & 87.2 & 92.3 & 1.0 & 5.2 \\
                                        & + DIS & 63.5 & 87.2 & 92.3 & 1.0 & 5.2 \\
                                        & + SN & 64.4 & 87.6 & 92.5 & 1.0 & 4.9 \\
                \midrule
                \multirow{2}{*}{val~\&~val} & + DualIS & 63.6 & 87.1 & 92.3 & 1.0 & 5.2 \\
                                        & + DBSN & 64.6 & 88.0 & 92.7 & 1.0 & 4.8 \\
                \midrule
                \multirow{3}{*}{train~\&~-} & + IS & 65.1 & 87.5 & 92.8 & 1.0 & 4.7 \\
                                            & + DIS & 65.1 & 87.5 & 92.8 & 1.0 & 4.7 \\
                                            & + SN & 66.3 & 88.1 & 93.1 & 1.0 & 4.7 \\
                \midrule
                \multirow{2}{*}{train~\&~train}   & + DualIS & 65.3 & 87.4 & 92.9 & 1.0 & 4.7 \\
                                                & + DBSN & 67.1 & 88.7 & 93.5 & 1.0 & 4.4 \\
                % \midrule
                % \multirow{2}{*}{LLaVA~\&~-}  & + IS & 63.5 & 87.2 & 92.3 & 1.0 & 5.2 \\
                %                             & + DIS & 63.5 & 87.2 & 92.3 & 1.0 & 5.2 \\
                %                             & + SN & 60.2 & 83.0 & 89.4 & 1.0 & 7.8 \\
                \midrule
                \multicolumn{7}{@{\;}c}{\bf CLIP4Clip on MSR-VTT(1k split) for Text$\rightarrow$Video} \\
                \midrule
                \multirow{3}{*}{test~\&~-} & -    & 43.9 & 70.6 & 80.7 & 2.0 & 16.0   \\
                                                     & + IS & 46.4 & 72.5 & 82.8 & 2.0 & 13.2   \\
                                                     & + SN & \textbf{49.2} & \textbf{75.4} & \textbf{83.8} & \textbf{2.0} & \textbf{11.9}  \\
                \midrule
                \multirow{3}{*}{train~\&~-} & + IS & 44.8 & 71.1 & 81.3 & 2.0 & 15.0 \\
                                            & + DIS & 44.8 & 71.1 & 81.3 & 2.0 & 15.0 \\
                                            & + SN & 44.5 & 70.9 & 80.9 & 2.0 & 15.4  \\
                \midrule
                \multirow{2}{*}{train~\&~train}   & + DualIS & 44.5 & 71.0 & 80.9 & 2.0 & 15.4 \\
                                                & + DBSN & 45.2 & 71.8 & 81.9 & 2.0 & 15.5 \\
                \midrule
                \multirow{2}{*}{jsfusion~\&~-}   & + IS & 45.1 & 70.5 & 81.2 & 2.0 & 15.6 \\
                                                & + SN & 46.2 & 71.1 & 81.8 & 2.0 & 15.0  \\
	       \bottomrule
	    \end{tabular}
	}
\end{table}

\subsection{Ablation.} 

\begin{figure}[tbp]
    \centering
    \includegraphics[width=\linewidth]{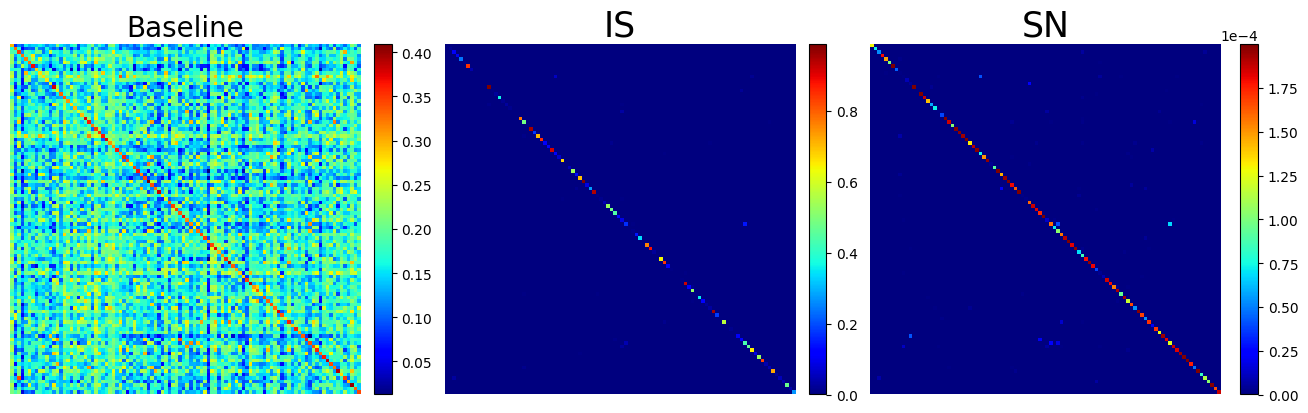} 
    \includegraphics[width=\linewidth]{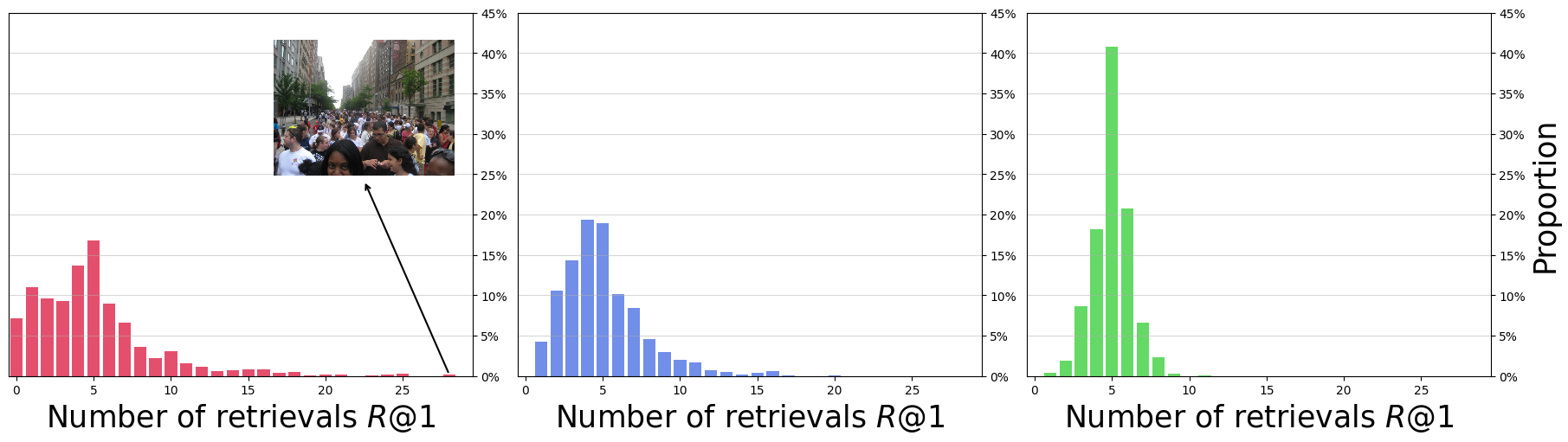} 
    \caption{Impact of normalization strategies on similarity matrix and k-occurrence distribution on Flickr30k.}
    \label{fig:img_sim&mat}

    \includegraphics[width=\linewidth]{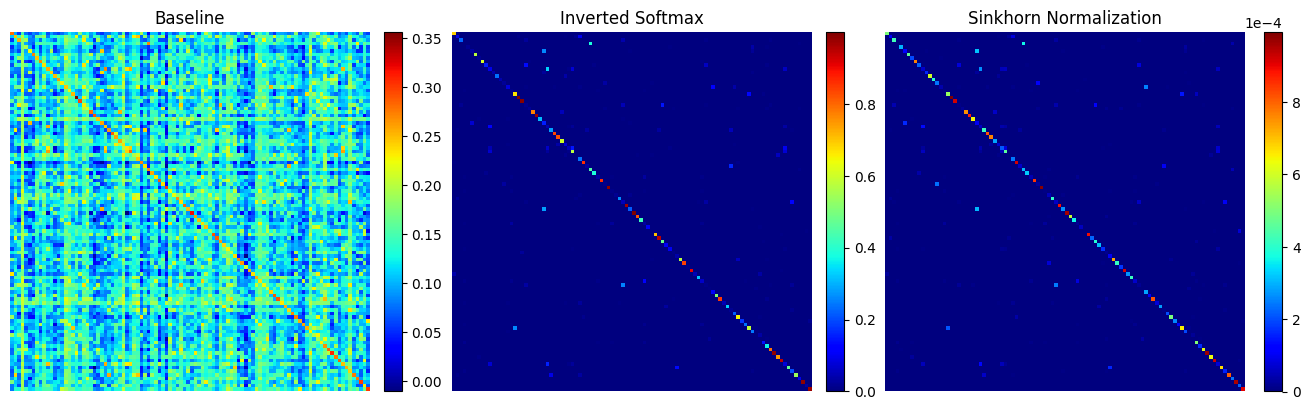} 
    \includegraphics[width=\linewidth]{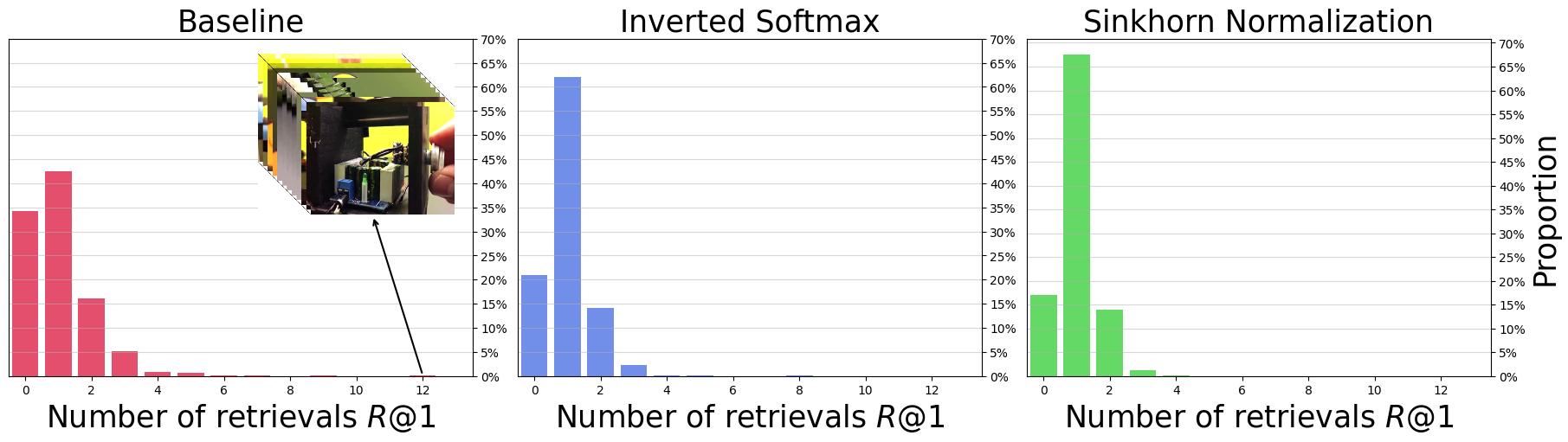} 
    \caption{Impact of normalization strategies on similarity matrix and k-occurrence distribution on MSR-VTT.}
    \label{fig:vid_sim&mat}
\end{figure}

\mypara{Skewness.\label{subsec:skewness}} The \emph{skewness of the k-occurrence distribution} is widely regarded as a critical indicator of \emph{hubness in embedding spaces}, which models the asymmetry in nearest-neighbor relationships; see~\cite{radovanovic2010hubs} for theoretical analysis. We present the comparative skewness analysis between SN and IS across multiple datasets in Table~\ref{tab:skewness}. Our proposed SN significantly reduces skewness, thereby mitigating hubness, compared to both the baseline and IS. To further investigate this phenomenon, Figure~\ref{fig:img_sim&mat} and Figure~\ref{fig:vid_sim&mat} visualize the similarity matrix and k-occurrence distributions before and after normalization. The contrast reveals that SN suppresses the long tail of high k-occurrence instances more effectively, whereas IS fails to adequately constrain dominant hubs.
\begin{table}[t]
    \centering
    \small
    \caption{Impact of SN on skewness across various datasets.}
    \label{tab:skewness}
    {
        \tabcolsep 1 pt
	    \begin{tabular}{ ccc  ccc  ccc  ccc}
	        \addlinespace
	        \toprule
            \multicolumn{3}{c}{Flickr30k} & \multicolumn{3}{c}{MS-COCO} & \multicolumn{3}{c}{MSR-VTT} & \multicolumn{3}{c}{Didemo}\\ \cmidrule(lr){1-3} \cmidrule(lr){4-6} \cmidrule(lr){7-9} \cmidrule(lr){10-12}
            zs CLIP & +IS & +SN & ft CLIP & +IS & +SN & CLIP4Clip & +IS & +SN & X-Pool & +IS & +SN \\
            \hline
            % 2.67 & 0.02 & 0.77 & 1.00 & -0.06 & 0.62 & 1.27 & -0.03 & 0.53 & 1.30 & 0.07 & 0.76 \\ %this results is the skewness-10
            2.24 & 0.50 & 0.07 & 2.03 & 1.61 & 1.05 & 1.87 & 0.87 & 0.46 & 1.18 & 0.67 & 0.36 \\ %this results is the skewness-1
	       \bottomrule
	    \end{tabular}
	}
\end{table}

\begin{figure}[tbp]
    \centering
    \includegraphics[width=1\linewidth]{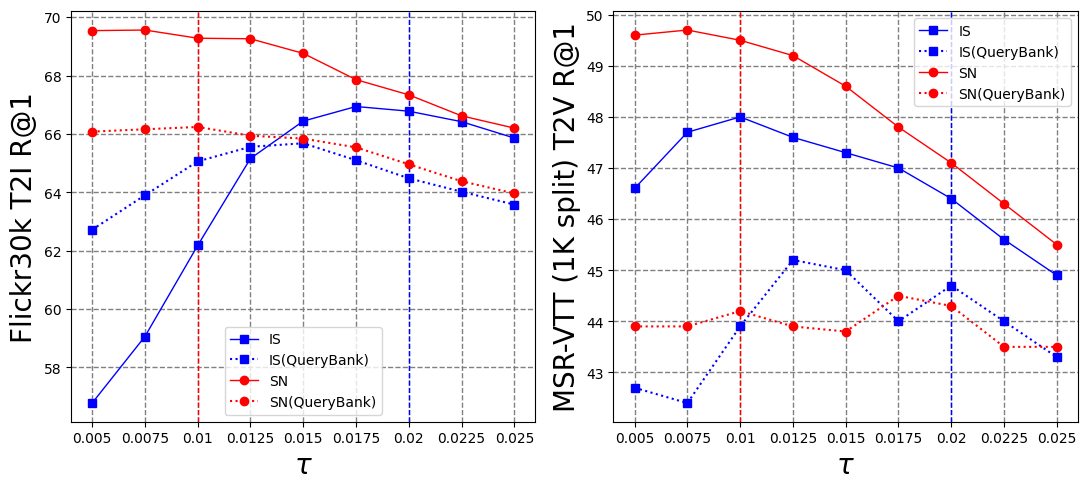}
    \caption{The influence of $\tau$ for IS and SN.}
    \label{fig:tau}
\end{figure}

\mypara{Hyper-parameter Sensitivity.} We assess the sensitivity of the hyper-parameter $\tau$ on the Flickr30k and MSR-VTT datasets. As shown in Figure~\ref{fig:tau}, the performance of IS initially improves and then declines as $\tau$ decreases, peaking at $\tau = 0.0175$ when using testing queries and at $\tau = 0.015$ when using the query bank. This behavior suggests that IS provides a biased estimation of target hubness, requiring precise $\tau$ calibration to minimize this estimation bias. In contrast, as $\tau$ decreases, the performance gap between SN and IS widens, with SN gradually converging to its optimal performance. This indicates that SN is more robust to variations in the $\tau$ parameter. Based on this observation, we choose $\tau = 0.02$ for IS, in line with current practices, and $\tau = 0.01$ for SN to balance effectiveness with efficiency. The fluctuations observed in the performance curves on MSR-VTT with the query bank setting have been explained in \S~\ref{sec:4.1-sn}.

\begin{figure*}[tbp]
    \centering
    \includegraphics[width=1\linewidth]{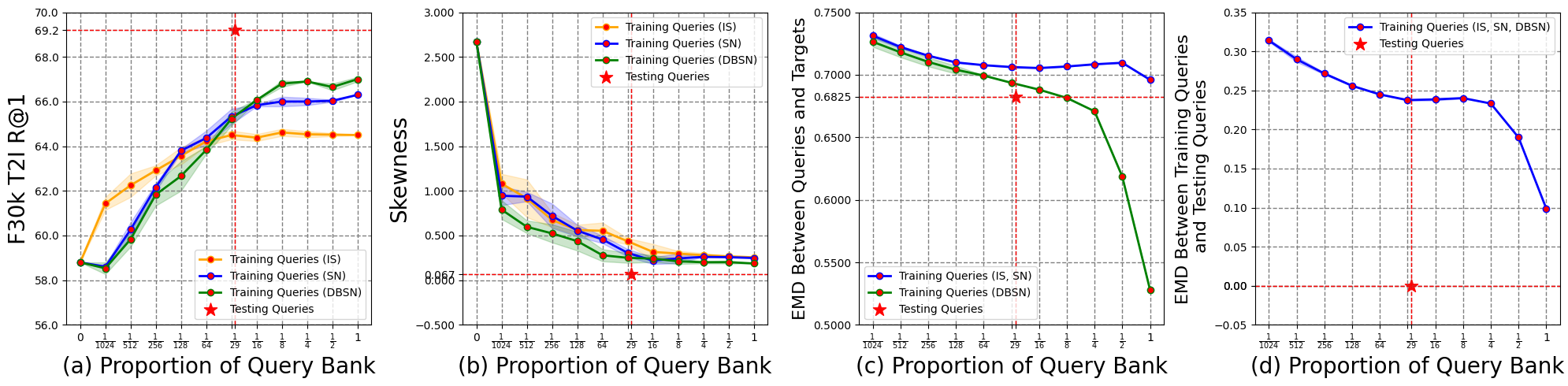}
    \caption{The influence of query bank sizes on Flickr30k. The x-axis is in logarithmic scale, denoting the proportion of the subset relative to the entire query set. }
    \label{fig:ratio}
\end{figure*}

\begin{figure}[tbp]
    \centering
    \includegraphics[width=1\linewidth]{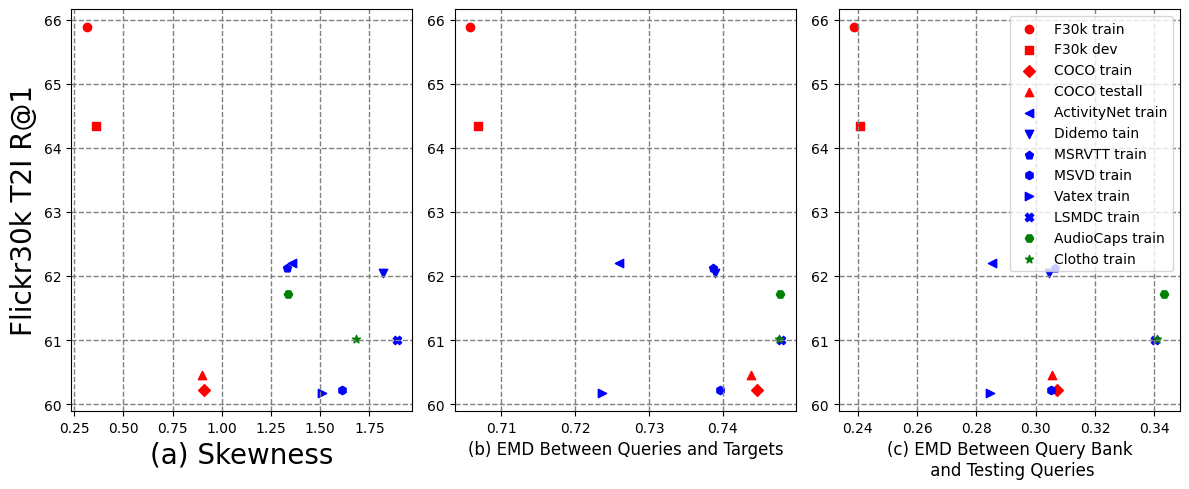}
    \caption{The influence of query bank source distributions.}
    \label{fig:querytype}
\end{figure}
\begin{table}[t]
    \centering
    \small
    \caption{The influence of normalization variants.}
    % \vspace{-10pt}
    \label{tab:norm_type}
    {
        \tabcolsep 5 pt
	    \begin{tabular}{ l ccccc c}
	        \addlinespace
	        \toprule
                Normalization & R@1$\uparrow$ & R@5$\uparrow$ & R@10$\uparrow$ & MdR$\downarrow$ & MnR$\downarrow$ & sparsity\\ 
                \midrule
                \multicolumn{7}{@{\;}c}{\bf zero-shot CLIP on Flicr30K for Text$\rightarrow$Image} \\
                \midrule
                % - & 58.8 & 83.4 & 90.1 & 1.0 & 6.0 & 0 \\
                % IS & 66.8 & 88.7 & 93.5 & 1.0 & 4.5 & 0 \\
                SN & \textbf{69.3} & \textbf{90.2} & \textbf{94.5} & \textbf{1.0} & \textbf{3.7} & 0 \\
                OTN & 69.1 & 73.1 & 73.2 & 1.0 & 134.5 & 0.99880 \\
                L2N & 66.3 & 66.6 & 66.8 & 1.0 & 168.7 & 0.99903 \\
                % M1N & 66.2 & 66.3 & 66.5 & 1.0 & 170.3 \\
                % M5N & 65.9 & 66.2 & 66.3 & 1.0 & 170.5 & 0.99903 \\
                HN & 16.2 & 16.5 & 16.9 & 402.0 & 418.6 & 0.99980 \\
                \midrule
                \multicolumn{7}{@{\;}c}{\bf CLIP4Clip on MSR-VTT(1k split) for Text$\rightarrow$Video} \\
                \midrule
                % - & 43.9 & 70.6 & 80.7 & 2.0 & 16.0 & 0 \\
                % IS & 46.4 & 72.5 & 82.8 & 2.0 & 13.2 & 0 \\
                SN & \textbf{49.2} & \textbf{75.4} & \textbf{83.8} & \textbf{2.0} & \textbf{11.9} & 0 \\
                OTN & 48.0 & 48.3 & 48.3 & 53.0 & 263.0 & 0.99900 \\
                L2N & 47.8 & 49.2 & 49.2 & 26.0 & 256.7 & 0.99891 \\
                % M1N & 47.8 & 49.2 & 49.2 & 26.0 & 256.7 \\
                % M5N & 47.9 & 49.3 & 49.3 & 26.0 & 256.6 & 0.99889 \\
                HN & 48.1 & 48.4 & 48.4 & 51.0 & 262.2 & 0.99900 \\
	       \bottomrule
	    \end{tabular}
	}
\end{table}

\begin{figure}[tbp]
    \textit{\textbf{Query}: Two blond women sit outside as people walk by wearing casual clothing and some wearing bookbags}.\\
    \begin{minipage}{\textwidth}
        \begin{minipage}{0.02\textwidth}
            \rotatebox{90}{\textbf{Baseline}}
        \end{minipage}
        \begin{minipage}{0.47\textwidth}
            \includegraphics[width=\textwidth]{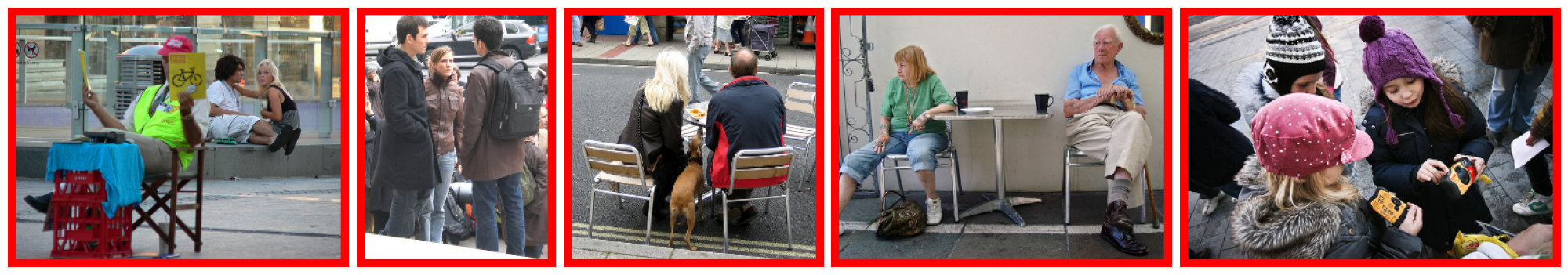}
        \end{minipage}
    \end{minipage}

    \begin{minipage}{\textwidth}
        \begin{minipage}{0.025\textwidth}
            \rotatebox{90}{\textbf{IS}}
        \end{minipage}%
        \begin{minipage}{0.47\textwidth}
            \includegraphics[width=\textwidth]{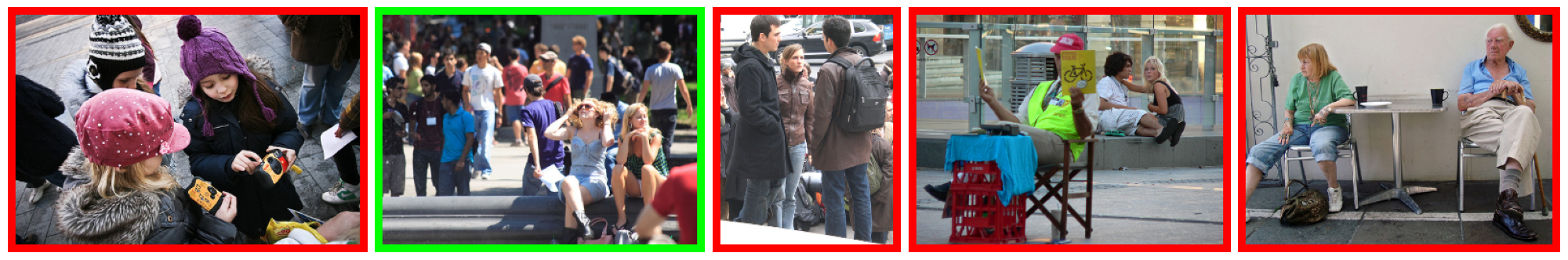}
        \end{minipage}
    \end{minipage}

    \begin{minipage}{\textwidth}
        \begin{minipage}{0.02\textwidth}
            \rotatebox{90}{\textbf{SN}}
        \end{minipage}
        \begin{minipage}{0.47\textwidth}
            \includegraphics[width=\textwidth]{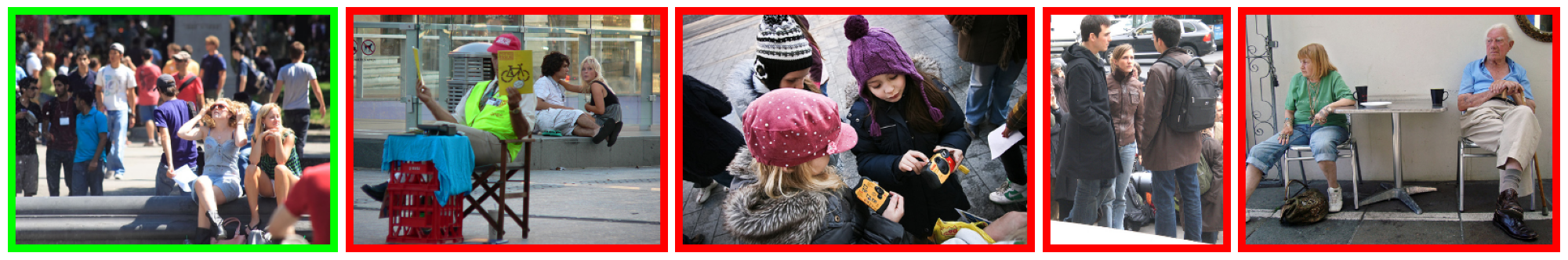}
        \end{minipage}
    \end{minipage}
    \caption{Visual comparisons of Top-5 text-to-image retrieval results on Flickr30k. \textcolor[rgb]{1.0, 0.0, 0.0}{Red} and \textcolor[rgb]{0.0, 1.0, 0.0}{green} boxes indicate incorrect and correct recalls, respectively.}
    \label{fig:t2i_comp}

    \textit{\textbf{Query}: Some one talking about top ten movies of the year}.\\
    \begin{minipage}{\textwidth}
        \begin{minipage}{0.02\textwidth}
            \rotatebox{90}{\textbf{Baseline}}
        \end{minipage}
        \begin{minipage}{0.47\textwidth}
            \includegraphics[width=\textwidth]{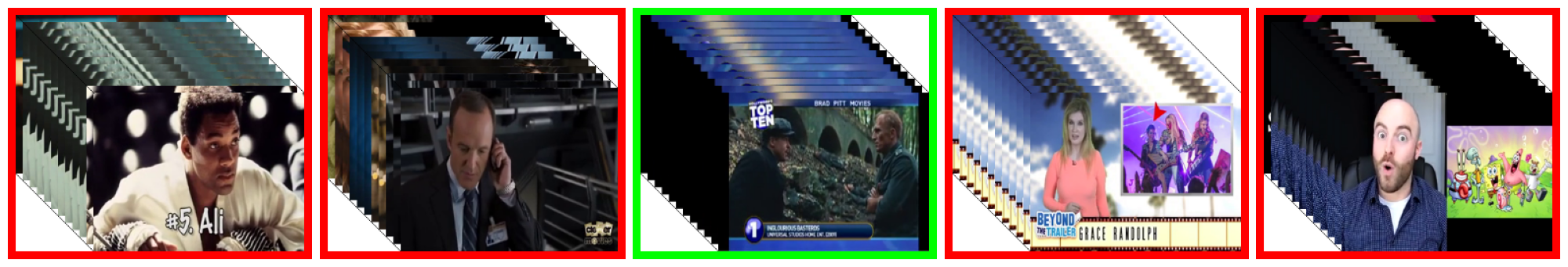}
        \end{minipage}
    \end{minipage}
    \begin{minipage}{\textwidth}
        \begin{minipage}{0.025\textwidth}
            \rotatebox{90}{\textbf{IS}}
        \end{minipage}%
        \begin{minipage}{0.47\textwidth}
            \includegraphics[width=\textwidth]{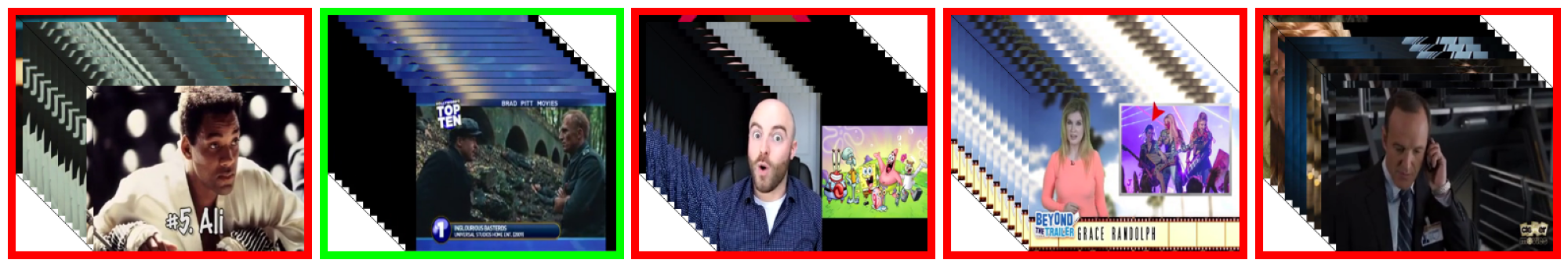}
        \end{minipage}
    \end{minipage}
    \begin{minipage}{\textwidth}
        \begin{minipage}{0.02\textwidth}
            \rotatebox{90}{\textbf{SN}}
        \end{minipage}
        \begin{minipage}{0.47\textwidth}
            \includegraphics[width=\textwidth]{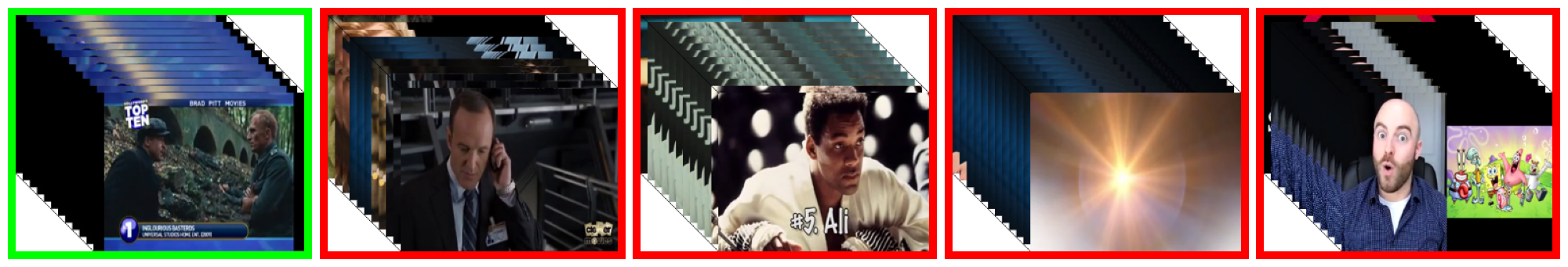}
        \end{minipage}
    \end{minipage}
    \caption{Visual comparisons of Top-5 text-to-video retrieval results on MSR-VTT. \textcolor[rgb]{1.0, 0.0, 0.0}{Red} and \textcolor[rgb]{0.0, 1.0, 0.0}{green} boxes indicate incorrect and correct recalls, respectively.}
    \label{fig:t2v_comp}
\end{figure}

\mypara{Query Bank Size.} We evaluate the impact of query bank size on four metrics: text-to-image retrieval performance (R@1), skewness, and Earth Mover’s Distance (EMD) between the query bank and both the target and testing query distributions. The results, shown in Figure~\ref{fig:ratio}, reveal that as the query bank size decreases, R@1 declines, while skewness and the two EMD measures generally increase. Figure~\ref{fig:ratio}(a) illustrates that for smaller query banks (e.g., smaller than the testing query set), IS outperforms DBSN, which can even underperform SN. This trend reverses as the query bank increases, demonstrating that SN and DBSN depend on sufficiently large query banks to be effective. Figure~\ref{fig:ratio}(b) shows that skewness is strictly correlated with R@1; DBSN further reduces skewness, thus achieving better retrieval performance than SN. IS consistently shows larger skewness than SN and DBSN, suggesting that skewness is a reliable indicator of retrieval performance. Figure~\ref{fig:ratio}(c) validates our Proposition~\ref{prop:dbsn}, showing that DBSN effectively narrows the distribution gap between the query bank and the target distribution, thereby reducing bias in target hubness estimation. Finally, Figure~\ref{fig:ratio}(d) shows that as the query bank size decreases, the distribution gap between the query bank and testing queries widens, leading to a corresponding drop in retrieval performance.

\mypara{Query Bank Source.} Figure~\ref{fig:querytype} illustrates the relationship between retrieval performance (R@1) and three other metrics when using query banks derived from different datasets. A noticeable trend is that using query banks from external datasets results in a significant distributional discrepancy between the query bank and both the testing queries and targets. This discrepancy leads to significantly lower retrieval performance (R@1) compared to when the training or validation queries from the same dataset are used as the query bank. Additionally, no strong statistical correlation is observed between R@1 and the other three metrics across different query banks. We attribute this to the high bias in target hubness estimation when using SN in scenarios with substantial distributional misalignment. This finding further reinforces that SN is effective when the distributional divergence between queries and targets is minimal.

\mypara{Normalization Type.} We conduct experiments to assess the density property of SN in comparison to other sparse normalization variants. Specifically, under the same marginal constraint outlined in Equation~\ref{eq:uab}, we refer to the solution of Equation~\ref{eq:proj} as Optimal Transport Normalization~(OTN) and the solution of Equation~\ref{eq:sot}, with an appropriate coefficient, as L2-constrained Normalization~(L2N). Additionally, we include Hungarian Normalization~(HN), which utilizes the Hungarian algorithm~\cite{kuhn1955hungarian} for Normalization, as discussed in~\cite{li2023progressive}. Table~\ref{tab:norm_type} shows that SN consistently outperforms the other methods across all settings. Although sparse normalization methods yield slightly lower R@1 scores than SN, their R@5 and R@10 results are nearly comparable. This can be attributed to the extreme sparsity of the resulting matrices, where 99.9\% of entries are zero. Notably, HN underperforms on Flickr30k due to its reliance on matching scenarios where the number of queries does not exceed the number of targets. Surplus query rows in the normalized matrix are filled with zeros, leading to performance degradation.

\subsection{Visualization.}
To qualitatively validate the effectiveness of the proposed SN, we present visual comparisons of the text-to-image retrieval task on Flickr30k in Figure~\ref{fig:t2i_comp} and the text-to-video retrieval task on MSR-VTT in Figure~\ref{fig:t2v_comp}. The results demonstrate that both IS and SN mitigate target hubness, thereby correcting erroneous retrievals of challenging samples observed in the baseline. Notably, SN outperforms both the baseline and IS in these visualizations, retrieving results with stronger semantic alignment to the textual queries and demonstrating its ability to address hubness-induced retrieval errors. For further examples and more detailed analysis, see Appendix.

\section{Conclusion}
\label{sec:5-conclu}
In this work, we examine the mechanism of Inverted Softmax (IS) and propose a probabilistic balancing framework to address the hubness problem in cross-modal retrieval. Within this framework, we introduce Sinkhorn Normalization (SN) to balance both target and query probabilities. To further address the limitations of single-bank SN in scenarios with unknown queries, we propose Dual Bank Sinkhorn Normalization (DBSN), which utilizes an additional target bank for more accurate target estimation. Comprehensive evaluations across various cross-modal retrieval tasks demonstrate the effectiveness of SN and DBSN.
\begin{acks}
     This work was supported by the National Key R\&D Program of China (2022ZD0160703), and the National Natural Science Foundation of China (62202422 and 62372408).
\end{acks}

\begin{appendix}
{\par\par\centering\fontsize{14}{16}\selectfont\textbf{Appendix}\par}

\newcommand{\beginsupplement}{%
    \setcounter{table}{0}
    \renewcommand{\thetable}{A\arabic{table}}%
    \setcounter{figure}{0}
    \renewcommand{\thefigure}{A\arabic{figure}}%
    \setcounter{algorithm}{0}
    \renewcommand{\thealgorithm}{A\arabic{algorithm}}%
    \setcounter{section}{0}
    \renewcommand{\thesection}{A}%
}
\beginsupplement

\section{Proofs}
\subsection{Proof of Proposition 1: IS functions to balance target probabilities.}
Recall the problem defined in Proposition~1:
\begin{align}
    \centering
    \begin{split}
        \vpi^\star &= \mathop{arg~max}\limits_{\vpi\in\mPi(\vb)}<\sS,\vpi>+\tau\mH(\vpi) \\
        \text{subject to}~~~~&\mPi(\vb)=\{\vpi\in\R_+^{m\times n}\mid\vpi^\top\ones_m=\vb\}
    \end{split}
    \label{eq:eot_r}
\end{align}
where  $\vb=\ones_n$ represents a $n$-dimensional normalized probabilistic vector of targets. We introduce the dual variable $ \vf \in \R^n$, and the Lagrangian of the Equation~\ref{eq:eot} is:
\begin{align}
    \centering
    \begin{split}
        \sL(\vpi, \vf) = <\sS,\vpi> + \tau\mH(\vpi) - <\vf,\vpi^\top\ones_m-\vb>
    \end{split}
    \label{eq:leot} % Lagrangian of entropy-constraint optimal transport
\end{align}
The first-order conditions are given by:
\begin{align}
    \centering
    \begin{split}
        \frac{ \partial \sL(\vpi, \vf) }{ \partial \vpi_{i,j} } = \sS_{i,j}-\tau\log(\vpi_{i,j}) - \vf_j = 0
    \end{split}
    \label{eq:fo_leot} % Lagrangian of entropy-constraint optimal transport
\end{align}

Thus we have $\vpi_{i,j} = \exp(\frac{\sS_{i,j}-\vf_j}{\tau})$ for every $i$ and $j$, for the optimal coupling $\vpi$ in the entropy-considered problem. Due to $\sum_{i}^{m}{\vpi_{i,j}}=1 $ for every $j$, we can calculate the Lagrangian parameter $\vf_j$ and the solution of the coupling is given by:
\begin{align}
    \centering
    \begin{split}
        % \frac{ } = \sS_{i,j}-\tau\log(\vpi_{i,j}) - \vf_j = 0
        \vpi_{i,j} = \frac{\exp(\frac{\sS_{i,j}}{\tau})}{\sum_{i}^{m}\exp(\frac{\sS_{i,j}}{\tau})}
    \end{split}
    \label{eq:solution} %
\end{align}
To this end, we demonstrate that Equation~1 in our main page gives the solution to the problem defined in Equation~\ref{eq:eot}. This confirms that the IS mechanism inherently maintains target probability normalization.

\subsection{Proof of Proposition 2: DBSN narrows the query-target gap.}
By constructing the auxiliary distribution $[0;B_t]$ as an intermediate state, according to the triangle inequality of the Earth Mover's Distance (EMD), we have, 
\begin{align}
    \centering
    EMD(\sB_q,\sB_t)&=EMD(\sB_q,[\mathbf{0};\sB_t]) \notag \\ 
            &<EMD(\sB_q,[\sT;\sB_t]) \\ 
            &< EMD(\sB_q,[\sT;\mathbf{0}])=EMD(\sB_q,\sT) \notag
    \label{eq:tri_equ}
\end{align}
Therefore, DBSN narrows the query-target gap.

\renewcommand{\thesection}{B}%
\section{Theoretical Comparison Between SN/DBSN and DIS/DualIS.}
Recall the Dynamic Inverted Softmax~(DIS) that
\begin{equation}
    \centering
    \mathop{DIS}(\sS_{i,j}) = 
        \begin{cases}
            \frac{\exp(\frac{\vq_i^\top\vt_j}{\tau})}{\sum_{u}\exp(\frac{\widehat{\vq_u}^\top\vt_j}{\tau})} & \text{if}~\vt_j \in \sT_c \\
            \vq_i^\top\vt_j & \text{otherwise}
        \end{cases}
    \label{eq:dis}
\end{equation}
where $\sT_c$ is a subset selected from the original target set $\sT$. The goal of forming such a subset is to minimize the distributional distance between the subset $\sT_c$ and the query bank $\sB_q$. For this purpose, \cite{bogolin2022cross} design a heuristic approach that composes $\sT$ by choosing the target located in the $k$-nearest neighbors of any query $\hat{\vq_i}$ in query bank $\sB_q=\{\widehat{\vq_u}\in\R^d\mid\|\vq_u\|_2=1,u=[1,\cdots,|\sB_q|]\}$. Formally, $\sT_c$ defined in \cite{bogolin2022cross} can be reformulated as:
 \begin{align}
    \centering
    \sT_c = \{ \vt_j\mid\vt_j\in \mathop{knn}(\widehat{\vq_u}), \forall u\in[1,\cdots,|\sB_q|]\}
    \label{eq:tc} %
\end{align}
here $k$ is set as a hyper-parameter. In practice, \cite{bogolin2022cross} directly sets $k=1$ as they observe that increasing $k$ does not lead to significant performance improvements. We further investigate the influence of $k$ for DIS. \cite{bogolin2022cross} demonstrates that even when setting $k=1$, $\sT_c$ encompasses more than 99\% of the targets in $\sT$, resulting in the distributional distance between $\sT_c$ and $\sB_q$ being nearly the same as that between $\sT$ and $\sB_q$. This observation reveals the intrinsic limitation of DIS; that is, subset selection is a computationally non-trivial yet marginally effective operation for hubness reduction.
% TODO: table: 

Authors in \cite{wang2023balance} suppose that the `hub' target will be frequently retrieved by both queries in another modality and targets within the same modality. Based on this, they further utilize an additional target bank $\sB_t=\{\widehat{\vt_v}\in\R^d\mid\|\vt_v\|_2=1,v=[1,\cdots,|\sB_t|]\}$ to assist the query bank $\sB_q$ in reducing the hubness of targets in $\sT$. Specifically, they introduce Dual Inverted Softmax~(DualIS) as:
\begin{align}
    \centering
    \begin{split}
        \mathop{DualIS}(\sS_{i,j}) = \frac{\mathop{exp}(\frac{\vq_i^\top\vt_j}{\tau_1})}{\sum_{u}\mathop{exp}(\frac{\widehat{\vq_u}^\top\vt_j}{\tau_1})}* \frac{\mathop{exp}(\frac{\vq_i^\top\vt_j}{\tau_2})}{\sum_{v}\mathop{exp}(\frac{\widehat{\vt_v}^\top\vt_j}{\tau_2})}
    \end{split}
    \label{eq:dualis}
\end{align}
Similar to Equation 7 in our main paper, DualIS can be reformulated as:
\begin{align}
    \centering
    \begin{split}
        \mathop{DualIS}(\sS_{i,j}) &= \exp(\frac{\vq_i^\top\vt_j - \hbar_{\sB_q}(\vt_j)-\hbar_{\sB_t}(\vt_j)}{\lambda}) \\
        &\text{subject to}~~\lambda = \frac{\tau_1\tau_2}{\tau_1+\tau_2} \\
        \hbar_{\sB_q}(\vt_j) &= \lambda\mathop{LogSumExp}\limits_{u}(\frac{\vq_u^\top\vt_j}{\tau_1}) \\
        \hbar_{\sB_t}(\vt_j) &= \lambda\mathop{LogSumExp}\limits_{v}(\frac{\vt_v^\top\vt_j}{\tau_2}) \\
    \end{split}
    \label{eq:dualis_bias}
\end{align}

The mechanism behind Equation~\ref{eq:dualis_bias} is that, due to the validity of Inequality $EMD(\sB_t,\sT)<EMD(\sQ,\sT)<EMD(\sB_q,\sT)$, the estimated target hubness approximately meets $\hbar_{\sB_q}(\vt_j)<\hbar_{Q}(\vt_j)<\hbar_{\sB_t}(\vt_j)$. Therefore, one can approximate $\hbar_{Q}(\vt_j)$ by the weighted sum of $\hbar_{\sB_q}(\vt_j)$ and $\hbar_{\sB_t}(\vt_j)$, the weighting coefficients $\tau_1$ and $\tau_2$. However, in practice, due to the significant modality gap~\cite{liang2022mind}, $EMD(\sB_t,\sT)<<EMD(\sQ,\sT)$, even with carefully tuned $\tau_1$ and $\tau_2$, DualIS does not achieve a significant performance improvement over IS. This demonstrates the sensitivity of DualIS to modality discrepancies, which is empirically validated by our ablation studies (see \S~4.2 in our main page and Table~\ref{tab:zsi2t_dbsn}-\ref{tab:xpool_dbsn}).

\renewcommand{\thesection}{C}%
\section{Algorithms}

Algorithms~\ref{alg:a1_sn} and~\ref{alg:a2_dbsn} provide the implementation details of SN and DBSN, respectively, for practical retrieval scenarios.

\renewcommand{\thesection}{D}%
\section{More Comparisons}
Extended results in Tables~\ref{tab:activity&lsmdc}-\ref{tab:sn_ic} complement Figure ~\ref{fig:radar}'s visualization, showing that SN outperforms state-of-the-art methods in query-aware scenarios. Results in Tables~\ref{tab:zsi2t_dbsn}-\ref{tab:xpool_dbsn} complement Table~\ref{tab:dbsn}, showing that DBSN improves SN in query-agnostic scenarios.

\begin{table*}[htbp]
    \centering
    \small
    \caption{Text-to-Video comparisons on Activitynet~\cite{caba2015activitynet} and LSMDC~\cite{rohrbach2015dataset}. All methods employ the same backbone CLIP VIT-B/32. $\ddag$ denotes results from our implementation.}
    \label{tab:activity&lsmdc}
    {
        \tabcolsep 5 pt
	    \begin{tabular}{ lc cccc cccc  cccc cccc}
	        \addlinespace
	        \toprule
                \multicolumn{1}{c}{\multirow{3}{*}{\symtext{Method}}} & \multicolumn{1}{c}{\multirow{3}{*}{\symtext{Norm}}}  & \multicolumn{8}{c}{Activitynet} & \multicolumn{8}{c}{LSMDC} \\ \cmidrule(lr){3-10} \cmidrule(lr){11-18}
                & & \multicolumn{4}{c}{Text$\rightarrow$Video} & \multicolumn{4}{c}{Video$\rightarrow$Text} & \multicolumn{4}{c}{Text$\rightarrow$Video} & \multicolumn{4}{c}{Video$\rightarrow$Text} \\ \cmidrule(lr){3-6} \cmidrule(lr){7-10} \cmidrule(lr){11-14} \cmidrule(lr){15-18}
                 &  &  R@1 & R@5 & R@10 & MnR$\downarrow$  &  R@1 & R@5 & R@10 & MnR$\downarrow$ &  R@1 & R@5 & R@10 & MnR$\downarrow$ &  R@1 & R@5 & R@10 & MnR$\downarrow$ \\ 
                \midrule
                \multicolumn{2}{l}{\textcolor{gray}{CLIP4Clip~\cite{luo2022clip4clip}}} & \textcolor{gray}{40.5} & \textcolor{gray}{72.4} & \textcolor{gray}{98.1} & \textcolor{gray}{7.4} & \textcolor{gray}{42.5} & \textcolor{gray}{74.1} & \textcolor{gray}{85.8} & \textcolor{gray}{6.6} & \textcolor{gray}{20.7} & \textcolor{gray}{38.9} & \textcolor{gray}{47.2} & \textcolor{gray}{65.3} & \textcolor{gray}{20.6} & \textcolor{gray}{39.4} & \textcolor{gray}{47.5} & \textcolor{gray}{56.7}  \\
                \multicolumn{2}{l}{\textcolor{gray}{CLIP2Video~\cite{fang2021clip2video}}} & \textcolor{gray}{45.6} & \textcolor{gray}{72.6} & \textcolor{gray}{81.7} & \textcolor{gray}{14.6} & \textcolor{gray}{43.5} & \textcolor{gray}{72.3} & \textcolor{gray}{82.1} & \textcolor{gray}{10.2} & \textcolor{gray}{-} & \textcolor{gray}{-} & \textcolor{gray}{-} & \textcolor{gray}{-} & \textcolor{gray}{-} & \textcolor{gray}{-} & \textcolor{gray}{-} & \textcolor{gray}{-} \\
                \multicolumn{2}{l}{\textcolor{gray}{X-CLIP~\cite{ma2022x}}} & \textcolor{gray}{44.3} & \textcolor{gray}{74.1} & \textcolor{gray}{-} & \textcolor{gray}{7.9} & \textcolor{gray}{43.9} & \textcolor{gray}{73.9} & \textcolor{gray}{-} & \textcolor{gray}{7.6} & \textcolor{gray}{23.3} & \textcolor{gray}{43.0} & \textcolor{gray}{-} & \textcolor{gray}{56.0} & \textcolor{gray}{22.5} & \textcolor{gray}{42.2} & \textcolor{gray}{-} & \textcolor{gray}{50.7} \\
                \multicolumn{2}{l}{\textcolor{gray}{DRL~\cite{wang2022disentangled}}} & \textcolor{gray}{44.2} & \textcolor{gray}{74.5} & \textcolor{gray}{86.1} & \textcolor{gray}{-} & \textcolor{gray}{42.2} & \textcolor{gray}{74.0} & \textcolor{gray}{86.2} & \textcolor{gray}{-} & \textcolor{gray}{24.9} & \textcolor{gray}{45.7} & \textcolor{gray}{55.3} & \textcolor{gray}{-} & \textcolor{gray}{24.9} & \textcolor{gray}{44.1} & \textcolor{gray}{53.8} & \textcolor{gray}{-} \\
                \multicolumn{2}{l}{\textcolor{gray}{X-Pool~\cite{gorti2022x}}}& \textcolor{gray}{-} & \textcolor{gray}{-} & \textcolor{gray}{-} & \textcolor{gray}{-} & \textcolor{gray}{-} & \textcolor{gray}{-} & \textcolor{gray}{-} & \textcolor{gray}{-} & \textcolor{gray}{25.2} & \textcolor{gray}{43.7} & \textcolor{gray}{53.5} & \textcolor{gray}{53.2} & \textcolor{gray}{22.7} & \textcolor{gray}{42.6} & \textcolor{gray}{51.2} & \textcolor{gray}{47.4} \\
                \multicolumn{2}{l}{\textcolor{gray}{TS2-Net~\cite{liu2022ts2}}} & \textcolor{gray}{41.0} & \textcolor{gray}{73.6} & \textcolor{gray}{84.5} & \textcolor{gray}{8.4} & \textcolor{gray}{-} & \textcolor{gray}{-} & \textcolor{gray}{-} & \textcolor{gray}{-} & \textcolor{gray}{23.4} & \textcolor{gray}{42.3} & \textcolor{gray}{50.9} & \textcolor{gray}{56.9} & \textcolor{gray}{-} & \textcolor{gray}{-} & \textcolor{gray}{-} & \textcolor{gray}{-} \\
                \multicolumn{2}{l}{\textcolor{gray}{UATVR~\cite{fang2023uatvr}}} & \textcolor{gray}{47.5} & \textcolor{gray}{73.9} & \textcolor{gray}{83.5} & \textcolor{gray}{12.3} & \textcolor{gray}{46.9} & \textcolor{gray}{73.8} & \textcolor{gray}{83.8} & \textcolor{gray}{8.6} & \textcolor{gray}{43.1} & \textcolor{gray}{71.8} & \textcolor{gray}{82.3} & \textcolor{gray}{15.1} & \textcolor{gray}{-} & \textcolor{gray}{-} & \textcolor{gray}{-} & \textcolor{gray}{-} \\
                \multicolumn{2}{l}{\textcolor{gray}{ProST~\cite{li2023progressive}}} & \textcolor{gray}{-} & \textcolor{gray}{-} & \textcolor{gray}{-} & \textcolor{gray}{-} & \textcolor{gray}{-} & \textcolor{gray}{-} & \textcolor{gray}{-} & \textcolor{gray}{-} & \textcolor{gray}{24.1} & \textcolor{gray}{42.5} & \textcolor{gray}{51.6} & \textcolor{gray}{54.6} & \textcolor{gray}{-} & \textcolor{gray}{-} & \textcolor{gray}{-} & \textcolor{gray}{-} \\
                \midrule
                \multicolumn{2}{l}{CLIP4Clip~\cite{luo2022clip4clip}$\ddag$} & 41.0 & 73.3 & 85.2 & 6.8 & 42.5 & 75.2 & 87.1 & 6.1 & 20.3 & 38.9 & 47.0 & 54.1 & 19.9 & 38.8 & 48.5 & 64.9 \\
                & + IS & 51.9 & 80.1 & 89.6 & 5.4 & 51.3 & 79.9 & 89.4 & 5.0 & 22.9 & 41.9 & 50.3 & 51.5 & 20.7 & 41.4 & 50.5 & 62.2 \\
                                            \rowcolor{gray} & + SN & 54.7 & 82.3 & 91.3 & 4.6 & 55.8 & 82.5 & 91.3 & 4.4 & 22.9 & 41.7 & 49.9 & 51.1 & 21.1 & 42.1 & 50.7 & 58.9  \\
                \midrule
                \multicolumn{2}{l}{DRL~\cite{wang2022disentangled}$\ddag$} & 41.9 & 73.9 & 86.2 & 6.2 & 43.0 & 76.1 & 87.7 & 5.7 & 20.7 & 40.2 & 48.5 & 63.6 & 21.0 & 39.4 & 48.9 & 53.9  \\
                                      & + IS & 54.5 & 82.4 & 90.5 & 5.1 & 54.5 & 81.5 & 90.5 & 4.9 & 22.0 & 41.6 & 50.1 & 61.7 & 23.7 & 42.2 & 51.1 & 50.2 \\
                                      \rowcolor{gray} & + SN & 57.7 & 84.4 & 92.0 & 4.3 & 57.6 & 83.9 & 92.1 & 4.2 & 22.1 & 42.8 & 50.7 & 58.4 & 24.7 & 42.9 & 51.8 & 50.0 \\
                \midrule
                \multicolumn{2}{l}{X-Pool~\cite{gorti2022x}$\ddag$} & 41.5 & 72.6 & 85.5 & 7.0 & 40.7 & 74.4 & 86.4 & 6.1 & 22.3 & 40.1 & 49.4 & 53.3 & 23.1 & 41.5 & 49.7 & 59.5  \\
                                        & + IS & 50.2 & 78.6 & 89.7 & 5.2 & 49.9 & 79.9 & 89.7 & 4.9 & 24.0 & 43.3 & 52.4 & 49.9 & 23.3 & 42.2 & 51.3 & 55.6  \\
                                        \rowcolor{gray} & + SN & 53.6 & 82.2 & 90.9 & 4.4 & 53.5 & 81.9 & 90.8 & 4.5 & 24.9 & 43.0 & 52.3 & 49.1 & 23.7 & 43.5 & 52.3 & 52.5 \\
	       \bottomrule
	    \end{tabular}
	}
\end{table*}

\begin{table*}[htbp]
    \centering
    \small
    \caption{Text-to-Video comparisons on Vatex~\cite{wang2019vatex} and MSVD~\cite{wu2017deep}. All methods employ the same backbone CLIP VIT-B/32. $\ddag$ denotes results from our implementation.}
    % We refer to the \SM for extensions of this table with more baselines and COCO 5K results.
    % \resizebox{\textwidth}{!}
    \label{tab:vatex&msvd}
    {
        \tabcolsep 5 pt
	    \begin{tabular}{ lc cccc cccc  cccc cccc}
	        \addlinespace
	        \toprule
                \multicolumn{1}{c}{\multirow{3}{*}{\symtext{Method}}} & \multicolumn{1}{c}{\multirow{3}{*}{\symtext{Norm}}}  & \multicolumn{8}{c}{Vatex} & \multicolumn{8}{c}{MSVD} \\ \cmidrule(lr){3-10} \cmidrule(lr){11-18}
                & & \multicolumn{4}{c}{Text$\rightarrow$Video} & \multicolumn{4}{c}{Video$\rightarrow$Text} & \multicolumn{4}{c}{Text$\rightarrow$Video} & \multicolumn{4}{c}{Video$\rightarrow$Text} \\ \cmidrule(lr){3-6} \cmidrule(lr){7-10} \cmidrule(lr){11-14} \cmidrule(lr){15-18}
                 &  &  R@1 & R@5 & R@10 & MnR$\downarrow$  &  R@1 & R@5 & R@10 & MnR$\downarrow$ &  R@1 & R@5 & R@10 & MnR$\downarrow$ &  R@1 & R@5 & R@10 & MnR$\downarrow$ \\ 
                \midrule
                \multicolumn{2}{l}{\textcolor{gray}{CLIP4Clip~\cite{luo2022clip4clip}}} & \textcolor{gray}{-} & \textcolor{gray}{-} & \textcolor{gray}{-} & \textcolor{gray}{-} & \textcolor{gray}{-} & \textcolor{gray}{-} & \textcolor{gray}{-} & \textcolor{gray}{-} & \textcolor{gray}{46.2} & \textcolor{gray}{76.1} & \textcolor{gray}{84.6} & \textcolor{gray}{10.0} & \textcolor{gray}{56.6} & \textcolor{gray}{79.7} & \textcolor{gray}{84.3} & \textcolor{gray}{7.6} \\
                \multicolumn{2}{l}{\textcolor{gray}{CLIP2Video~\cite{fang2021clip2video}}} & \textcolor{gray}{-} & \textcolor{gray}{-} & \textcolor{gray}{-} & \textcolor{gray}{-} & \textcolor{gray}{-} & \textcolor{gray}{-} & \textcolor{gray}{-} & \textcolor{gray}{-} & \textcolor{gray}{47.0} & \textcolor{gray}{76.8} & \textcolor{gray}{85.9} & \textcolor{gray}{9.6} & \textcolor{gray}{58.7} & \textcolor{gray}{85.6} & \textcolor{gray}{91.6} & \textcolor{gray}{4.3} \\
                \multicolumn{2}{l}{\textcolor{gray}{X-CLIP~\cite{ma2022x}}} & \textcolor{gray}{-} & \textcolor{gray}{-} & \textcolor{gray}{-} & \textcolor{gray}{-} & \textcolor{gray}{-} & \textcolor{gray}{-} & \textcolor{gray}{-} & \textcolor{gray}{-} & \textcolor{gray}{47.1} & \textcolor{gray}{77.8} & \textcolor{gray}{-} & \textcolor{gray}{9.5} & \textcolor{gray}{60.9} & \textcolor{gray}{87.8} & \textcolor{gray}{-} & \textcolor{gray}{4.7} \\
                \multicolumn{2}{l}{\textcolor{gray}{DRL~\cite{wang2022disentangled}}} & \textcolor{gray}{63.5} & \textcolor{gray}{91.7} & \textcolor{gray}{96.5} & \textcolor{gray}{-} & \textcolor{gray}{77.0} & \textcolor{gray}{98.0} & \textcolor{gray}{99.4} & \textcolor{gray}{-} & \textcolor{gray}{48.3} & \textcolor{gray}{79.1} & \textcolor{gray}{87.3} & \textcolor{gray}{-} & \textcolor{gray}{62.3} & \textcolor{gray}{86.3} & \textcolor{gray}{92.2} & \textcolor{gray}{-} \\
                \multicolumn{2}{l}{\textcolor{gray}{X-Pool~\cite{gorti2022x}}}& \textcolor{gray}{-} & \textcolor{gray}{-} & \textcolor{gray}{-} & \textcolor{gray}{-} & \textcolor{gray}{-} & \textcolor{gray}{-} & \textcolor{gray}{-} & \textcolor{gray}{-} & \textcolor{gray}{25.2} & \textcolor{gray}{43.7} & \textcolor{gray}{53.5} & \textcolor{gray}{53.2} & \textcolor{gray}{22.7} & \textcolor{gray}{42.6} & \textcolor{gray}{51.2} & \textcolor{gray}{47.4} \\
                \multicolumn{2}{l}{\textcolor{gray}{TS2-Net~\cite{liu2022ts2}}} & \textcolor{gray}{59.1} & \textcolor{gray}{90.0} & \textcolor{gray}{95.2} & \textcolor{gray}{3.5} & \textcolor{gray}{-} & \textcolor{gray}{-} & \textcolor{gray}{-} & \textcolor{gray}{-} & \textcolor{gray}{-} & \textcolor{gray}{-} & \textcolor{gray}{-} & \textcolor{gray}{-} & \textcolor{gray}{-} & \textcolor{gray}{-} & \textcolor{gray}{-} & \textcolor{gray}{-} \\
                % \textcolor{gray}{CAMoE}       &         & \textcolor{gray}{47.3} & \textcolor{gray}{74.2} & \textcolor{gray}{84.5} & \textcolor{gray}{2.0}   & \textcolor{gray}{11.9}        \\
                \multicolumn{2}{l}{\textcolor{gray}{UATVR~\cite{fang2023uatvr}}} & \textcolor{gray}{61.3} & \textcolor{gray}{91.0} & \textcolor{gray}{95.6} & \textcolor{gray}{3.3} & \textcolor{gray}{-} & \textcolor{gray}{-} & \textcolor{gray}{-} & \textcolor{gray}{-} & \textcolor{gray}{46.0} & \textcolor{gray}{76.3} & \textcolor{gray}{85.1} & \textcolor{gray}{10.4} & \textcolor{gray}{-} & \textcolor{gray}{-} & \textcolor{gray}{-} & \textcolor{gray}{-} \\ 
                \multicolumn{2}{l}{\textcolor{gray}{ProST~\cite{li2023progressive}}} & \textcolor{gray}{60.6} & \textcolor{gray}{90.5} & \textcolor{gray}{95.4} &  \textcolor{gray}{3.4} & \textcolor{gray}{-} & \textcolor{gray}{-} & \textcolor{gray}{-} & \textcolor{gray}{-} & \textcolor{gray}{-} & \textcolor{gray}{-} & \textcolor{gray}{8-} & \textcolor{gray}{-} & \textcolor{gray}{-} & \textcolor{gray}{-} & \textcolor{gray}{-} & \textcolor{gray}{-} \\ 
                \midrule
                \multicolumn{2}{l}{CLIP4Clip~\cite{luo2022clip4clip}$\ddag$} & 57.7 & 89.4 & 94.8 & 3.6 & 75.4 & 95.0 & 97.5 & 2.0 & 46.2 & 75.2 & 84.4 & 10.1 & 62.4 & 89.1 & 93.6 & 3.4  \\
                                            & + IS & 59.1 & 89.5 & 94.7 & 3.8 & 83.2 & 96.7 & 98.7 & 1.6 & 47.9 & 76.9 & 85.1 & 10.0 & 75.4 & 92.8 & 96.5 & 2.3 \\
                                            \rowcolor{gray} & + SN & 64.0 & 91.9 & 96.2 & 3.0 & 85.6 & 97.5 & 99.3 & 1.5 & 50.7 & 78.7 & 86.3 & 9.5 & 74.7 & 92.7 & 96.9 & 2.2 \\
                \midrule
                \multicolumn{2}{l}{DRL~\cite{wang2022disentangled}$\ddag$} & 57.6 & 90.1 & 95.4 & 3.4 & 75.1 & 95.3 & 98.1 & 2.0 & 46.3 & 75.2 & 84.6 & 10.1 & 64.6 & 90.1 & 95.4 & 3.1  \\
                                      & + IS & 61.1 & 90.7 & 95.5 & 3.4 & 84.3 & 97.5 & 99.1 & 1.5 & 47.8 & 76.7 & 85.4 & 10.0 & 72.6 & 93.9 & 96.2 & 2.2  \\
                                      \rowcolor{gray} & + SN & 65.2 & 92.5 & 96.5 & 2.9 & 86.1 & 97.9 & 99.2 & 1.4 & 50.9 & 78.8 & 86.6 & 9.5 & 74.7 & 93.7 & 97.2 & 2.0 \\
                \midrule
                \multicolumn{2}{l}{X-Pool~\cite{gorti2022x}$\ddag$} & 59.4 & 90.3 & 95.5 & 3.3 & 75.1 & 94.5 & 98.2 & 1.9 & 47.1 & 76.3 & 84.8 & 9.9 & 65.2 & 92.3 & 95.9 & 2.9 \\
                                        & + IS &60.6 & 90.4 & 95.3 & 3.4 & 84.0 & 96.8 & 98.8 & 1.5 & 48.2 & 77.7 & 85.8 & 9.8 & 75.4 & 93.5 & 96.4 & 2.3  \\
                                        \rowcolor{gray} & + SN & 64.5 & 92.4 & 96.5 & 2.9 & 85.7 & 98.1 & 99.3 & 1.4 & 51.4 & 79.4 & 86.7 & 9.3 & 75.2 & 92.0 & 95.6 & 2.4 \\
	       \bottomrule
	    \end{tabular}
	}
\end{table*}

\begin{table}[tbp]
    \centering
    \small
    \caption{Text-to-Video comparisons of CE-based models~\cite{liu2019use,croitoru2021teachtext} on MSR-VTT and Didemo.}
    \label{tab:ce}
    {
        \tabcolsep 4 pt
	    \begin{tabular}{ lc ccccc }
	        \toprule
                Method &  Normalization &  R@1$\uparrow$ & R@5$\uparrow$ & R@10$\uparrow$ & MdR$\downarrow$ & MnR$\downarrow$ \\  
                \midrule
                \multicolumn{7}{@{\;}c}{\bf MSR-VTT~(full split)~~Text$\rightarrow$Video} \\
                \midrule
                \textcolor{gray}{RoME} & - & \textcolor{gray}{10.7} & \textcolor{gray}{29.6} & \textcolor{gray}{41.2} & \textcolor{gray}{17.0} & \textcolor{gray}{-} \\
                \textcolor{gray}{Frozen} & - & \textcolor{gray}{32.5} & \textcolor{gray}{61.5} & \textcolor{gray}{71.2} & \textcolor{gray}{-} & \textcolor{gray}{-}     \\
                \multicolumn{2}{l}{\textcolor{gray}{T2VLAD~\cite{wang2021t2vlad}}} & \textcolor{gray}{12.7} & \textcolor{gray}{34.8} & \textcolor{gray}{47.1} & \textcolor{gray}{12.0} & \textcolor{gray}{-}     \\
                \midrule
                \multicolumn{2}{l}{CE+~\cite{liu2019use}} & 13.7 & 36.4 & 49.2 & 11.0 & 68.6  \\
                                                         & + IS & 14.9 & 38.3 & 50.8 & 10.0 & 68.2   \\
                                                         \rowcolor{gray}& + SN & 15.9 & 39.8 & 52.4 & 9.0 & 64.5  \\
                \midrule
                \multicolumn{2}{l}{TT-CE+~\cite{croitoru2021teachtext}} & 14.6 & 37.8 & 50.8 & 10.0 & 63.5   \\
                                                             & + IS & 16.1 & 39.8 & 52.7 & 9.0 & 63.8   \\
                                                             \rowcolor{gray}& + SN & \textbf{17.1} & \textbf{41.6} & \textbf{54.3} & \textbf{8.0} & \textbf{60.1}  \\
                \midrule
                \multicolumn{7}{@{\;}c}{\bf Didemo~~Text$\rightarrow$Video} \\
                \midrule
                \multicolumn{2}{l}{CE+~\cite{liu2019use}} & 17.0 & 43.2 & 56.1 & 8.0 & 46.8 \\
                                             & + IS & 18.5 & 44.4 & 55.5 & 7.0 & 45.8  \\
                                             \rowcolor{gray}& + SN & 20.9 & 47.5 & 60.0 & 6.0 & 42.5  \\
                                                    \midrule
                \multicolumn{2}{l}{TT-CE+~\cite{croitoru2021teachtext}} & 21.3 & 49.6 & 61.4 & 6.0 & 38.6  \\
                                         & + IS & 24.3 & 52.8 & 64.5 & 5.0 & 34.5 \\
                                        \rowcolor{gray} & + SN & 25.9 & 53.0 & 64.6 & 4.0 & 34.6 \\
	       \bottomrule
	    \end{tabular}
	}
\end{table}

\begin{table*}[htbp]
    \centering
    \small
    \caption{Medical Image-Text Retrieval Results of PLIP~\cite{huang2023visual} on PubMed~\cite{gamper2021multiple} and BookSet~\cite{gamper2021multiple}.}
    % We refer to the \SM for extensions of this table with more baselines and COCO 5K results.
    % \resizebox{\textwidth}{!}
    \label{tab:plip}
    {
        \tabcolsep 5 pt
	    \begin{tabular}{ lc cccc cccc  cccc cccc}
	        \addlinespace
	        \toprule
                \multicolumn{1}{c}{\multirow{3}{*}{\symtext{Method}}} & \multicolumn{1}{c}{\multirow{3}{*}{\symtext{Norm}}}  & \multicolumn{8}{c}{PubMed} & \multicolumn{8}{c}{BookSet} \\ \cmidrule(lr){3-10} \cmidrule(lr){11-18}
                & & \multicolumn{4}{c}{Text$\rightarrow$Image} & \multicolumn{4}{c}{Image$\rightarrow$Text} & \multicolumn{4}{c}{Text$\rightarrow$Image} & \multicolumn{4}{c}{Image$\rightarrow$Text} \\ \cmidrule(lr){3-6} \cmidrule(lr){7-10} \cmidrule(lr){11-14} \cmidrule(lr){15-18}
                 &  &  R@1 & R@5 & R@10 & MnR$\downarrow$  &  R@1 & R@5 & R@10 & MnR$\downarrow$ &  R@1 & R@5 & R@10 & MnR$\downarrow$ &  R@1 & R@5 & R@10 & MnR$\downarrow$ \\ 
                \midrule
                \multicolumn{2}{l}{zs PLIP~\cite{huang2023visual}} & 1.2 & 5.2 & 7.9 & 538.7 & 1.4 & 5.1 & 8.2 & 654.0 & 0.7 & 2.8 & 4.8 & 692.9 & 0.8 & 3.2 & 5.7 & 676.2 \\
                            & + IS & 1.4 & 5.1 & 8.2 & 654.0 & 1.8 & 5.9 & 9.5 & 509.8&0.8 & 3.2 & 5.7 & 676.2 & 0.9 & 3.5 & 6.3 & 628.4 \\
                            \rowcolor{gray} & + SN & 1.6 & 6.3 & 10.0 & 485.3 & 1.6 & 6.1 & 10.1 & 478.7 & 1.5 & 4.8 & 7.7 & 548.3 &1.4 & 5.0 & 7.3 & 555.1 \\
	       \bottomrule
	    \end{tabular}
	}
\end{table*}
\begin{table*}[htbp]
    \centering
    \small
    \caption{Imgae-to-Image comparisons on  CUB-200-2011 (CUB)~\cite{welinder2010caltech}, Cars-196 (Cars)~\cite{krause20133d}, Stanford Online Product (SOP)~\cite{oh2016deep} and In-shop Clothes Retrieval (In-Shop)~\cite{liu2016deepfashion}. Results of ResNet-50, DeiT-S, DINO and ViT-S are copied from~\cite{ermolov2022hyperbolic}. Note that the results on the Cars dataset are generally lower than those reported in~\cite{ermolov2022hyperbolic}.}
    % \vspace{-7pt}
    \label{tab:i2i}
    {
        \tabcolsep 5 pt
	    \begin{tabular}{ lc cccc cccc cccc cccc}
	        % \addlinespace
	        \toprule
                \multicolumn{1}{c}{\multirow{2}{*}{\symtext{Method}}} & \multicolumn{1}{c}{\multirow{2}{*}{\symtext{Norm}}} & \multicolumn{4}{c}{CUB} & \multicolumn{4}{c}{Cars} & \multicolumn{4}{c}{SOP} & \multicolumn{4}{c}{In-Shop} \\ \cmidrule(lr){3-6}\cmidrule(lr){7-10} \cmidrule(lr){11-14} \cmidrule(lr){15-18} 
                & & R@1 & R@2 & R@4 & R@8 & R@1 & R@2 & R@4 & R@8 & R@1 & R@10 & R@100 & R@1000 & R@1 & R@10 & R@20 & R@30 \\ 
                \midrule
                % \textcolor{gray}{RoME} & - & \textcolor{gray}{10.7} & \textcolor{gray}{29.6} & \textcolor{gray}{41.2} & \textcolor{gray}{17.0} & \textcolor{gray}{-} \\
                % \textcolor{gray}{Frozen} & - & \textcolor{gray}{32.5} & \textcolor{gray}{61.5} & \textcolor{gray}{71.2} & \textcolor{gray}{-} & \textcolor{gray}{-}     \\
                \multicolumn{2}{l}{\textcolor{gray}{ResNet-50~\cite{he2016deep}}} & \textcolor{gray}{41.2} & \textcolor{gray}{53.8} & \textcolor{gray}{66.3} & \textcolor{gray}{77.5} & \textcolor{gray}{41.4} & \textcolor{gray}{53.6} & \textcolor{gray}{66.1} & \textcolor{gray}{76.6} & \textcolor{gray}{50.6} & \textcolor{gray}{66.7} & \textcolor{gray}{80.7} & \textcolor{gray}{93.0} & \textcolor{gray}{25.8} & \textcolor{gray}{49.1} & \textcolor{gray}{56.4} & \textcolor{gray}{60.5} \\
                \multicolumn{2}{l}{\textcolor{gray}{DeiT-S~\cite{touvron2021training}}} & \textcolor{gray}{70.6} & \textcolor{gray}{81.3} & \textcolor{gray}{88.7} & \textcolor{gray}{93.5} & \textcolor{gray}{52.8} & \textcolor{gray}{65.1} & \textcolor{gray}{76.2} & \textcolor{gray}{85.3} & \textcolor{gray}{58.3} & \textcolor{gray}{73.9} & \textcolor{gray}{85.9} & \textcolor{gray}{95.4} & \textcolor{gray}{37.9} & \textcolor{gray}{64.7} & \textcolor{gray}{72.1} & \textcolor{gray}{75.9} \\
                \multicolumn{2}{l}{\textcolor{gray}{DINO~\cite{caron2021emerging}}} & \textcolor{gray}{70.8} & \textcolor{gray}{81.1} & \textcolor{gray}{88.8} & \textcolor{gray}{93.5} & \textcolor{gray}{42.9} & \textcolor{gray}{53.9} & \textcolor{gray}{64.2} & \textcolor{gray}{74.4} & \textcolor{gray}{63.4} & \textcolor{gray}{78.1} & \textcolor{gray}{88.3} & \textcolor{gray}{96.0} & \textcolor{gray}{46.1} & \textcolor{gray}{71.1} & \textcolor{gray}{77.5} & \textcolor{gray}{81.1} \\
                \multicolumn{2}{l}{\textcolor{gray}{ViT-S~\cite{dosovitskiy2020image}}} & \textcolor{gray}{83.1} & \textcolor{gray}{90.4} & \textcolor{gray}{94.4} & \textcolor{gray}{96.5} & \textcolor{gray}{47.8} & \textcolor{gray}{60.2} & \textcolor{gray}{72.2} & \textcolor{gray}{82.6} & \textcolor{gray}{62.1} & \textcolor{gray}{77.7} & \textcolor{gray}{89.0} & \textcolor{gray}{96.8} & \textcolor{gray}{43.2} & \textcolor{gray}{70.2} & \textcolor{gray}{76.7} & \textcolor{gray}{80.5} \\
                \midrule
                \multicolumn{2}{l}{DINO~\cite{caron2021emerging}} 
                        & 69.83 & 80.54 & 88.30 & 92.88 & 33.97 & 43.48 & 54.51 & 65.88 & 63.41 & 78.07 & 88.27 & 95.96 & 45.9 & 71.0 & 77.4 & 81.0 \\
                        & + IS  & 69.04 & 80.22 & 87.78 & 93.26 & 33.80 & 44.71 & 56.76 & 67.62 & 62.18 & 78.53 & 88.99 & 96.16 & 46.8 & 72.8 & 79.2 & 82.7 \\
                        \rowcolor{gray}& + SN & 70.41 & 81.48 & 89.40 & 94.07 & 34.63 & 45.46 & 57.96 & 69.23 & 63.38 & 79.24 & 89.36 & 96.38 & 49.3 & 75.4 & 81.3 & 84.3 \\
	       \bottomrule
	    \end{tabular}
	}
\end{table*}

\begin{table}[tbp]
    \centering
    \small
    \caption{Image classification comparisons on Imagenet~\cite{deng2009imagenet}.$\ddag$ denotes results from our implementation.}
    \label{tab:sn_ic}
    {
        \tabcolsep 5 pt
	    \begin{tabular}{ lc cccc }
	        % \addlinespace
	        \toprule
                Method &  Normalization &  acc@1(\%) & acc@5(\%) & acc@10(\%) \\  
                \midrule
                \multicolumn{2}{l}{Resnet-50~\cite{he2016deep}}  & \textcolor{gray}{76.15} & \textcolor{gray}{92.87} & \textcolor{gray}{95.83} \\
                \midrule
                \multicolumn{2}{l}{Resnet-50 w/o norm \& bias$\ddag$} & 75.07 & 92.68 & 95.74 \\
                     & + IS & 75.72 & 92.81 & 95.69 \\
                     \rowcolor{gray}& + SN & \textbf{76.44} & \textbf{93.21} & \textbf{95.91} \\
                \midrule
                \multicolumn{2}{l}{zs CLIP ViT-B/32} 
                                            & 63.34 & 88.80 & 93.66 \\
                                     & + IS & 63.49 & 88.70 & 93.55 \\
                     \rowcolor{gray}& + SN  & 65.64 & 89.88 & 94.28 \\
                \midrule
                \multicolumn{2}{l}{zs CLIP ViT-L/14@336px} 
                                            & 83.58 & 96.53 & 99.16 \\
                                     & + IS & 83.77 & 96.59 & 99.18 \\
                     \rowcolor{gray}& + SN  & 83.90 & 96.51 & 99.22 \\
	        \bottomrule
	    \end{tabular}
	}
\end{table}

\renewcommand{\algorithmicrequire}{\textbf{Input:}}
\renewcommand{\algorithmicensure}{\textbf{Output:}}
\begin{algorithm}[htb]
	\caption{Target Hubness Estimation with SN.}
	\label{alg:a1_sn}
	\begin{algorithmic}[1]
            \REQUIRE Query-target similarity $\sS\in\R^{m\times n}$, parameter~$\tau$.
            \ENSURE Target hubness $\hat{\hbar}(\sT)$.
  		\STATE Initialize $\mxi=\exp(\frac{\sS}{\tau})$, $\va=\frac{1}{m}\ones_m$,~$\vb=\frac{1}{n}\ones_n$, 
            \STATE Initialize $\vbeta^{(0)}=\ones_n$, $T=10$.
            \FOR{$t=1,\cdots, T$}
		\STATE Update $\valpha^{(t)}=\frac{\va}{\mxi\vbeta^{(t-1)}}$. 
            \STATE Update  $\vbeta^{(t)}=\frac{\vb}{\mxi^\top\valpha^{(t)}}$.
            \ENDFOR
            \STATE Calculate target hubness $\hat{\hbar}(\sT)=-\tau\log(\vbeta^{(T)})$.
            \RETURN $\hat{\hbar}(\sT)$
	\end{algorithmic}  
\end{algorithm}
\renewcommand{\algorithmicrequire}{\textbf{Input:}}
\renewcommand{\algorithmicensure}{\textbf{Output:}}
\begin{algorithm}[htb]
	\caption{Dual Bank Sinhorn Normalization (DBSN).}
	\label{alg:a2_dbsn}
	\begin{algorithmic}[1]
            \REQUIRE Query-target similarity $\sS\in\R^{m\times n}$, parameter~$\tau$,\\
            querybank-target similarity $\sS_{bt}\in\R^{|\sB_q|\times n}$,\\
            querybank-targetbank similarity $\sS_{bb}\in\R^{|\sB_q|\times |\sB_t|}$.
            \ENSURE Normalized similarity $\tilde{\sS}$.
            \STATE Calculate joint target hubness using Algorithm~\ref{alg:a1_sn}: \\
            ~~~~~~~~~~~~~$[\tilde{\hbar}(\sT); \tilde{\hbar}(\sB_t)]=\emph{SN}([\sS_{bt}; \sS_{bb}], ~\tau)$.
            \STATE Calculate $\tilde{\sS}=\sS-\tilde{\hbar}(\sT)$
            \RETURN $\tilde{\sS}$
	\end{algorithmic}  
\end{algorithm}
\begin{table}[htbp]
    \centering
    \small
    \caption{Comparisons between DBSN and other querybank normalization methods on Flickr 30K and MSR-VTT.}
    % \vspace{-10pt}
    \label{tab:zsi2t_dbsn}
    {
        \tabcolsep 3 pt
	    \begin{tabular}{ ll ccccc}
	        \addlinespace
	        \toprule
                $\sB_q$~\&~$\sB_t$ & Normalization & R@1$\uparrow$ & R@5$\uparrow$ & R@10$\uparrow$ & MdR~$\downarrow$ & MnR~$\downarrow$ \\ 
                \midrule
                \multicolumn{7}{@{\;}c}{\bf zero-shot CLIP on Flicr30K for Text$\rightarrow$Image} \\
                \midrule
                \multirow{3}{*}{test~\&~-}  & -    & 58.8 & 83.4 & 90.1 & 1.0 & 6.0 \\
                                        & + IS & 66.8 & 88.7 & 93.5 & 1.0 & 4.5 \\
                                        & + SN & \textbf{69.3} & \textbf{90.2} & \textbf{94.5} & \textbf{1.0} & \textbf{3.7} \\
                \midrule
                \multirow{3}{*}{val~\&~-} & + IS & 63.5 & 87.2 & 92.3 & 1.0 & 5.2 \\
                                        & + DIS & 63.5 & 87.2 & 92.3 & 1.0 & 5.2 \\
                                        & + SN & 64.4 & 87.6 & 92.5 & 1.0 & 4.9 \\
                \midrule
                \multirow{2}{*}{val~\&~val} & + DualIS & 63.6 & 87.1 & 92.3 & 1.0 & 5.2 \\
                                        & + DBSN & 64.6 & 88.0 & 92.7 & 1.0 & 4.8 \\
                \midrule
                \multirow{3}{*}{train~\&~-} & + IS & 65.1 & 87.5 & 92.8 & 1.0 & 4.7 \\
                                            & + DIS & 65.1 & 87.5 & 92.8 & 1.0 & 4.7 \\
                                            & + SN & 66.3 & 88.1 & 93.1 & 1.0 & 4.7 \\
                \midrule
                \multirow{2}{*}{train~\&~train}   & + DualIS & 65.3 & 87.4 & 92.9 & 1.0 & 4.7 \\
                                                & + DBSN & 67.1 & 88.7 & 93.5 & 1.0 & 4.4 \\
                \midrule
                \multicolumn{7}{@{\;}c}{\bf zero-shot CLIP on MSCOCO for Text$\rightarrow$Image} \\
                \midrule
                \multirow{3}{*}{test~\&~-}  & -    & 30.5 & 56.0 & 66.8 & 4.0 & 24.5 \\
                                        & + IS & 38.6 & 64.0 & 74.1 & 2.0 & 20.0 \\
                                        & + SN & \textbf{40.8} & \textbf{66.4} & \textbf{76.1} & \textbf{2.0} & \textbf{17.8} \\
                \midrule
                \multirow{3}{*}{val~\&~-} & + IS & 36.9 & 62.7 & 72.8 & 3.0 & 21.4 \\
                                        & + DIS & 36.3 & 61.5 & 71.8 & 3.0 & 21.2 \\
                                        & + SN & 37.0 & 63.0 & 73.2 & 3.0 & 20.6 \\
                \midrule
                \multirow{2}{*}{val~\&~val} & + DualIS & 36.8 & 62.6 & 72.7 & 3.0 & 21.5 \\
                                        & + DBSN & 37.1 & 63.2 & 73.3 & 3.0 & 20.5 \\
                \midrule
                \multirow{3}{*}{train~\&~-} & + IS & 37.5 & 63.3 & 73.4 & 3.0 & 21.1 \\
                                            & + DIS & 37.6 & 63.3 & 73.4 & 3.0 & 21.0 \\
                                            & + SN & 38.9 & 64.6 & 74.7 & 2.0 & 19.5 \\
                \midrule
                \multirow{2}{*}{train~\&~train}   & + DualIS & 37.5 & 63.2 & 73.3 & 3.0 & 21.2 \\
                                                & + DBSN & 39.4 & 64.9 & 74.8 & 2.0 & 19.1 \\
	       \bottomrule
	    \end{tabular}
	}
\end{table}
\begin{table}[htbp]
    \centering
    \small
    \caption{Comparisons between DBSN and other querybank normalization methods on Flickr 30K and MSR-VTT.}
    % \vspace{-10pt}
    \label{tab:fti2t_dbsn}
    {
        \tabcolsep 3 pt
	    \begin{tabular}{ ll ccccc}
	        \addlinespace
	        \toprule
                $\sB_q$~\&~$\sB_t$ & Normalization & R@1$\uparrow$ & R@5$\uparrow$ & R@10$\uparrow$ & MdR~$\downarrow$ & MnR~$\downarrow$ \\ 
                \midrule
                \multicolumn{7}{@{\;}c}{\bf fine-tuned CLIP on Flickr30K for Text$\rightarrow$Image} \\
                \midrule
                \multirow{3}{*}{test~\&~-} & -         & 74.2 & 93.4 & 96.7 & 1.0 & 2.8   \\
                                                & + IS & 76.6 & 93.9 & 97.1 & 1.0 & 2.5   \\
                                                & + SN & \textbf{79.2} & \textbf{94.8} & \textbf{97.5} & \textbf{1.0} & \textbf{2.2}  \\
                \midrule
                \multirow{3}{*}{val~\&~-} & + IS & 73.9 & 92.5 & 96.2 & 1.0 & 3.0 \\
                                            & + DIS & 73.9 & 92.5 & 96.2 & 1.0 & 3.0 \\
                                            & + SN & 73.5 & 92.5 & 96.0 & 1.0 & 3.0  \\
                \midrule
                \multirow{2}{*}{val~\&~val} & + DualIS & 73.9 & 92.5 & 96.2 & 1.0 & 3.0 \\
                                        & + DBSN & 73.6 & 92.7 & 96.0 & 1.0 & 2.9 \\
                \midrule
                \multirow{3}{*}{train~\&~-} & + IS & 74.9 & 93.0 & 96.6 & 1.0 & 2.9 \\
                                            & + DIS & 74.9 & 93.0 & 96.6 & 1.0 & 2.9 \\
                                            & + SN & 75.0 & 93.1 & 96.7 & 1.0 & 2.8 \\
                \midrule
                \multirow{2}{*}{train~\&~train}   & + DualIS & 74.8 & 92.9 & 96.6 & 1.0 & 2.9 \\
                                                & + DBSN & 76.0 & 93.6 & 96.7 & 1.0 & 2.7 \\
                \midrule
                \multicolumn{7}{@{\;}c}{\bf fine-tuned CLIP on MS COCO for Text$\rightarrow$Image} \\
                \midrule
                \multirow{3}{*}{test~\&~-}  & -    & 47.5 & 74.1 & 83.2 & 2.0 & 11.3 \\
                                        & + IS & 50.1 & 75.9 & 84.3  & 1.0 & 10.8 \\
                                        & + SN & \textbf{51.6} & \textbf{77.0} & \textbf{85.3} & \textbf{1.0} & \textbf{10.0} \\
                \midrule
                \multirow{3}{*}{val~\&~-} & + IS & 46.9 & 73.6 & 82.7 & 2.0 & 12.2 \\
                                        & + DIS & 47.2 & 73.8 & 82.9 & 2.0 & 11.8 \\
                                        & + SN & 44.8 & 71.7 & 81.0 & 2.0 & 13.1 \\
                \midrule
                \multirow{2}{*}{val~\&~val} & + DualIS & 46.8 & 73.5 & 82.7 & 2.0 & 12.3 \\
                                        & + DBSN & 45.9 & 72.8 & 81.9 & 2.0 & 12.7 \\
                \midrule
                \multirow{3}{*}{train~\&~-} & + IS & 48.6 & 74.9 & 83.6 & 2.0 & 11.7 \\
                                            & + DIS & 48.6 & 74.9 & 83.6 & 2.0 & 11.7 \\
                                            & + SN & 49.0 & 75.1 & 83.7 & 2.0 & 11.6 \\
                \midrule
                \multirow{2}{*}{train~\&~train}   & + DualIS & 48.6 & 74.9 & 83.6 & 2.0 & 11.7 \\
                                                & + DBSN & 49.3 & 75.6 & 84.1 & 2.0 & 11.2 \\
	       \bottomrule
	    \end{tabular}
	}
\end{table}
\begin{table}[htbp]
    \centering
    \small
    \caption{Comparisons between DBSN and other querybank normalization methods on Flickr 30K and MSR-VTT.}
    % \vspace{-10pt}
    \label{tab:c4c_dbsn}
    {
        \tabcolsep 3 pt
	    \begin{tabular}{ ll ccccc}
	        \addlinespace
	        \toprule
                $\sB_q$~\&~$\sB_t$ & Normalization & R@1$\uparrow$ & R@5$\uparrow$ & R@10$\uparrow$ & MdR~$\downarrow$ & MnR~$\downarrow$ \\ 
                \midrule
                \multicolumn{7}{@{\;}c}{\bf CLIP4Clip on MSR-VTT(1k split) for Text$\rightarrow$Video} \\
                \midrule
                \multirow{3}{*}{test~\&~-} & -    & 43.9 & 70.6 & 80.7 & 2.0 & 16.0   \\
                                                     & + IS & 46.4 & 72.5 & 82.8 & 2.0 & 13.2   \\
                                                     & + SN & \textbf{49.2} & \textbf{75.4} & \textbf{83.8} & \textbf{2.0} & \textbf{11.9}  \\
                \midrule
                \multirow{3}{*}{train~\&~-} & + IS & 44.8 & 71.1 & 81.3 & 2.0 & 15.0 \\
                                            & + DIS & 44.8 & 71.1 & 81.3 & 2.0 & 15.0 \\
                                            & + SN & 44.5 & 70.9 & 80.9 & 2.0 & 15.4  \\
                \midrule
                \multirow{2}{*}{train~\&~train}   & + DualIS & 44.5 & 71.0 & 80.9 & 2.0 & 15.4 \\
                                                & + DBSN & 45.2 & 71.8 & 81.9 & 2.0 & 15.5 \\
                \midrule
                \multirow{2}{*}{jsfusion~\&~-}   & + IS & 45.1 & 70.5 & 81.2 & 2.0 & 15.6 \\
                                                & + SN & 46.2 & 71.1 & 81.8 & 2.0 & 15.0  \\
                \midrule
                \multicolumn{7}{@{\;}c}{\bf CLIP4Clip on Didemo for Text$\rightarrow$Video} \\
                \midrule
                \multirow{3}{*}{test~\&~-}  & -    & 40.8 & 69.7 & 80.3 & 2.0 & 18.4  \\
                & + IS &  46.0 & 72.8 & 81.1 & 2.0 &  15.8 \\
                & + SN & \textbf{48.0} & \textbf{74.6} & \textbf{82.9} & \textbf{2.0} & \textbf{13.6} \\
                \midrule
                \multirow{3}{*}{val~\&~-} & + IS & 39.8 & 68.9 & 79.3 & 2.0 & 17.1 \\
                                & + DIS & 39.7 & 69.0 & 79.1 & 2.0 & 17.2  \\
                                & + SN & 38.4 & 66.9 & 78.7 & 2.0 & 19.1 \\
                \midrule
                \multirow{2}{*}{val~\&~val} & + DualIS & 39.9 & 69.3 & 79.3 & 2.0 & 17.1 \\
                                & + DBSN & 39.8 & 69.1 & 79.4 & 2.0 & 17.6 \\
                \midrule
                \multirow{3}{*}{train~\&~-} & + IS & 40.7 & 70.0 & 80.8 & 2.0 & 16.8 \\
                                    & + DIS & 40.7 & 70.0 & 80.8 & 2.0 & 16.8 \\
                                    & + SN & 39.6 & 68.8 & 79.1 & 2.0 & 19.1 \\
                \midrule
                \multirow{2}{*}{train~\&~train}   & + DualIS & 40.7 & 69.9 & 80.8 & 2.0 & 16.8 \\
                                        & + DBSN & 41.3 & 70.1 & 81.0 & 2.0 & 17.1 \\
	       \bottomrule
	    \end{tabular}
	}
\end{table}
\begin{table}[htbp]
    \centering
    \small
    \caption{Comparisons between DBSN and other querybank normalization methods on Flickr 30K and MSR-VTT.}
    % \vspace{-10pt}
    \label{tab:xpool_dbsn}
    {
        \tabcolsep 3 pt
	    \begin{tabular}{ ll ccccc}
	        \addlinespace
	        \toprule
                $\sB_q$~\&~$\sB_t$ & Normalization & R@1$\uparrow$ & R@5$\uparrow$ & R@10$\uparrow$ & MdR~$\downarrow$ & MnR~$\downarrow$ \\ 
                \midrule
                \multicolumn{7}{@{\;}c}{\bf X-Pool on MSR-VTT(1k split) for Text$\rightarrow$Video} \\
                \midrule
                \multirow{3}{*}{test~\&~-} & -    & 48.0 & 73.1 & 83.2 & 2.0 & 14.0   \\
                                                     & + IS & 50.8 & 77.2 & 86.5 & 1.0 & 10.5   \\
                                                     & + SN & \textbf{52.7} & \textbf{78.4} & \textbf{86.5} & \textbf{1.0} & \textbf{10.1}  \\
                \midrule
                \multirow{3}{*}{train~\&~-} & + IS & 48.9 & 74.0 & 83.9 & 2.0 & 13.7 \\
                                            & + DIS & 48.9 & 74.0 & 83.9 & 2.0 & 13.7 \\
                                            & + SN & 48.1 & 73.8 & 83.4 & 2.0 & 13.3  \\
                \midrule
                \multirow{2}{*}{train~\&~train}   & + DualIS & 48.9 & 74.0 & 83.9 & 2.0 & 13.7 \\
                                                & + DBSN & 49.2 & 74.3 & 84.3 & 2.0 & 13.5 \\
                \midrule
                \multicolumn{7}{@{\;}c}{\bf X-Pool on Didemo for Text$\rightarrow$Video} \\
                \midrule
                \multirow{3}{*}{test~\&~-}  & -    & 47.3 & 73.5 & 82.8 & 2.0 & 14.8 \\
                & + IS & 50.9 & 76.3 & 85.1 & 1.0 & 11.4 \\
                & + SN & \textbf{53.1} & \textbf{78.6} & \textbf{86.8} & \textbf{1.0} & \textbf{9.6} \\
                \midrule
                \multirow{3}{*}{val~\&~-} & + IS & 46.0 & 73.4 & 82.0 & 2.0 & 14.4 \\
                                & + DIS & 46.5 & 73.5 & 82.8 & 2.0 & 14.2 \\
                                & + SN & 42.9 & 73.7 & 81.9 & 2.0 & 15.6\\
                \midrule
                \multirow{2}{*}{val~\&~val} & + DualIS & 46.2 & 73.5 & 83.9 & 2.0 & 13.4 \\
                                & + DBSN & 45.0 & 73.7 & 82.5 & 2.0 & 14.4 \\
                \midrule
                \multirow{3}{*}{train~\&~-} & + IS & 46.5 & 74.4 & 83.3 & 2.0 & 13.6  \\
                                    & + DIS & 46.7 & 74.2 & 83.6 & 2.0 & 12.9 \\
                                    & + SN & 45.9 & 74.0 & 83.4 & 2.0 & 14.9  \\
                \midrule
                \multirow{2}{*}{train~\&~train}   & + DualIS & 46.6 & 74.1 & 83.3 & 2.0 & 12.9 \\
                                        & + DBSN & 47.8 & 74.4 & 83.3 & 2.0 & 12.8  \\
	       \bottomrule
	    \end{tabular}
	}
\end{table}

\renewcommand{\thesection}{E}%
\section{Visualizations}
\begin{figure}[tbp]
    \textit{\textbf{Query}: A couple of people sit outdoors at a table with an umbrella and talk}.\\
    \begin{minipage}{\textwidth}
        \begin{minipage}{0.02\textwidth}
            \rotatebox{90}{\textbf{Baseline}}
        \end{minipage}
        \begin{minipage}{0.47\textwidth}
            \includegraphics[width=\textwidth]{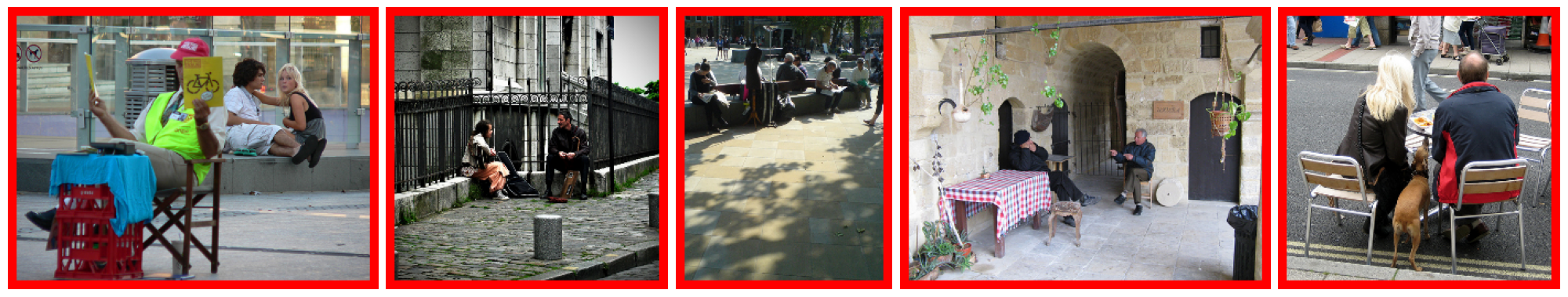}
        \end{minipage}
    \end{minipage}

    \begin{minipage}{\textwidth}
        \begin{minipage}{0.025\textwidth}
            \rotatebox{90}{\textbf{IS}}
        \end{minipage}%
        \begin{minipage}{0.47\textwidth}
            \includegraphics[width=\textwidth]{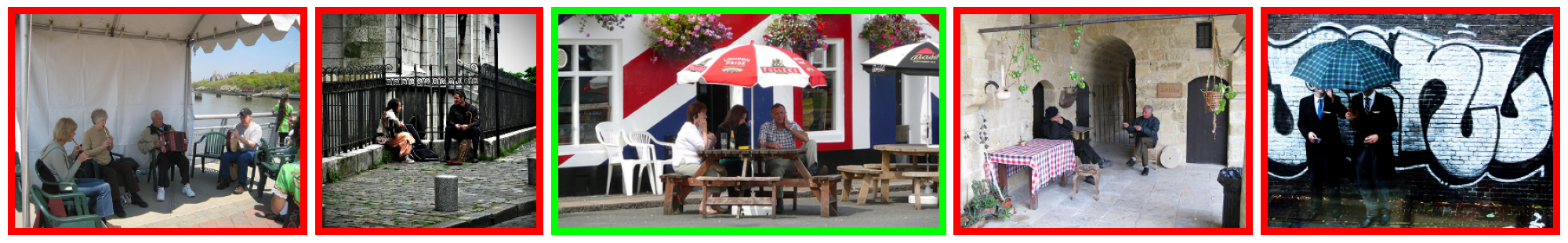}
        \end{minipage}
    \end{minipage}

    \begin{minipage}{\textwidth}
        \begin{minipage}{0.02\textwidth}
            \rotatebox{90}{\textbf{SN}}
        \end{minipage}
        \begin{minipage}{0.47\textwidth}
            \includegraphics[width=\textwidth]{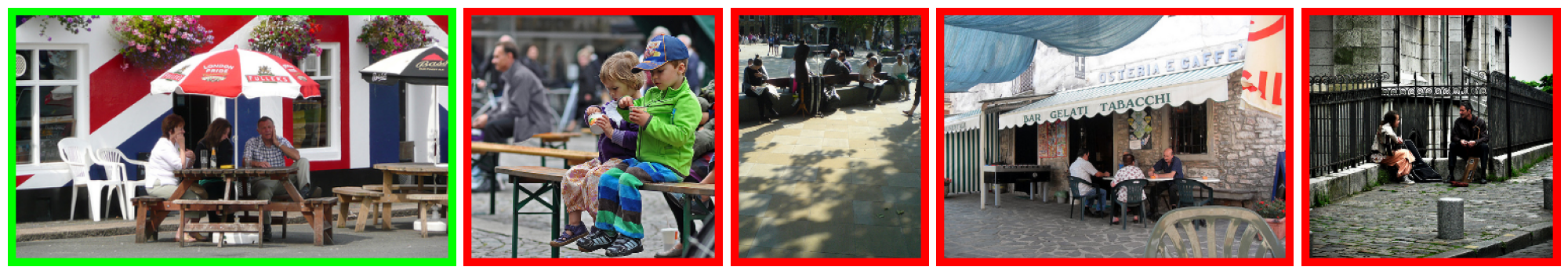}
        \end{minipage}
    \end{minipage}
    
    \caption{Visual comparisons of Top-5 text-to-image retrieval results on Flickr30k. \textcolor[rgb]{1.0, 0.0, 0.0}{Red} and \textcolor[rgb]{0.0, 1.0, 0.0}{green} boxes indicate incorrect and correct recalls, respectively.}
    \label{fig:t2i_comp_2}
\end{figure}

\begin{figure}[tbp]
    \textit{\textbf{Query}: A man in a red shirt and blue pants is going into a building while a dog watches him}.\\
    \begin{minipage}{\textwidth}
        \begin{minipage}{0.02\textwidth}
            \rotatebox{90}{\textbf{Baseline}}
        \end{minipage}
        \begin{minipage}{0.47\textwidth}
            \includegraphics[width=\textwidth]{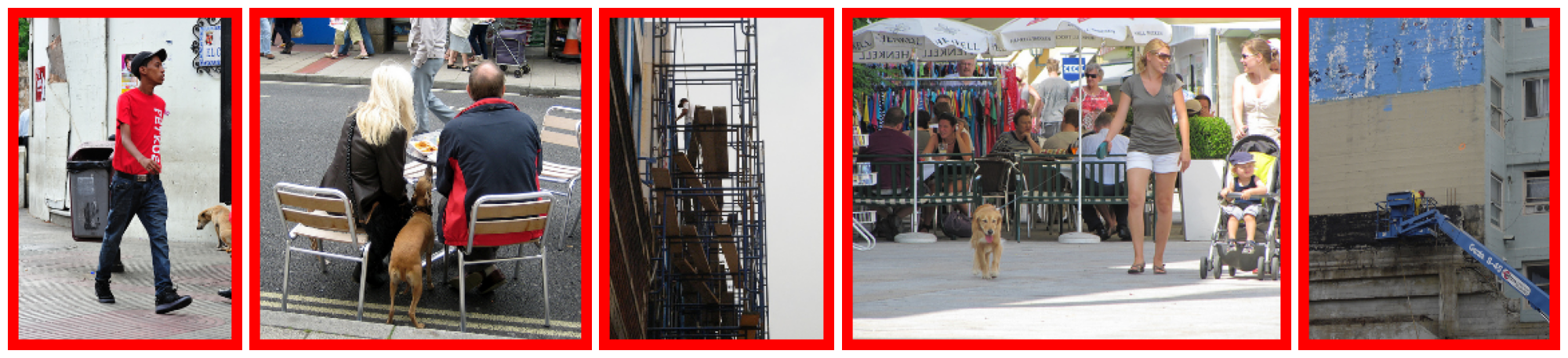}
        \end{minipage}
    \end{minipage}

    \begin{minipage}{\textwidth}
        \begin{minipage}{0.025\textwidth}
            \rotatebox{90}{\textbf{IS}}
        \end{minipage}%
        \begin{minipage}{0.47\textwidth}
            \includegraphics[width=\textwidth]{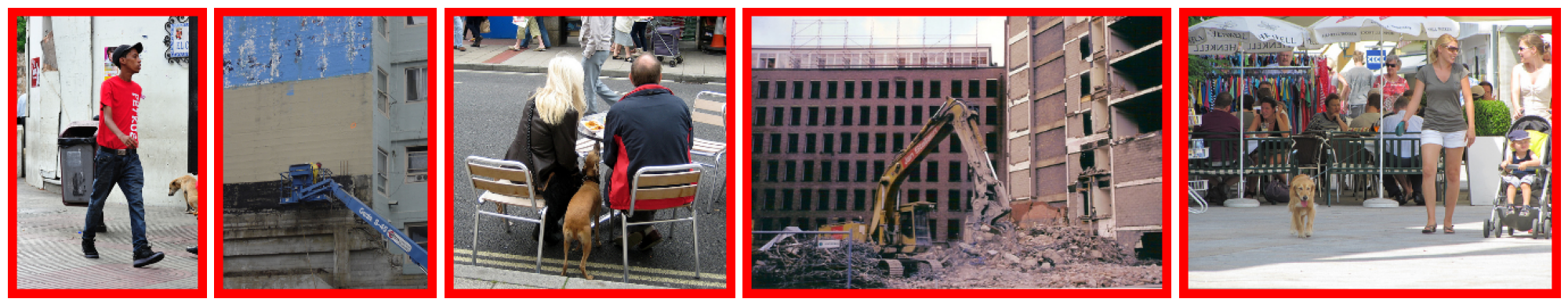}
        \end{minipage}
    \end{minipage}

    \begin{minipage}{\textwidth}
        \begin{minipage}{0.02\textwidth}
            \rotatebox{90}{\textbf{SN}}
        \end{minipage}
        \begin{minipage}{0.47\textwidth}
            \includegraphics[width=\textwidth]{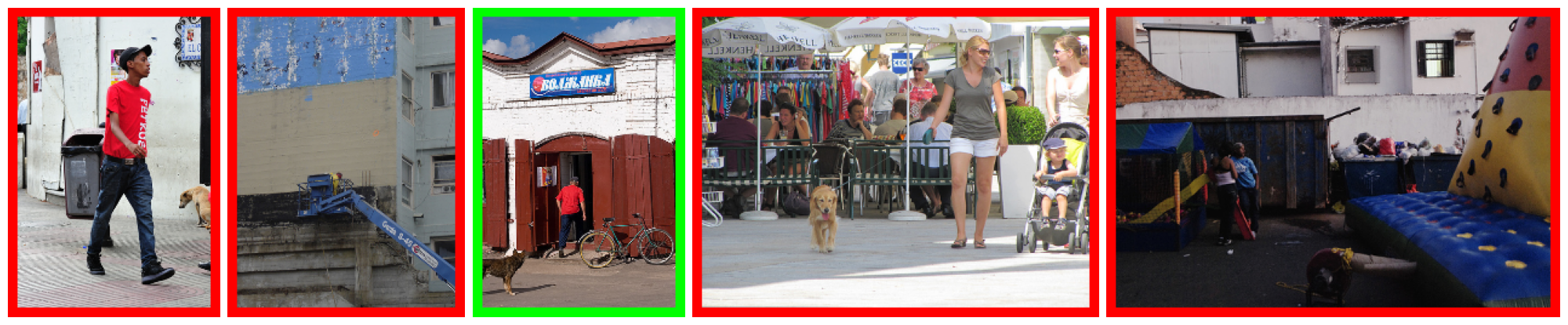}
        \end{minipage}
    \end{minipage}
    
    \caption{Visual comparisons of Top-5 text-to-image retrieval results on Flickr30k. \textcolor[rgb]{1.0, 0.0, 0.0}{Red} and \textcolor[rgb]{0.0, 1.0, 0.0}{green} boxes indicate incorrect and correct recalls, respectively.}
    \label{fig:t2i_comp_3}
\end{figure}

\begin{figure}[tbp]
    \textit{\textbf{Query}: A man is cooking on a stove in a kitchen , using wooden utensil}.\\
    \begin{minipage}{\textwidth}
        \begin{minipage}{0.02\textwidth}
            \rotatebox{90}{\textbf{Baseline}}
        \end{minipage}
        \begin{minipage}{0.47\textwidth}
            \includegraphics[width=\textwidth]{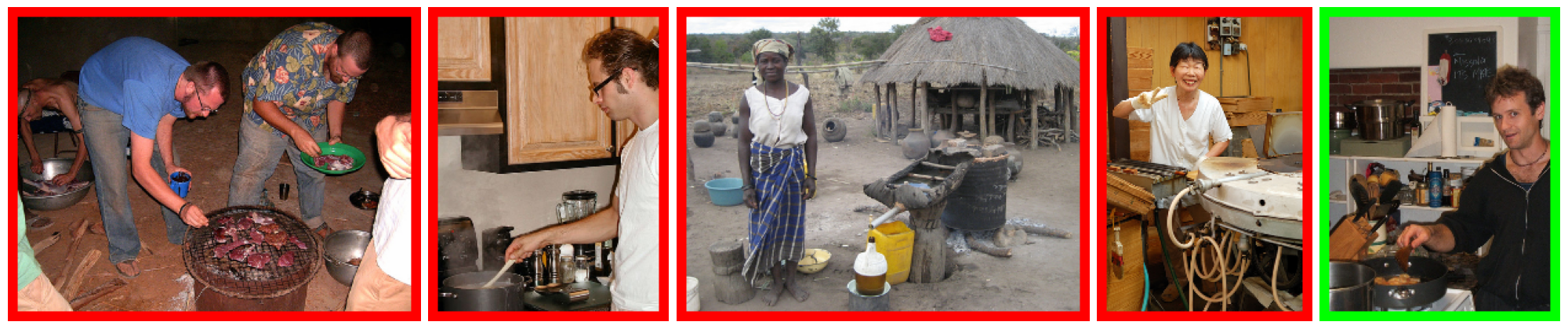}
        \end{minipage}
    \end{minipage}

    \begin{minipage}{\textwidth}
        \begin{minipage}{0.025\textwidth}
            \rotatebox{90}{\textbf{IS}}
        \end{minipage}%
        \begin{minipage}{0.47\textwidth}
            \includegraphics[width=\textwidth]{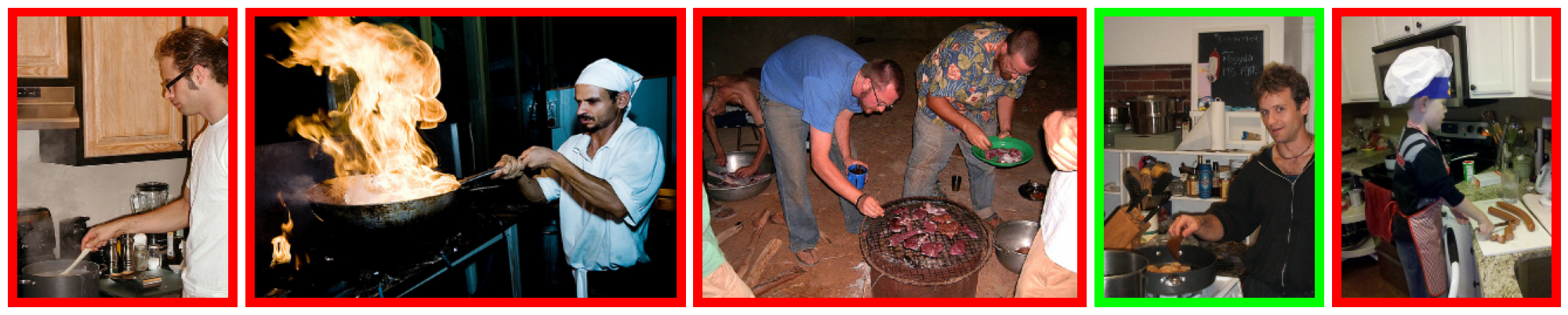}
        \end{minipage}
    \end{minipage}

    \begin{minipage}{\textwidth}
        \begin{minipage}{0.02\textwidth}
            \rotatebox{90}{\textbf{SN}}
        \end{minipage}
        \begin{minipage}{0.47\textwidth}
            \includegraphics[width=\textwidth]{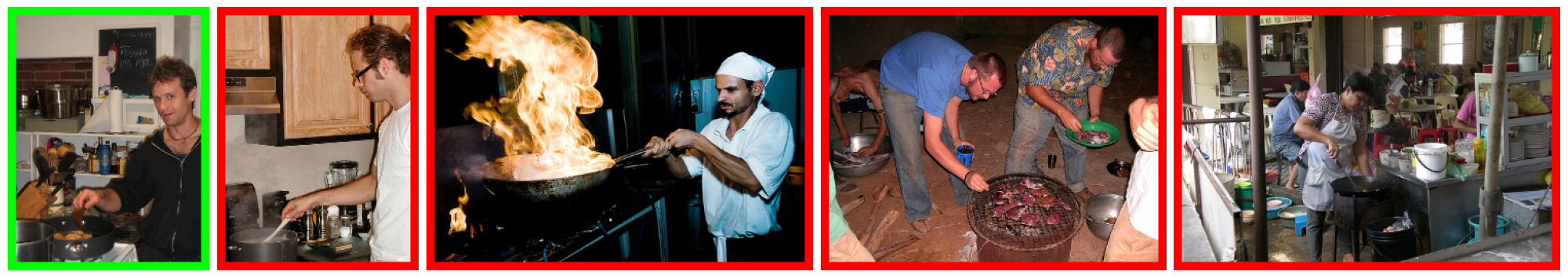}
        \end{minipage}
    \end{minipage}
    
    \caption{Visual comparisons of Top-5 text-to-image retrieval results on Flickr30k. \textcolor[rgb]{1.0, 0.0, 0.0}{Red} and \textcolor[rgb]{0.0, 1.0, 0.0}{green} boxes indicate incorrect and correct recalls, respectively.}
    \label{fig:t2i_comp_4}
\end{figure}

\begin{figure}[tbp]
    \textit{\textbf{Query}: A person did a side flip while water boarding}.\\
    \begin{minipage}{\textwidth}
        \begin{minipage}{0.02\textwidth}
            \rotatebox{90}{\textbf{Baseline}}
        \end{minipage}
        \begin{minipage}{0.47\textwidth}
            \includegraphics[width=\textwidth]{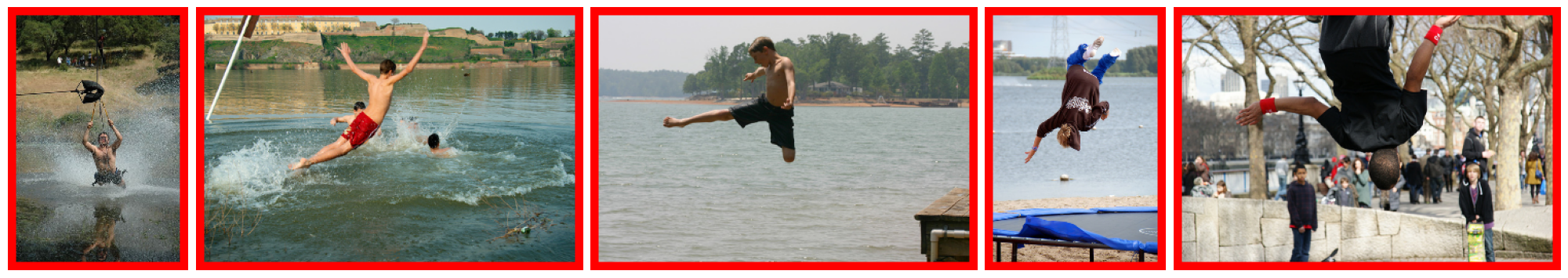}
        \end{minipage}
    \end{minipage}

    \begin{minipage}{\textwidth}
        \begin{minipage}{0.025\textwidth}
            \rotatebox{90}{\textbf{IS}}
        \end{minipage}%
        \begin{minipage}{0.47\textwidth}
            \includegraphics[width=\textwidth]{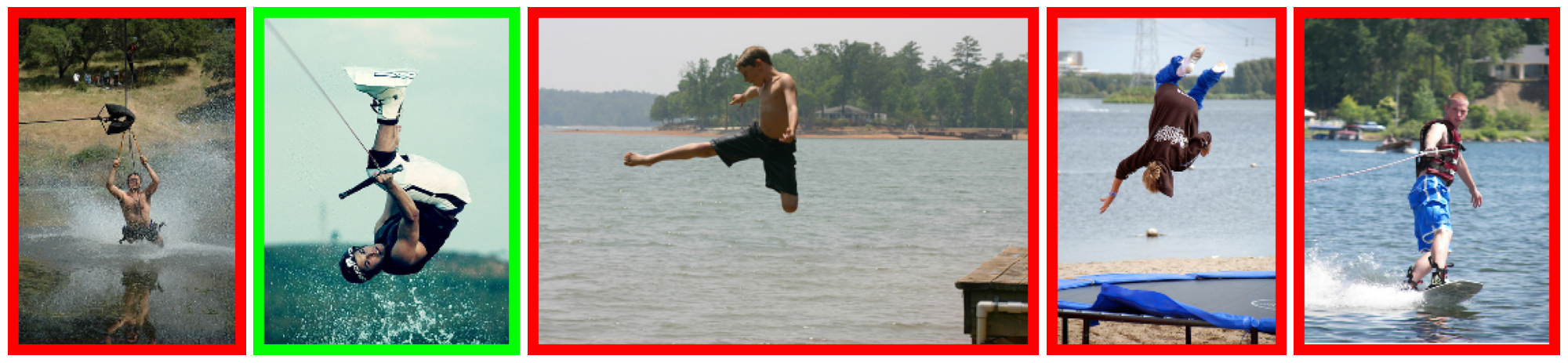}
        \end{minipage}
    \end{minipage}

    \begin{minipage}{\textwidth}
        \begin{minipage}{0.02\textwidth}
            \rotatebox{90}{\textbf{SN}}
        \end{minipage}
        \begin{minipage}{0.47\textwidth}
            \includegraphics[width=\textwidth]{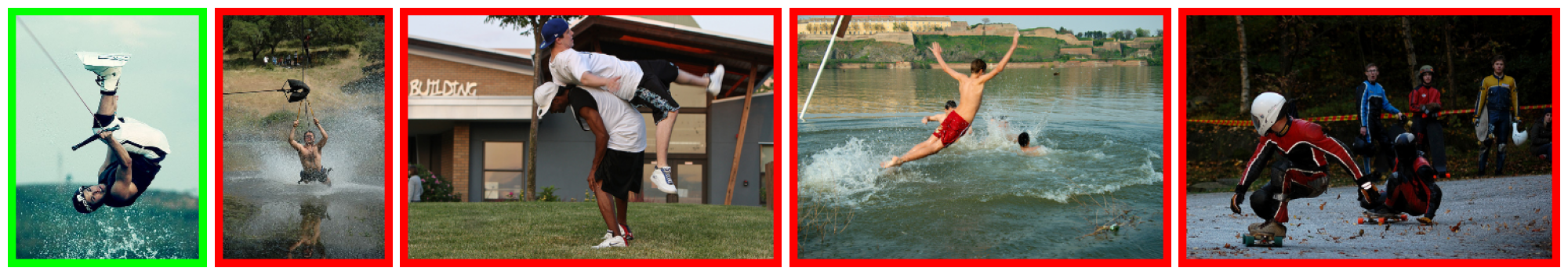}
        \end{minipage}
    \end{minipage}
    
    \caption{Visual comparisons of Top-5 text-to-image retrieval results on Flickr30k. \textcolor[rgb]{1.0, 0.0, 0.0}{Red} and \textcolor[rgb]{0.0, 1.0, 0.0}{green} boxes indicate incorrect and correct recalls, respectively.}
    \label{fig:t2i_comp_5}
\end{figure}

\begin{figure}[tbp]
    \textit{\textbf{Query}: People are cheering at a stadium}.\\
    \begin{minipage}{\textwidth}
        \begin{minipage}{0.02\textwidth}
            \rotatebox{90}{\textbf{Baseline}}
        \end{minipage}
        \begin{minipage}{0.47\textwidth}
            \includegraphics[width=\textwidth]{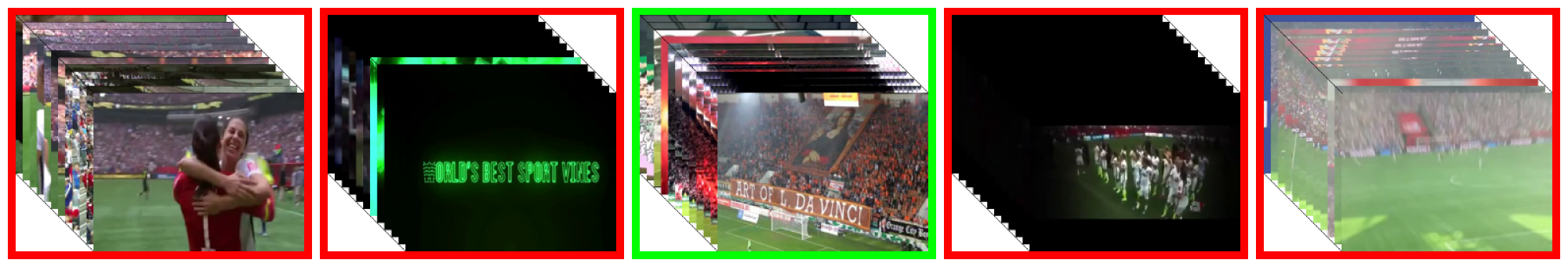}
        \end{minipage}
    \end{minipage}

    \begin{minipage}{\textwidth}
        \begin{minipage}{0.025\textwidth}
            \rotatebox{90}{\textbf{IS}}
        \end{minipage}%
        \begin{minipage}{0.47\textwidth}
            \includegraphics[width=\textwidth]{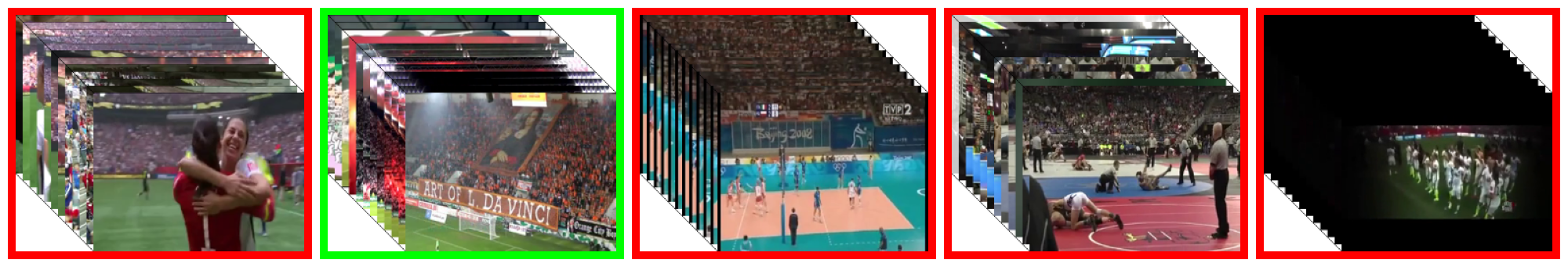}
        \end{minipage}
    \end{minipage}

    \begin{minipage}{\textwidth}
        \begin{minipage}{0.02\textwidth}
            \rotatebox{90}{\textbf{SN}}
        \end{minipage}
        \begin{minipage}{0.47\textwidth}
            \includegraphics[width=\textwidth]{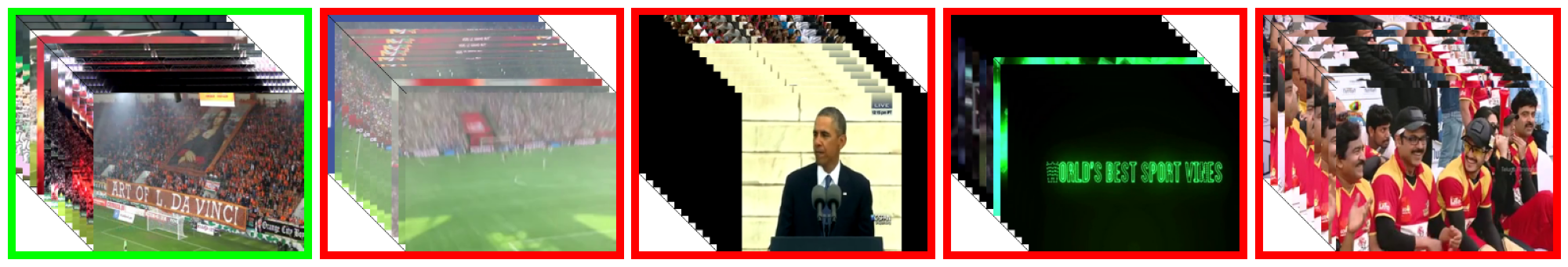}
        \end{minipage}
    \end{minipage}
    
    \caption{Visual comparisons of Top-5 text-to-video retrieval results on MSR-VTT. \textcolor[rgb]{1.0, 0.0, 0.0}{Red} and \textcolor[rgb]{0.0, 1.0, 0.0}{green} boxes indicate incorrect and correct recalls, respectively.}
    \label{fig:t2v_comp_2}
\end{figure}

\begin{figure}[tbp]
    \textit{\textbf{Query}: A news reader is reading the news and asking question to some people}.\\
    \begin{minipage}{\textwidth}
        \begin{minipage}{0.02\textwidth}
            \rotatebox{90}{\textbf{Baseline}}
        \end{minipage}
        \begin{minipage}{0.47\textwidth}
            \includegraphics[width=\textwidth]{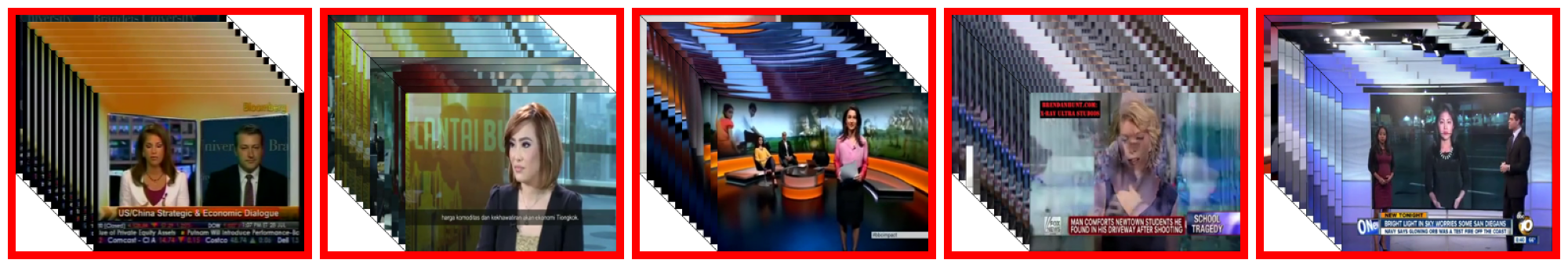}
        \end{minipage}
    \end{minipage}

    \begin{minipage}{\textwidth}
        \begin{minipage}{0.025\textwidth}
            \rotatebox{90}{\textbf{IS}}
        \end{minipage}%
        \begin{minipage}{0.47\textwidth}
            \includegraphics[width=\textwidth]{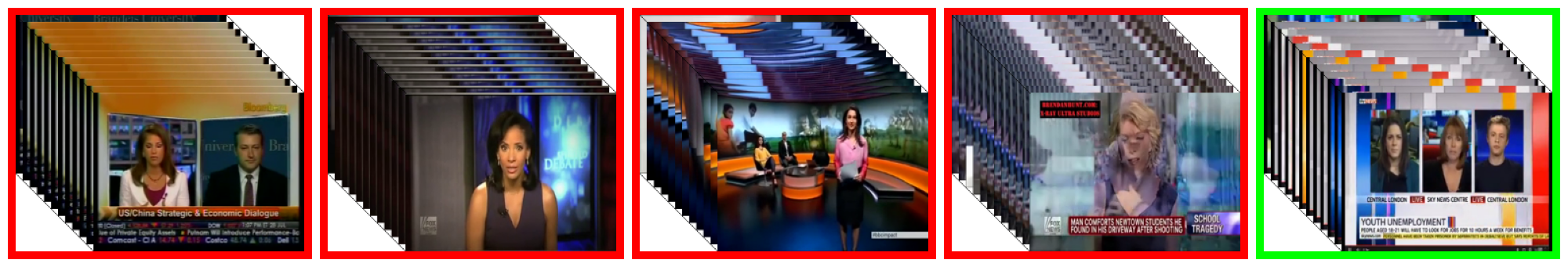}
        \end{minipage}
    \end{minipage}

    \begin{minipage}{\textwidth}
        \begin{minipage}{0.02\textwidth}
            \rotatebox{90}{\textbf{SN}}
        \end{minipage}
        \begin{minipage}{0.47\textwidth}
            \includegraphics[width=\textwidth]{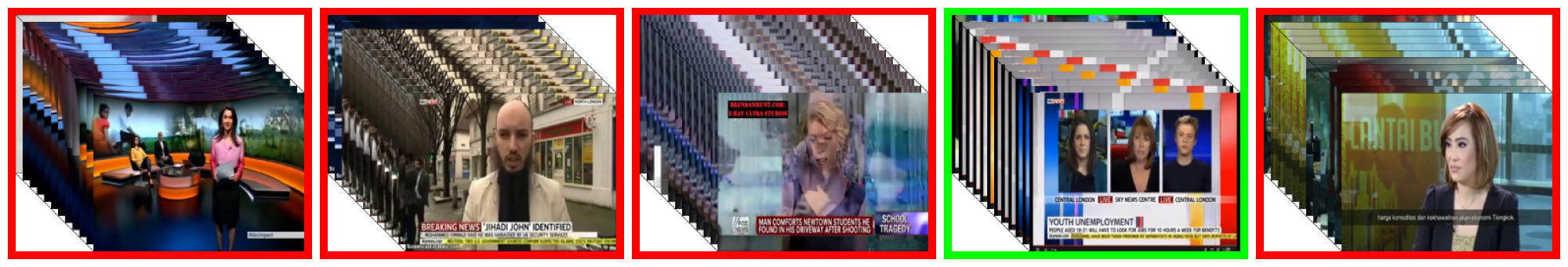}
        \end{minipage}
    \end{minipage}
    
    \caption{Visual comparisons of Top-5 text-to-video retrieval results on MSR-VTT. \textcolor[rgb]{1.0, 0.0, 0.0}{Red} and \textcolor[rgb]{0.0, 1.0, 0.0}{green} boxes indicate incorrect and correct recalls, respectively.}
    \label{fig:t2v_comp_3}
\end{figure}

\begin{figure}[tbp]
    \textit{\textbf{Query}: This is a jigsaw puzzle video}.\\
    \begin{minipage}{\textwidth}
        \begin{minipage}{0.02\textwidth}
            \rotatebox{90}{\textbf{Baseline}}
        \end{minipage}
        \begin{minipage}{0.47\textwidth}
            \includegraphics[width=\textwidth]{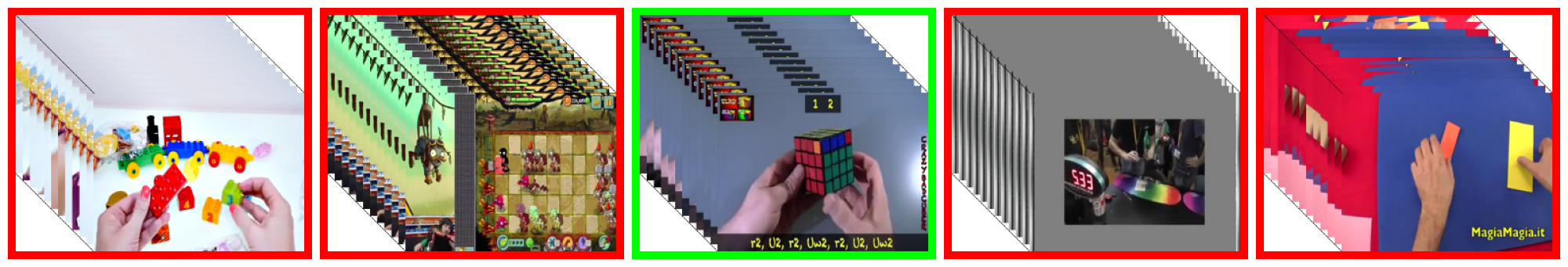}
        \end{minipage}
    \end{minipage}

    \begin{minipage}{\textwidth}
        \begin{minipage}{0.025\textwidth}
            \rotatebox{90}{\textbf{IS}}
        \end{minipage}%
        \begin{minipage}{0.47\textwidth}
            \includegraphics[width=\textwidth]{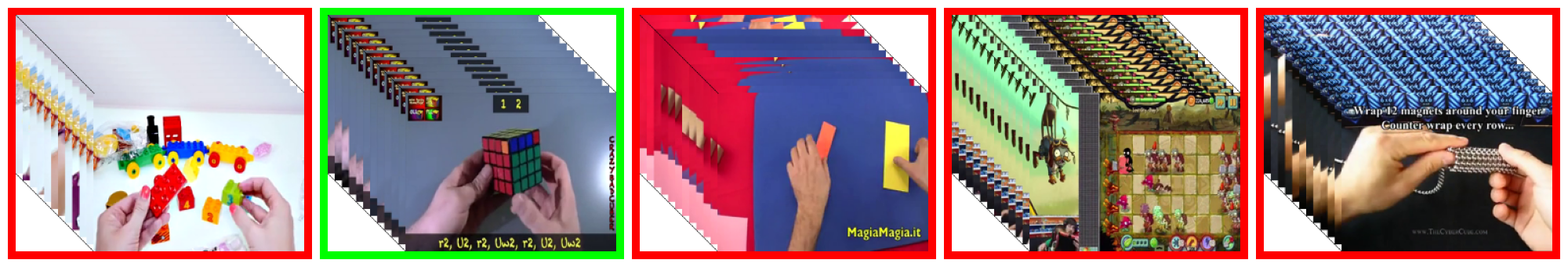}
        \end{minipage}
    \end{minipage}

    \begin{minipage}{\textwidth}
        \begin{minipage}{0.02\textwidth}
            \rotatebox{90}{\textbf{SN}}
        \end{minipage}
        \begin{minipage}{0.47\textwidth}
            \includegraphics[width=\textwidth]{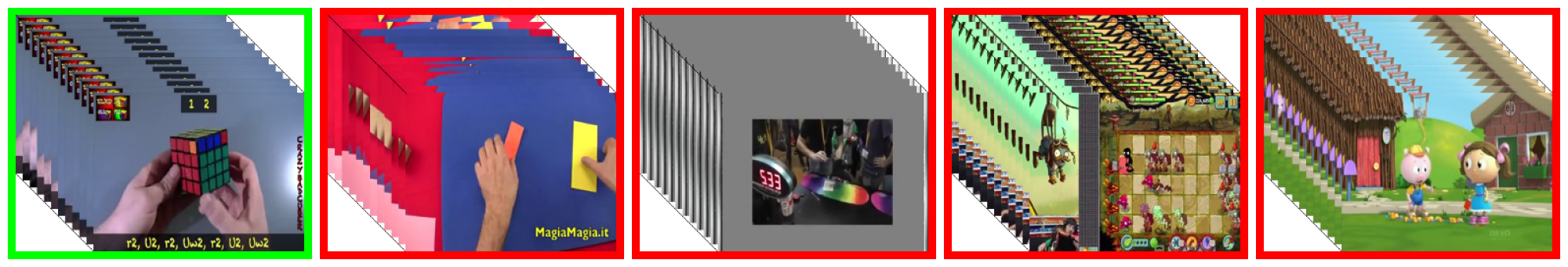}
        \end{minipage}
    \end{minipage}
    
    \caption{Visual comparisons of Top-5 text-to-video retrieval results on MSR-VTT. \textcolor[rgb]{1.0, 0.0, 0.0}{Red} and \textcolor[rgb]{0.0, 1.0, 0.0}{green} boxes indicate incorrect and correct recalls, respectively.}
    \label{fig:t2v_comp_4}
\end{figure}

\begin{figure}[tbp]
    \textit{\textbf{Query}: A man in a suit is talking on a television economy program}.\\
    \begin{minipage}{\textwidth}
        \begin{minipage}{0.02\textwidth}
            \rotatebox{90}{\textbf{Baseline}}
        \end{minipage}
        \begin{minipage}{0.47\textwidth}
            \includegraphics[width=\textwidth]{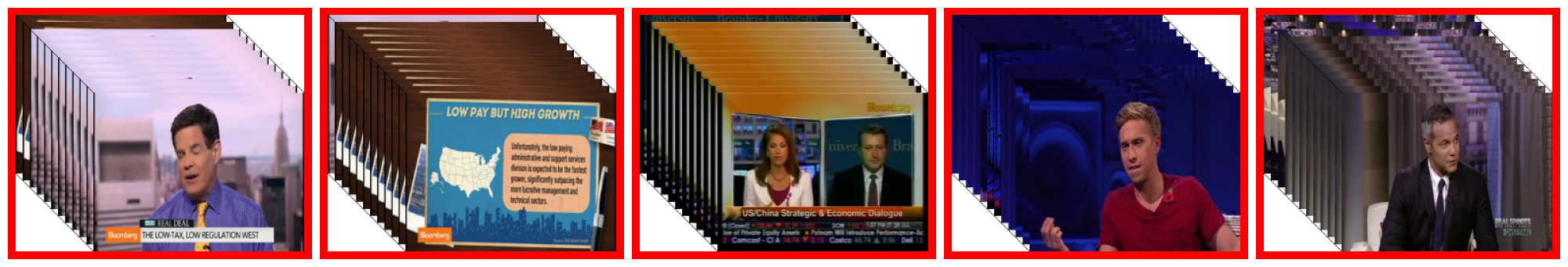}
        \end{minipage}
    \end{minipage}

    \begin{minipage}{\textwidth}
        \begin{minipage}{0.025\textwidth}
            \rotatebox{90}{\textbf{IS}}
        \end{minipage}%
        \begin{minipage}{0.47\textwidth}
            \includegraphics[width=\textwidth]{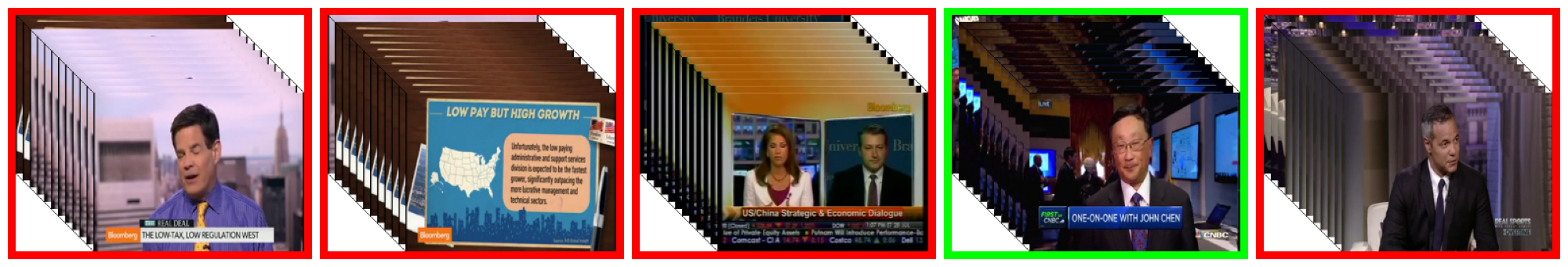}
        \end{minipage}
    \end{minipage}

    \begin{minipage}{\textwidth}
        \begin{minipage}{0.02\textwidth}
            \rotatebox{90}{\textbf{SN}}
        \end{minipage}
        \begin{minipage}{0.47\textwidth}
            \includegraphics[width=\textwidth]{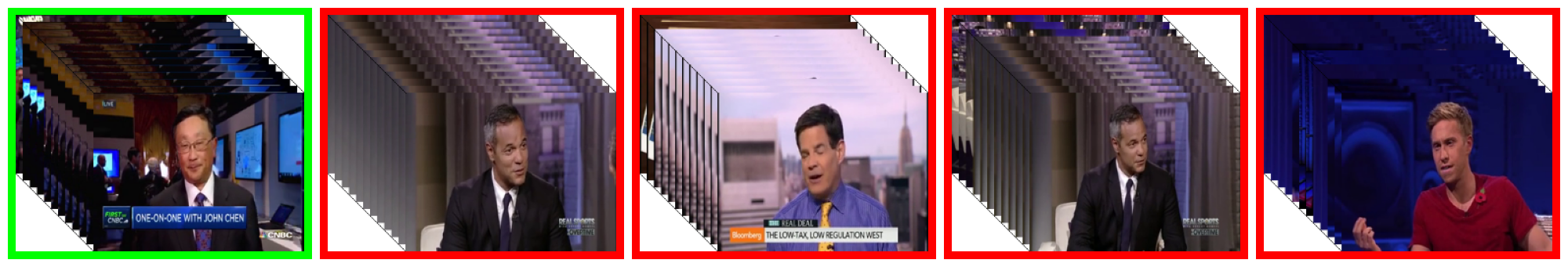}
        \end{minipage}
    \end{minipage}
    
    \caption{Visual comparisons of Top-5 text-to-video retrieval results on MSR-VTT. \textcolor[rgb]{1.0, 0.0, 0.0}{Red} and \textcolor[rgb]{0.0, 1.0, 0.0}{green} boxes indicate incorrect and correct recalls, respectively.}
    \label{fig:t2v_comp_5}
\end{figure}

\begin{figure}[tbp]
    \centering
    \begin{subfigure}[b]{0.235\textwidth}
        \centering
        \includegraphics[width=\textwidth]{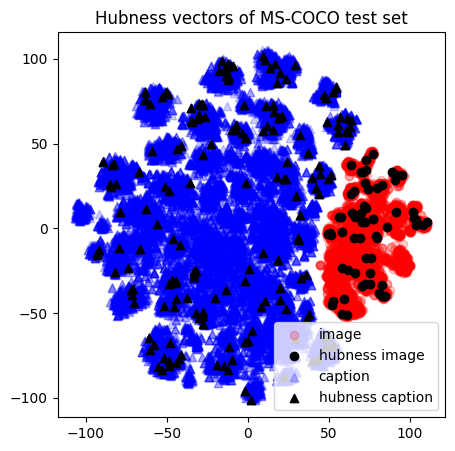}
        % \caption{Framework}
    \end{subfigure}
    \begin{subfigure}[b]{0.235\textwidth}
        \centering
        \includegraphics[width=\textwidth]{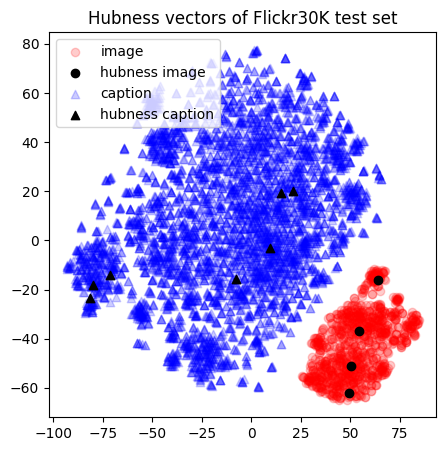}
        % \caption{Radar}
    \end{subfigure}
    \caption{t-sne visualization of \textit{`hub'} vectors in the MS-COCO and Flickr30K test set.}
    \label{fig:tsne}
\end{figure}

\begin{figure}[tbp]
    \centering
    \includegraphics[width=\linewidth]{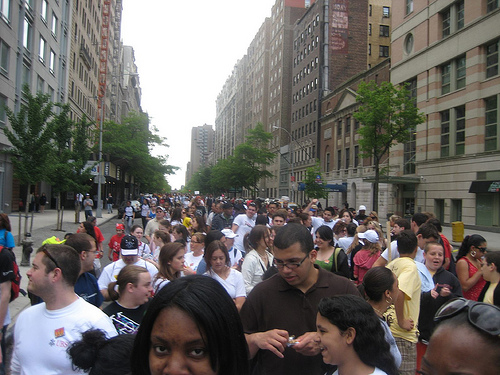} 
    \begin{minipage}{\linewidth}
        \centering
        \begin{tabular}{>{\raggedright\arraybackslash}p{\linewidth}} 
        \textbf{Query 1:} \textcolor{red}{People standing outside of a building.} \\
        \textbf{Query 2:} \textcolor{red}{Four police officers are swallowed up by the large crowd of people on the street.} \\
        \textbf{Query 3:} \textcolor{red}{A large crowd of people walking while being controlled by officers among them.} \\
        \textbf{Query 4:} \textcolor{red}{A large crowd of people are walking down the street.} \\
        \textbf{Query 5:} \textcolor{red}{A large group of people fill a street.} \\
        \textbf{Query 6:} \textcolor{red}{A large crowd of people are walking.} \\
        \textbf{Query 7:} \textcolor{red}{These people are walking in a crowd of people.} \\
        \textbf{Query 8:} \textcolor{red}{A crowd forms on a busy street to watch a street performer.} \\
        \textbf{Query 9:} \textcolor{red}{A crowd of people walking down the middle of a city street.} \\
        \textbf{Query 10:} \textcolor{green}{A large group of people walking down a city street.} \\
        \textbf{Query 11:} \textcolor{green}{A crowd of people walking in the street of a city.} \\
        \textbf{Query 12:} \textcolor{green}{A crowd is assembled in a street.} \\
        \textbf{Query 13:} \textcolor{green}{A view of a crowded city street.} \\
        \textbf{Query 14:} \textcolor{green}{People are gathered in a park.} \\
        \textbf{Query 15:} \textcolor{red}{A crowd on a busy daytime street.} \\
        \textbf{Query 16:} \textcolor{red}{A crowd is gathered in a large outdoor public space.} \\
        \textbf{Query 17:} \textcolor{red}{A woman in a white shirt and hat speaks to a large crowd of men and women using a megaphone.} \\
        \textbf{Query 18:} \textcolor{red}{Someone in a white shirt yelling through a megaphone to a crowd of people.} \\
        \textbf{Query 19:} \textcolor{red}{Many people stand in a line while a person in white talks on a megaphone.} \\
        \textbf{Query 20:} \textcolor{red}{An event with young adults.} \\
        \textbf{Query 21:} \textcolor{red}{Man in white shirts and khaki pants rests head in hand.} \\
        \textbf{Query 22:} \textcolor{red}{A musical concert with a large number of people.} \\
        \textbf{Query 23:} \textcolor{red}{A guy in a white shirt is walking with a drink in his hand.} \\
        \textbf{Query 24:} \textcolor{red}{Blond man crossing street, in white shirt and red t-shirt, carrying a white bag.} \\
        \textbf{Query 25:} \textcolor{red}{A group of men in white shirts perform in a parade.} \\
        % \textbf{Query 26:} \textcolor{red}{A group of people run a race or marathon while a crowd of people watch.} \\
        % \textbf{Query 27:} \textcolor{red}{People look on as participants in a marathon pass by.} \\
        % \textbf{Query 28:} \textcolor{red}{People at a protest regarding military personnel.} \\
        % \textbf{Query 29:} \textcolor{red}{A group of people stand in the park of a city.} \\
        % \textbf{Query 30:} \textcolor{red}{Someone looking at a mass of people in the street.} \\
        % \textbf{Query 31:} \textcolor{red}{Protesters joining on a city street.} \\
        % \textbf{Query 32:} \textcolor{red}{A middle-aged man standing in the middle of a crowd looking the other way.} \\
        % \textbf{Query 33:} \textcolor{red}{Three construction workers stand at the front of a large group of people all wearing white shirts.} \\
        % \textbf{Query 34:} \textcolor{red}{A group of men and a child in white shirts are standing in the road.} \\
        % \textbf{Query 35:} \textcolor{red}{A child stands outside with a crowd of adults in white.} \\
        % \textbf{Query 36:} \textcolor{red}{A crowd of people on the street gathering to watch several young men put on a show.} \\
        % \textbf{Query 37:} \textcolor{red}{A crowd of people are watch two guys play buckets.} \\
        % \textbf{Query 38:} \textcolor{red}{A busy street filled with random people underneath a bridge.} \\
        % \textbf{Query 39:} \textcolor{red}{A man and woman, dressed in white, embrace each other while their actions go unnoticed to the other people in the background.}
        \end{tabular}
    \end{minipage}
    \caption{Highest-frequency retrieved video in Flicr30K dataset with corresponding nearest-neighbor queries. \textcolor[rgb]{1.0, 0.0, 0.0}{Red} and \textcolor[rgb]{0.0, 1.0, 0.0}{green} lines indicate semantically matched and mismatched queries, respectively. (39 total neighbors, TOP25 shown.) }
    \label{fig:image_with_queries}
\end{figure}

\begin{figure}[htbp]
    \centering
    % 上半部分：图片
    \includegraphics[width=\linewidth]{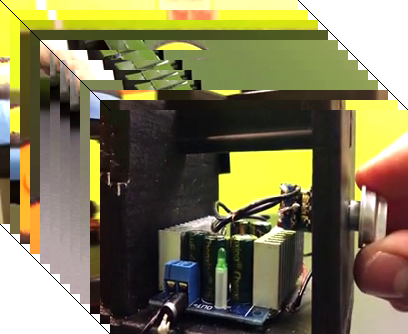}
    % 下半部分：文本查询
    \begin{minipage}{\linewidth}
        \centering
        \begin{tabular}{>{\raggedright\arraybackslash}p{\linewidth}} % 设置文字左对齐，行宽一致
        \textbf{Query 1:} \textcolor{red}{A student explains to his teacher about the sheep of another student.} \\
        \textbf{Query 2:} \textcolor{red}{There is a man shooting other people in a corridor.} \\
        \textbf{Query 3:} \textcolor{red}{A man is giving his commentary on a current event television show.} \\
        \textbf{Query 4:} \textcolor{green}{There was a resistor in the back.} \\
        \textbf{Query 5:} \textcolor{red}{Advertisement of seat basket.} \\
        \textbf{Query 6:} \textcolor{red}{A video game is played.} \\
        \textbf{Query 7:} \textcolor{red}{A scene from spongebob squarepants where the townspeople are carrying torches and chasing a giant squidward.} \\
        \textbf{Query 8:} \textcolor{red}{A woman applies makeup to her eyes in double speed.} \\
        \textbf{Query 9:} \textcolor{red}{A girl singing a song and her group were playing music.} \\
        \textbf{Query 10:} \textcolor{red}{Two guys are wrestling in a competition.} \\
        \textbf{Query 11:} \textcolor{red}{News of marijuana business having trouble growing.} \\
        \textbf{Query 12:} \textcolor{red}{Two people playing basketball and the one with a hat makes every shot.}
        \end{tabular}
    \end{minipage}
    \caption{Highest-frequency retrieved video in MSR-VTT dataset with corresponding nearest-neighbor queries. \textcolor[rgb]{1.0, 0.0, 0.0}{Red} and \textcolor[rgb]{0.0, 1.0, 0.0}{green} lines indicate semantically matched and mismatched queries, respectively.}
    \label{fig:video_with_queries}
\end{figure}

Figure \ref{fig:t2i_comp_2}-\ref{fig:video_with_queries} shows detailed results of our main paper. 

% \input{appendix/tables/dbsn}

% figure 1
%% we have to use this in the main tex file
% \begin{figure}[htbp]
%     \textit{\textbf{Query}: Two blond women sit outside as people walk by wearing casual clothing and some wearing bookbags}.\\
%     \begin{minipage}{\textwidth}
%         \begin{minipage}{0.02\textwidth}
%             \rotatebox{90}{\textbf{Baseline}}
%         \end{minipage}
%         \begin{minipage}{0.47\textwidth}
%             \includegraphics[width=\textwidth]{appendix/figures/t2i_1-1.png}
%         \end{minipage}
%     \end{minipage}

%     \begin{minipage}{\textwidth}
%         \begin{minipage}{0.025\textwidth}
%             \rotatebox{90}{\textbf{IS}}
%         \end{minipage}%
%         \begin{minipage}{0.47\textwidth}
%             \includegraphics[width=\textwidth]{appendix/figures/t2i_1-2.png}
%         \end{minipage}
%     \end{minipage}

%     \begin{minipage}{\textwidth}
%         \begin{minipage}{0.02\textwidth}
%             \rotatebox{90}{\textbf{SN}}
%         \end{minipage}
%         \begin{minipage}{0.47\textwidth}
%             \includegraphics[width=\textwidth]{appendix/figures/t2i_1-3.png}
%         \end{minipage}
%     \end{minipage}
    
%     \caption{Visual comparisons of Top-5 text-to-image retrieval results on Flickr30k. \textcolor[rgb]{1.0, 0.0, 0.0}{Red} and \textcolor[rgb]{0.0, 1.0, 0.0}{green} boxes indicate incorrect and correct recalls, respectively.}
%     \label{fig:t2i_comp_1}
% \end{figure}

% figure 2

% figure 5

% figure 5

% figure 5
% video 1
%% we have to use this in the main tex file
% \begin{figure}[htbp]
%     \textit{\textbf{Query}: Some one talking about top ten movies of the year}.\\
%     \begin{minipage}{\textwidth}
%         \begin{minipage}{0.02\textwidth}
%             \rotatebox{90}{\textbf{Baseline}}
%         \end{minipage}
%         \begin{minipage}{0.47\textwidth}
%             \includegraphics[width=\textwidth]{appendix/figures/t2v_1-1.png}
%         \end{minipage}
%     \end{minipage}

%     \begin{minipage}{\textwidth}
%         \begin{minipage}{0.025\textwidth}
%             \rotatebox{90}{\textbf{IS}}
%         \end{minipage}%
%         \begin{minipage}{0.47\textwidth}
%             \includegraphics[width=\textwidth]{appendix/figures/t2v_1-2.png}
%         \end{minipage}
%     \end{minipage}

%     \begin{minipage}{\textwidth}
%         \begin{minipage}{0.02\textwidth}
%             \rotatebox{90}{\textbf{SN}}
%         \end{minipage}
%         \begin{minipage}{0.47\textwidth}
%             \includegraphics[width=\textwidth]{appendix/figures/t2v_1-3.png}
%         \end{minipage}
%     \end{minipage}
    
%     \caption{Visual comparisons of Top-5 text-to-video retrieval results on MSR-VTT. \textcolor[rgb]{1.0, 0.0, 0.0}{Red} and \textcolor[rgb]{0.0, 1.0, 0.0}{green} boxes indicate incorrect and correct recalls, respectively.}
%     \label{fig:t2v_comp_1}
% \end{figure}

% video 2

% video 5

% figure 5

% figure 5

%% we have to use this in the main tex file
% \begin{figure}[htbp]
%     \centering
%     \includegraphics[width=\linewidth]{appendix/figures/f30k_mat.png} 
%     \includegraphics[width=\linewidth]{appendix/figures/f30k_dis.png} 
%     \caption{Impact of normalization strategies on similarity matrix and k-occurrence distribution on Flickr30K.}
%     \label{fig:img_sim&mat}
% \end{figure}

% \textcolor{red}{a student explains to his teacher about the sheep of another student.} \\
% \textcolor{red}{there is a man shooting other people in a corridor.} \\
% \textcolor{red}{a man is giving his commentary on a current event television show.} \\
% \textcolor{green}{there was a resistor in the back.} \\
% \textcolor{red}{advertisement of seat basket.} \\
% \textcolor{red}{a video game is played.} \\
% \textcolor{red}{a scene from spongebob squarepants where the townspeople are carrying torches and chasing a giant squidward.} \\
% \textcolor{red}{a woman applies makeup to her eyes in double speed.} \\
% \textcolor{red}{a girl singing a song and her group were playing music.} \\
% \textcolor{red}{two guys are wrestling in a competition.} \\
% \textcolor{red}{news of marijuana business having trouble growing.} \\
% \textcolor{red}{two people playing basketball and the one with a hat makes every shot.}

\end{appendix}

\clearpage
%%
%% The next two lines define the bibliography style to be used, and
%% the bibliography file.
\bibliographystyle{ACM-Reference-Format}
\bibliography{bibs/hubness}

%%
%% If your work has an appendix, this is the place to put it.
\end{document}